\documentclass[a4paper,12pt]{article}
\usepackage{jcappub}
\usepackage{amsthm,graphicx,multirow}
\usepackage{epsfig}
\usepackage{latexsym, amssymb} 
\usepackage{amsmath}
\hypersetup{
    unicode=false,          
    pdftoolbar=true,        
    pdfmenubar=true,        
    pdffitwindow=false,     
    pdfstartview={FitH},    
    pdftitle={My title},    
    pdfauthor={Author},     
    pdfsubject={Subject},   
    pdfcreator={Creator},   
    pdfproducer={Producer}, 
    pdfkeywords={keyword1} {key2} {key3}, 
    pdfnewwindow=true,      
    colorlinks=true,       
    linkcolor=red,          
    citecolor=cyan,        
    filecolor=magenta,      
    urlcolor=green,           
    linktocpage=true
}

\def\beq{\begin{equation}}
\def\eeq{\end{equation}}
\def\br{\begin{eqnarray}}
\def\er{\end{eqnarray}}
\def\benu{\begin{enumerate}}
\def\efnu{\end{enumerate}}

\def\deg{^{\circ}}

\begin{document}
\begin{center}
\title{Search for a direction in the forest of Lyman-$\alpha$} 
\end{center}
\author[a]{Dhiraj Kumar Hazra,}  
\author[b]{Arman Shafieloo}

\affiliation[a]{Asia Pacific Center for Theoretical Physics, Pohang, Gyeongbuk 790-784, Korea}
\affiliation[b]{Korea Astronomy and Space Science Institute 776, Daedeokdae-ro, Yuseong-gu, Daejeon 305-348, Korea}
\emailAdd{dhiraj@apctp.org, shafieloo@kasi.re.kr} 

\abstract{We report the first test of  isotropy of the Universe in the matter dominated epoch using
 the Lyman-$\alpha$ forest data from the high redshift quasars ($z>2$) from SDSS-III BOSS-DR9 datasets. 
Using some specified data cuts, we obtain the probability distribution function (PDF) of the 
Lyman-$\alpha$ forest transmitted flux and use the statistical moments of the PDF to address the 
isotropy of the Universe. In an isotropic Universe one would expect the transmitted flux  to have consistent
statistical characteristics in different parts of the sky. We trisect the total survey area of 3275 ${\rm deg}^2$ along the galactic latitude 
and using quadrant convention. We also make three subdivisions in the data for three different signal-to-noise-ratios (SNR).
Finally we obtain and compare the statistical moments in the mean redshifts of 2.3, 2.6 and 2.9. 
We find, that the moments from all patches agree at all redshifts and at all SNRs,   within 3$\sigma$ uncertainties. Since Lyman-$\alpha$
transmitted flux directly maps the neutral hydrogen distribution in the inter galactic medium (IGM), our results indicate, 
within the limited survey area and sensitivity of the data, the distribution
of the neutral hydrogen in the Universe is consistent with isotropic
distribution.  We should mention that we report few deviations from
isotropy in the data with low statistical significance. Increase in
survey area and larger amount of data are needed to make any strong
conclusion about these deviations.
}

\maketitle
\section{Introduction}
Our understanding of the physical cosmology strongly depends on the data that we observe and the model assumptions 
that we make. Assumptions in the standard model sometime pose problems in understanding the true nature of the 
Universe. Statistical isotropy is one of such assumptions which we assume in almost all the data analysis. In this assumption
we claim that the statistical properties of the observables in the Universe are same in all direction. While in parameter 
estimations from the datasets, isotropy is assumed, there have been a number of tests to confirm this assumption, using 
data from different cosmological probes, such as Cosmic Microwave Background (CMB)~\cite{Hinshaw:1996ut,Spergel:2003cb,
Copi:2010na,Ade:2013nlj,Akrami:2014eta,de OliveiraCosta:2003pu,Abramo:2006gw,Land:2005ad,Land:2006bn,Rakic:2007ve,
Samal:2007nw,Samal:2008nv,Eriksen:2007pc,Hoftuft:2009rq,Copi:2006tu,Copi:2005ff,Schwarz:2004gk,Souradeep:2006dz} and Large Scale 
Structure (LSS)~\cite{Fernandez-Cobos:2013fda,Cai:2013lja,Keenan:2009jh,Keenan:2012gr,Keenan:2013mfa,Whitbourn:2013mwa,
Frith:2003tb,Busswell:2003ta,Frith:2005az,Frith:2004wd,Frith:2004tw,Appleby:2014lra}. 
In this paper  we investigate isotropy  of the matter dominated Universe using the
 Lyman-$\alpha$ forest datasets in the redshift range $z \sim 2 -3$. Spectra of high redshift 
quasars contain absorption lines that trace the components of the IGM along the line of sight. In the case of Lyman-$\alpha$
forest, we find absorption lines from the first ionization state of the Hydrogen atoms. The wavelength of the absorption
identifies the redshift of the neutral hydrogen cloud. Analyzing these absorption lines, collectively from a number of 
quasars, we can constrain properties of the IGM. In particular, the Lyman-$\alpha$ transmitted
flux ($F$) (absorption {\it w.r.t} the estimated continuum spectrum) can be related to the dark matter overdensity 
($\delta$) as $F=\bar{F}\exp[-A(1+\delta)^{2-0.7(\gamma-1)}]$~\cite{fdelta}, where, $\bar{F}$ is the mean transmitted 
flux, $\gamma-1$ dictates the temperature density relation and $A$ is a redshift dependent constant. We shall address the statistics 
of the transmitted flux in the Lyman-$\alpha$ region at different redshifts. To have a theoretical model independent analysis, in this
work we only consider the statistical properties of the observational data. In the last decade, detection of
large number of quasar spectrum with high SNR enables us to perform this type of analysis. We use 
Baryon Oscillation Spectroscopic Survey (BOSS) DR9 from Sloan Digital Sky Survey (SDSS)-III~\cite{sdss,boss}. We perform our test 
in three different redshifts in the matter dominated epoch that enables us to track the isotropy along time. In each 
redshift, the isotropy is tested in medium, high and highest SNR of detection. We divide the sky
in three patches using two different patch selection criteria. Since SDSS is a ground base survey which only has partial sky 
coverage, our test of isotropy will be limited by the survey area. Throughout the analysis we closely follow the 
survey parameters for the selection of data pixels from the quasar spectrum in order to obtain the 
PDF of the transmitted flux.

The paper is organized as follows : In section~\ref{sec:formalism} we discuss the data from BOSS-DR9 and the selection 
criteria for the data pixels for our analysis. We also outline the error estimation procedure for the flux PDF. 
Afterwards we discuss and tabulate the properties of the sky patches selected. In the results section~\ref{sec:results}
we discuss the main outcome of our analysis. Finally in section~\ref{sec:conclusions} we conclude along with 
highlighting future prospects.

\section{Data analysis}~\label{sec:formalism}

We use the latest available BOSS DR9 quasar Lyman-$\alpha$ forest data~\cite{Lyman-sample}. The ninth data release
contains 54,468 spectra of Lyman-$\alpha$ quasars with redshifts more than 2.15. They provide Lyman-$\alpha$ forest data in the redshift
range $z\sim2-5.7$. However due to a smaller number of quasars with redshift higher than $z\sim3$ we restrict our analysis 
in the redshift $z\sim2-3$. BOSS-DR9 has a survey area of 3275 square degrees and hence our test of isotropy will be 
limited by this area.
As has been discussed in~\cite{Lyman-sample}, out of 87,822 total spectra, the set of 54,468 spectra 
was selected after removing low redshift quasars, quasars having broad absorption lines, too low SNR and negative continuum.
Note that the continuum to each quasar spectrum is estimated using mean-flux regulated principal component analysis (MF-PCA) 
techniques~\cite{Leeetal} and they are provided along with the data. We would like to mention that even at this stage not all the 
data are appropriate for the isotropy test due to low SNR, damped Lyman-$\alpha$ (DLA), limited exposures and few other
characteristics. Below we point out the data cuts that we have used to select a spectrum.


\subsection{Quasar selection, data cuts}
Our selection of data closely follows the BOSS criteria used to obtain constraints on the IGM~\cite{bossigm}. As we mentioned earlier, 
we test the isotropy of the sky in three redshift bins with the central redshifts being 2.3, 2.6 and 2.9. The redshift bins have a bin width of 
$\Delta z=0.3$. In each of the redshift bins we impose the following data cuts. Firstly since large number of low SNR quasars can have a systematic bias 
in our results, we reject all the quasars with SNR $<6$. The rest of the quasars are binned in {\it good} ($6\le{\rm SNR}<8$), 
{\it better} ($8\le{\rm SNR}<10$) and {\it best} (${\rm SNR}\ge10$) categories. In the quasar rest frame we use 
$1041-1185\mathring{A}$ as the Lyman-$\alpha$ domain. 

Lyman-$\alpha$ forest typically refers to neutral hydrogen atoms with column density of $10^{14}$ atoms per ${\rm cm}^2$ in the 
line of sight.  The presence of DLA in a spectrum indicates extremely high column density (${\cal O} (10^{20})$ atoms per 
${\rm cm}^2$).  Following~\cite{bossigm} we discard spectra with identified DLA in the sightlines. The provided data uses 
a DLA concordance catalogue for the identification of DLA's.  The detection efficiency of the catalogue decreases below 
neutral hydrogen column densities  less than $10^{20.3}/{\rm cm}^2$.  Hence, below this criteria, in our analysis, for each of the forest we 
correct the flux for the damping wings of DLA in the sightlines, provided with the data~\cite{Lyman-sample}.  
We leave the spectra with neutral hydrogen column density of more than $10^{20.3}/{\rm cm}^2$ out of our analysis.

As mentioned earlier, the survey provides the continuum of each spectrum estimated using MF-PCA technique. Note that 
the transmitted flux is the observed flux {\it w.r.t.} the estimated continuum and hence a bad continuum estimation 
shall bias the calculated flux~\footnote{The effects of different continuum estimations are discussed in~\cite{Busca:2012bu}}. 
Here too, following the survey criteria we reject the quasars which do not provide
good fit to the spectrum redwards to the Lyman-$\alpha$ forest. We also reject spectra which had 
less than three individual exposures.

Our final selection criteria depends on the resolution of the BOSS spectographs. As has been mentioned in~\cite{bossigm},
BOSS spectographs do not resolve the Lyman-$\alpha$ forest. To evade this systematic effect that can reflect in the 
transmitted flux, we use similar procedure used in~\cite{bossigm}. We stack the spectra according to the intrinsic 
wavelength dispersion ($\sigma_{\rm disp}$) at the Lyman-$\alpha$ wavelength of each central redshift bin. Note that in each pixel 
the $\sigma_{\rm disp}$ is provided along with the spectrum. From each redshift bin we then discard the 5$\%$ spectra
from below and 10$\%$ spectra from above.

Each quasar spectrum contain certain number of data pixels in the Lyman-$\alpha$ region. In a redshift bin (say, $2.15<z<2.45$)
some quasars may have all the Lyman-$\alpha$ pixels, some will have partial pixels from either blue end or red end of the 
Lyman-$\alpha$ domain. We here consider all the quasars satisfying the above criteria and also contributing more than 
30 pixels in the redshift window. 

\subsection{Selection of sky patches}
\begin{figure*}[!htb]
\begin{center} 
\resizebox{230pt}{120pt}{\includegraphics{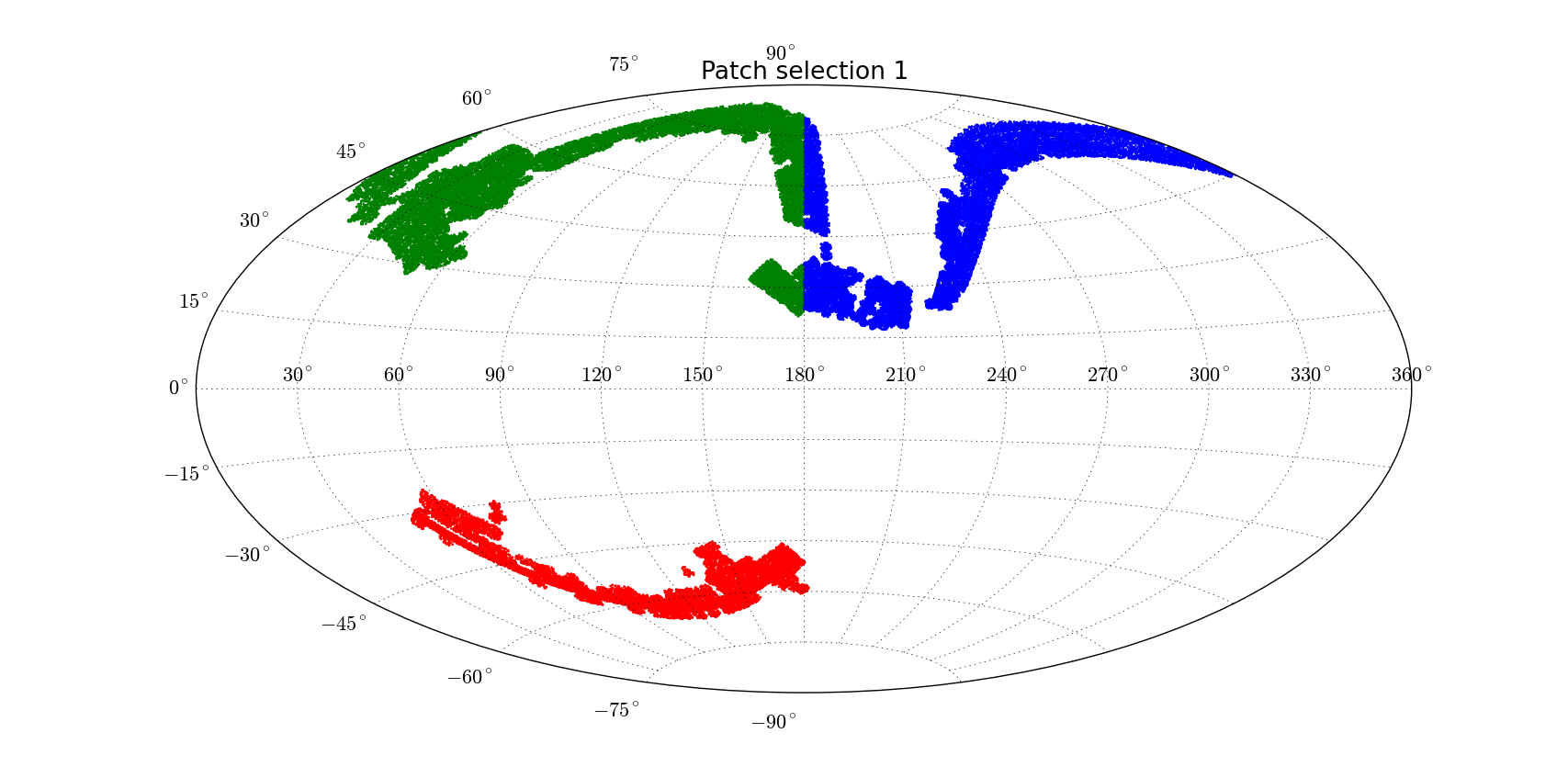}}\hskip -22 pt 
\resizebox{230pt}{120pt}{\includegraphics{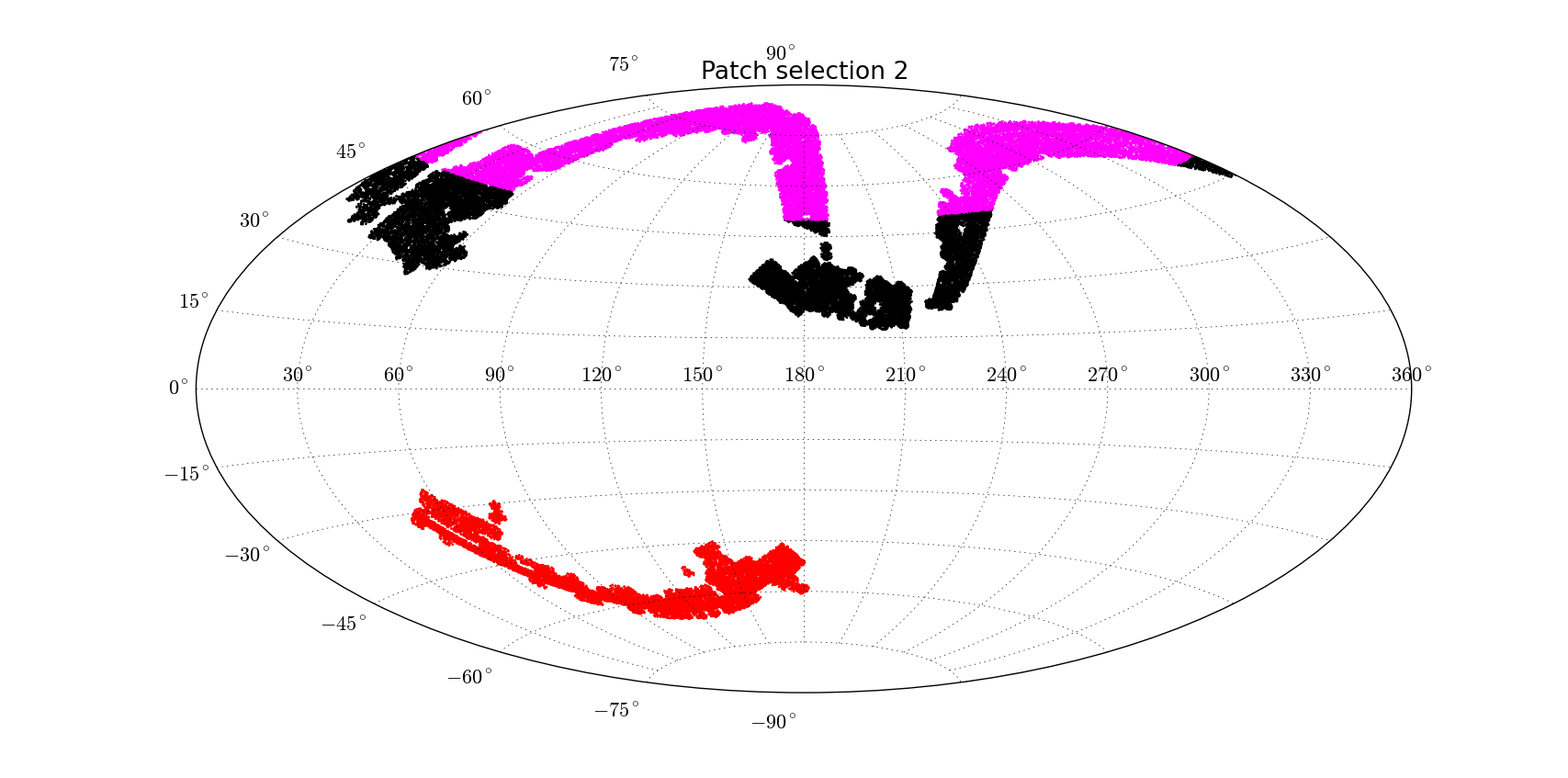}} 
\end{center}
\caption{\footnotesize\label{fig:skycut} The BOSS-DR9 survey area in galactic coordinates and our 
selection of sky patches for the test of isotropy. To accommodate comparable number of quasars in each 
patch we have divided the sky in two manner, namely patch selection 1 (left) and patch selection 2 (right).}
\end{figure*}

Since BOSS-DR9 surveys in 3275 ${\rm deg}^2$ area, we need to make patches in the sky in particular ways. First we 
convert the coordinate of the quasars from J2000 equatorial system to galactic coordinate ($l,b$) system. Below, in 
Fig.~\ref{fig:skycut} we provide our selection of patches with different colors. In the first system we divide the sky  
using quadrant convention.  Here we divide the sky in three patches where one patch remains in the southern 
hemisphere (red, $b<0\deg$) and the other two patches are in northern 
hemisphere ($b>0\deg$). The two patches in the northern hemisphere are divided in $l>180\deg$ (blue) and  
$l<180\deg$ (green) parts. In principle, the quadrant convention divides the southern hemisphere into two parts  
($l>180\deg$ and  $l<180\deg$) too, but since there are barely any data in  $b<0\deg,l>180\deg$, we take only one patch 
from southern hemisphere. In our other selection, we divide the sky according 
to galactic latitude. The red patch remains same as it stays in the southern hemisphere. The patches in the northern 
hemisphere are divided in one patch with $b>50\deg$ (magenta) and another with $0\deg<b\le50\deg$ (black). Our first selection 
of patches is provided in the left of Fig.~\ref{fig:skycut} and the other selection is shown to the right.  

In the following table~\ref{tab:patch-info} we tabulate the essential information about the selected patches and the 
overall sample in different redshift and different SNR. For different patches we provide the number of quasars and 
the number of pixels that pass all our data cuts discussed in the paragraph before. Note that apart from the red patch 
that stays in the southern hemisphere, other patches contain quasar number and the pixels that are comparable to each 
other. As we go towards higher redshifts we find lesser number of data pixels which is expected as we know there are not 
many high redshift quasars. This decrease in number of high redshift quasars limit our analysis to $z<3$. However, in 
each SNR bin we have enough pixels to perform a relatively robust statistical analysis. Also note that compared to $6\le {\rm SNR}<8$ 
bin, in the two higher SNR bins the numbers of quasars or pixels are approximately half. We should mention that this 
is not a concern since we assume the data from different SNR are independent and we compare the statistical properties
of the transmitted flux in each SNR separately. Moreover the lack of large number of quasars in higher SNR bins are 
compensated by the less dispersion in the data from higher SNR quasar samples.
\bgroup
\renewcommand{\arraystretch}{1.1}
\begin{table*}[!t]
\begin{scriptsize} 
\begin{center}
\begin{tabular}{c | c | c | c | c}
\hline\hline
Redshift range($z$) & SNR & Patch location & Number of quasars & Number of Lyman-$\alpha$ pixels\\
\cline{1-5}
& & Complete sky& 1091&285878 \\
& & $b<0^\circ$ (Red)& 226& 60631\\
& & $b>0^\circ,~l<180^\circ$ (Green)&390 &100289 \\
& $6-8$ & $b>0^\circ,~l>180^\circ$ (Blue)&475 &124958 \\
& & $0^\circ<b<50^\circ$ (Black)&401 &106772 \\
& & $b>50^\circ$ (Magenta)&464 &118475 \\
\cline{2-5}
& & Complete sky&498 &130242 \\
& & $b<0^\circ$ (Red)&106 &28686 \\
& & $b>0^\circ,~l<180^\circ$ (Green)&185 &48158 \\
$2.15-2.45$ ($\bar{z}=2.3$) & $8-10$ &$b>0^\circ,~l>180^\circ$ (Blue) &207 &53398\\
& & $0^\circ<b<50^\circ$ (Black)&168 &45290 \\
& & $b>50^\circ$ (Magenta)&224 &56266 \\
\cline{2-5}
& & Complete sky&567 &147679 \\
& & $b<0^\circ$ (Red)&122 &30899 \\
& & $b>0^\circ,~l<180^\circ$ (Green)& 187&47806 \\
& $>10$ &$b>0^\circ,~l>180^\circ$ (Blue)&258 &68974\\
& & $0^\circ<b<50^\circ$ (Black)& 228& 62655\\
& & $b>50^\circ$ (Magenta)&217 & 54125\\
\cline{1-5}
& & Complete sky&975 &230165 \\
& & $b<0^\circ$ (Red)& 221&53062 \\
& & $b>0^\circ,~l<180^\circ$ (Green)&373 &86995 \\
& $6-8$ & $b>0^\circ,~l>180^\circ$ (Blue)&381 &90108 \\
& & $0^\circ<b<50^\circ$ (Black)&348 & 80712\\
& & $b>50^\circ$ (Magenta)& 406&96391 \\
\cline{2-5}
& & Complete sky&475 &107539 \\
& & $b<0^\circ$ (Red)&110 &25099 \\
& & $b>0^\circ,~l<180^\circ$ (Green)& 181&41564 \\
$2.45-2.75$($\bar{z}=2.6$) & $8-10$ &$b>0^\circ,~l>180^\circ$ (Blue) &184 &40876\\
& & $0^\circ<b<50^\circ$ (Black)& 141& 32220\\
& & $b>50^\circ$ (Magenta)&224 &50220 \\
\cline{2-5}
& & Complete sky&607 &144797 \\
& & $b<0^\circ$ (Red)&139 &33690 \\
& & $b>0^\circ,~l<180^\circ$ (Green)&224 &54156 \\
& $>10$ &$b>0^\circ,~l>180^\circ$ (Blue)&244 &56951\\
& & $0^\circ<b<50^\circ$ (Black)&248 & 56434\\
& & $b>50^\circ$ (Magenta)&220 &54673 \\
\cline{1-5}
& & Complete sky&628 &139373 \\
& & $b<0^\circ$ (Red)&135 &30291 \\
& & $b>0^\circ,~l<180^\circ$ (Green)&250 &57698 \\
& $6-8$ & $b>0^\circ,~l>180^\circ$ (Blue)&243 & 51384\\
& & $0^\circ<b<50^\circ$ (Black)&225 &50412 \\
& & $b>50^\circ$ (Magenta)&268 &58670 \\
\cline{2-5}
& & Complete sky&372 &87257 \\
& & $b<0^\circ$ (Red)& 75&17879 \\
& & $b>0^\circ,~l<180^\circ$ (Green)& 150&35068 \\
$2.75-3.05$ ($\bar{z}=2.9$)& $8-10$ &$b>0^\circ,~l>180^\circ$ (Blue) &147 &34310\\
& & $0^\circ<b<50^\circ$ (Black)&112 &25351 \\
& & $b>50^\circ$ (Magenta)& 185& 44027\\
\cline{2-5}
& & Complete sky&432 &97826 \\
& & $b<0^\circ$ (Red)&111 & 25645\\
& & $b>0^\circ,~l<180^\circ$ (Green)&174 & 37713\\
& $>10$ &$b>0^\circ,~l>180^\circ$ (Blue)&147 &34468\\
& & $0^\circ<b<50^\circ$ (Black)&164 & 39660\\
& & $b>50^\circ$ (Magenta)&157 &32521 \\
\cline{1-5}
\end{tabular}
\end{center}
\caption{~\label{tab:patch-info}The number of quasars and the number of Lyman-$\alpha$ pixels that contribute 
in different bins and patches. We bin our samples in three different redshift bins with mean redshift being 
2.3, 2.6 and 2.9 and also in three signal-to-noise-ratio, namely $6\le{\rm SNR}<8$, $8\le{\rm SNR}<10$ and ${\rm SNR}\ge10$. The whole 
sky is divided in two different patch types (see, Fig.~\ref{fig:skycut}). Each type contains three different patches with different color
codes provided in this table.} 
\end{scriptsize}
\end{table*}
\egroup
\clearpage
\subsection{Error estimation}
In this paper, as we have mentioned earlier, we compare the statistical properties of the observed data in 
different direction of the sky without comparing with any theoretical model. To obtain the uncertainties/errors in the estimated PDF of the 
transmitted flux, we need to generate the covariance matrix associated with the flux PDF. Here too, we follow similar 
procedure as adopted in the BOSS analysis, in order to obtain the covariance matrix. We follow bootstrap resampling of 
the data in each bin. We have a total of 36 bins in both type of patch selection: 3 SNR bins, 3 redshift bins and 4 bins corresponding to 
each of the 3 patches and a complete sample. In each bin, we gather all the pixels that pass through our data cut and 
provided in Table~\ref{tab:patch-info}. We perform bootstrap resampling of chunks upto $100\mathring{A}$ wavelengths 
obtained from each quasar spectrum and generate 1000 realizations of the data. From these samples covariance matrix
of the flux PDF can be easily calculated. Using different realizations we did check the convergence of 
the diagonal terms of the covariance matrix and we could conclude that the choice of 1000 realizations has been conservative. 

We concentrate on the statistical properties of the PDF or the data. Hence we calculate different statistical moments 
such as mean, median, variance, skewness and kurtosis of the PDF. We report the moments of the distribution that we obtain 
from the original data as well as the upper and lower error bounds on each of the moments that we calculate from the bootstrap 
simulations. For distribution of each moment, we generate the empirical cumulative distribution function (ECDF) and report the 
(34.15\%) upper and lower error bounds, which we hereafter shall refer as 1$\sigma$ uncertainties in the distribution of the 
particular moment. 

Before discussing the isotropy test on patches in next section, in Table~\ref{tab:flux} we report the mean transmitted flux  
and 1$\sigma$ uncertainties of the flux PDF that we obtain from the complete sample in the three redshifts and in three SNR bins.
The number of quasars and the number of pixels that are used in our analysis are provided in Table~\ref{tab:patch-info} for 
the complete sample size. Note that as we go back in time (higher in redshift) the increase of neutral hydrogen
is evident from the decrease of the transmitted flux in all SNR. In each SNR, the error on the  
flux are comparable and that indicates, though at high SNR the signal is better, the less number of 
quasars in high SNR keeps the uncertainties comparable to the uncertainties at lower SNR which contain larger number of quasars 
(for example, compare the {\it good} and the {\it best} SNR cases). 
\bgroup
\renewcommand{\arraystretch}{1.2}
\begin{table*}[!htb]
\begin{center}
\vskip -15 pt
\begin{tabular}{c | c | c}
\hline\hline
Redshift range($z$) & SNR & $\bar{F}\pm\Delta F$\\
\cline{1-3}
& $6-8$&$0.826^{+0.154}_{-0.375}$\\
$2.15-2.45$ ($\bar{z}=2.3$)&$8-10$&$0.822^{+0.138}_{-0.405}$\\
& $>10$&$0.819^{+0.129}_{-0.487}$\\
\cline{1-3}
& $6-8$&$0.762^{+0.172}_{-0.39}$\\
$2.45-2.75$ ($\bar{z}=2.6$)&$8-10$&$0.758^{+0.159}_{-0.427}$\\
& $>10$&$0.756^{+0.152}_{-0.454}$\\
\cline{1-3}
& $6-8$&$0.69^{+0.191}_{-0.377}$\\
$2.75-3.05$ ($\bar{z}=2.9$)&$8-10$&$0.687^{+0.181}_{-0.396}$\\
& $>10$&$0.686^{+0.176}_{-0.413}$\\
\cline{1-3}
\end{tabular}
\end{center}
\caption{~\label{tab:flux}The flux statistics of the Lyman-$\alpha$ forest transmitted flux in different redshift
bins and different signal-to-noise bins, provided for the samples across the complete sky. The mean flux and the $1\sigma$ 
error bars on the PDF on both sides of the mean are provided.} 
\end{table*}
\egroup
\section{Analysis, results and discussions}~\label{sec:results}

In the previous section we mentioned that we trisect the survey area in two ways. 
One patch, that is in southern galactic hemisphere remains common 
to both the selection. For the 5 patches, that are color coded (red, green, blue, magenta and black) 
we obtain the PDF of transmitted flux. However, we compare the statistical properties of the PDF 
within each selection type.

\begin{figure*}[!htb]
\begin{center} 
\resizebox{140pt}{110pt}{\includegraphics{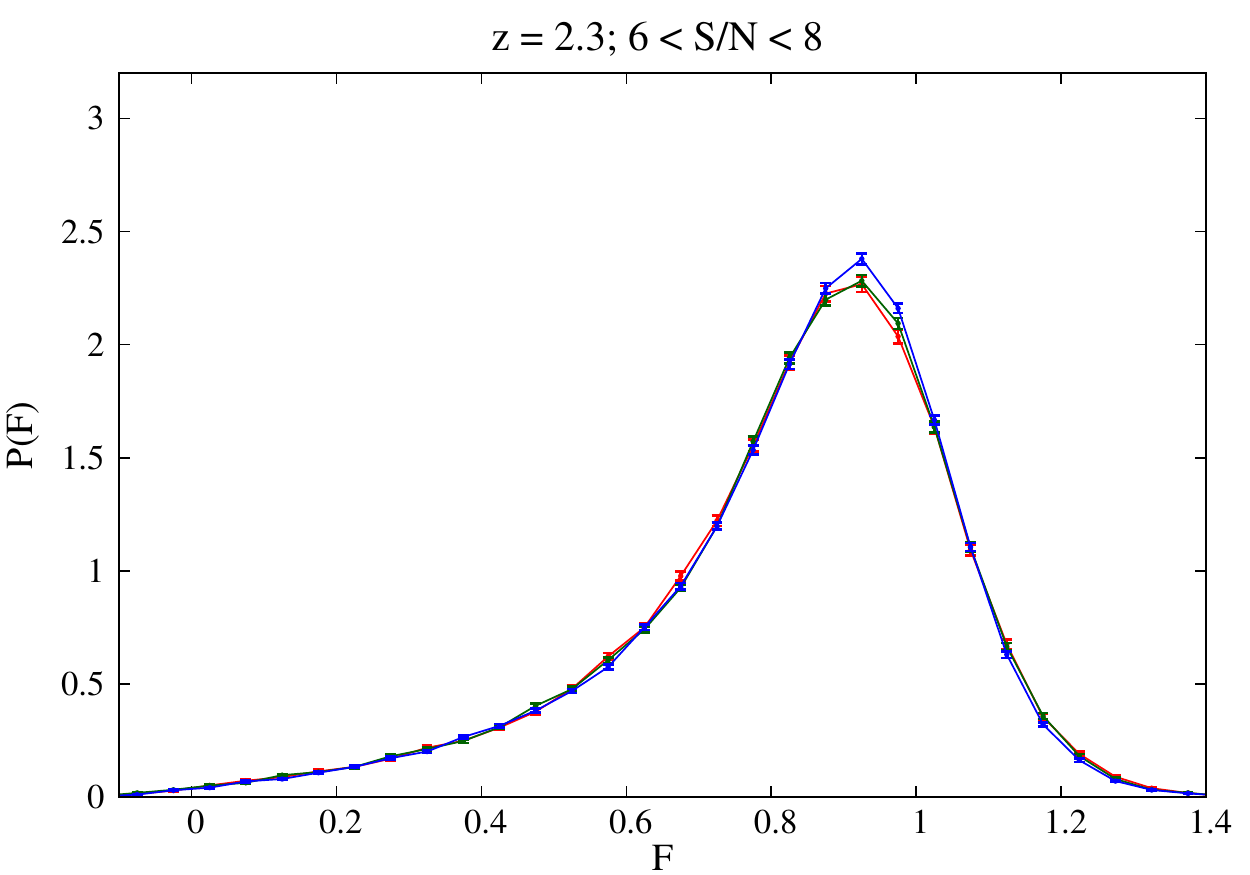}} 
\resizebox{140pt}{110pt}{\includegraphics{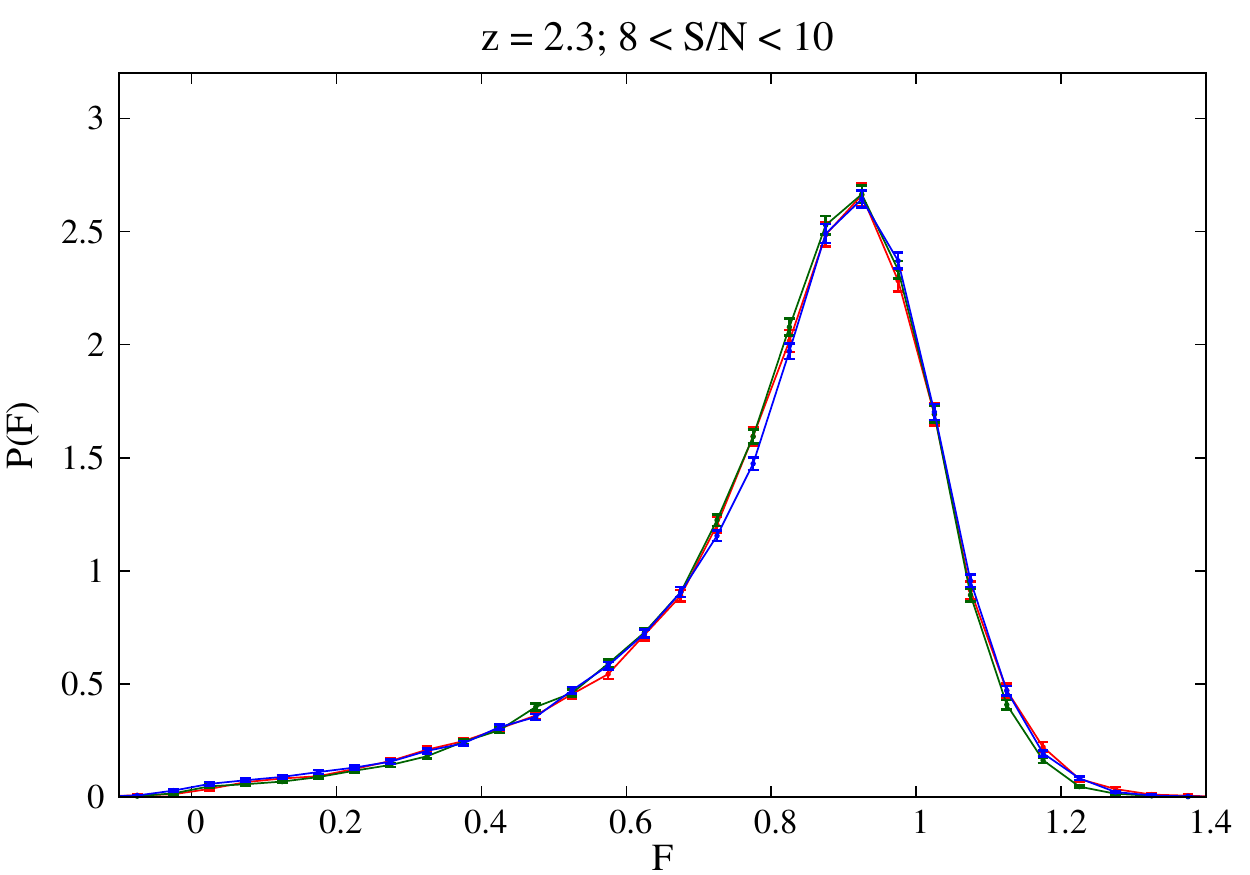}} 
\resizebox{140pt}{110pt}{\includegraphics{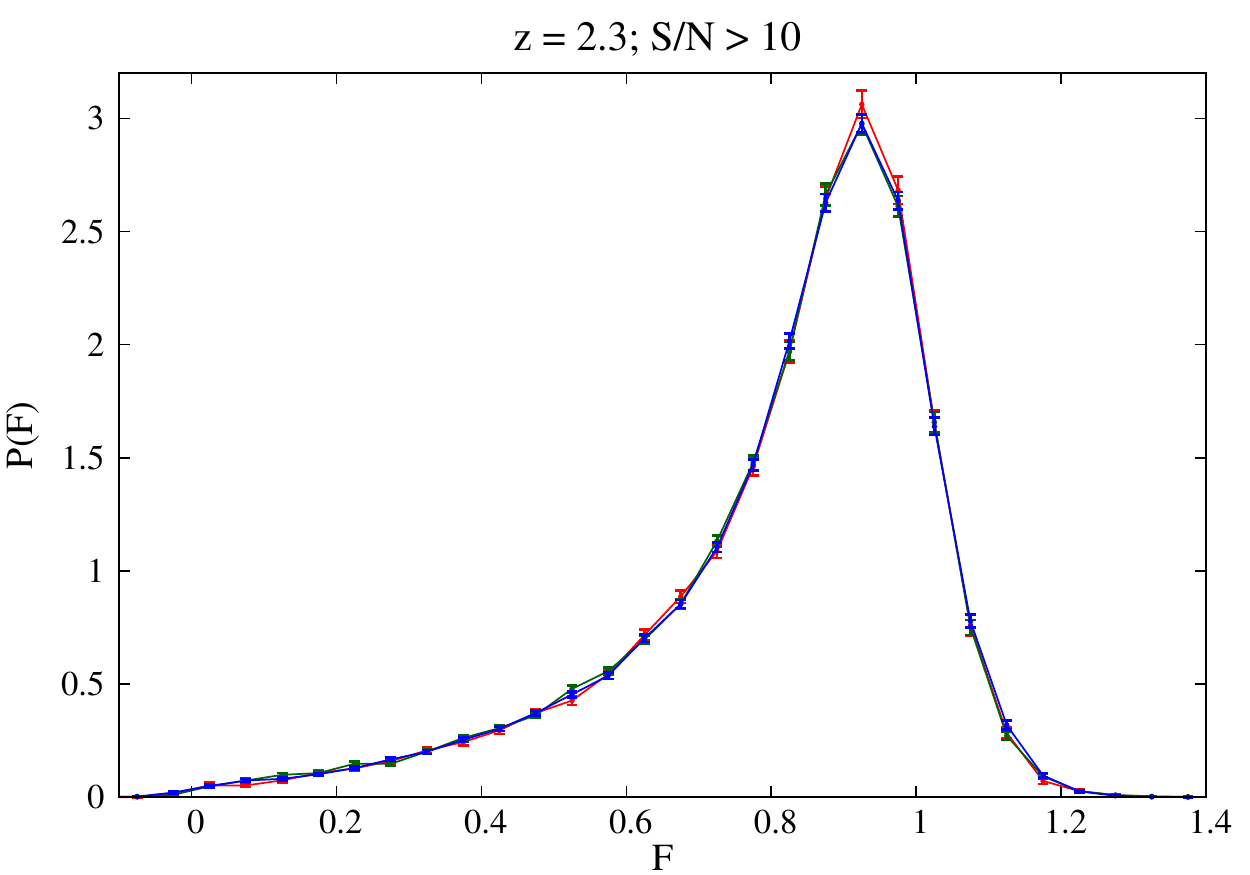}} 

\resizebox{140pt}{110pt}{\includegraphics{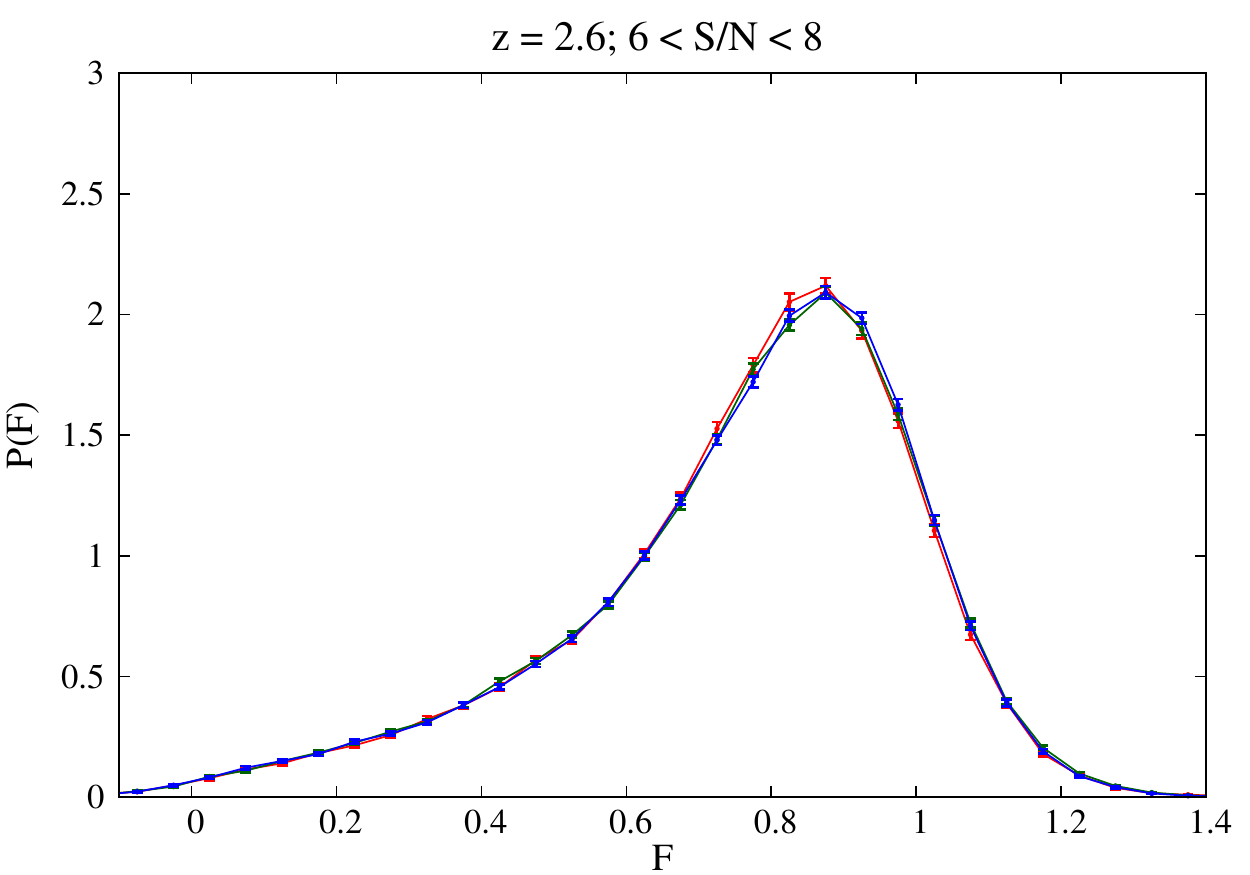}} 
\resizebox{140pt}{110pt}{\includegraphics{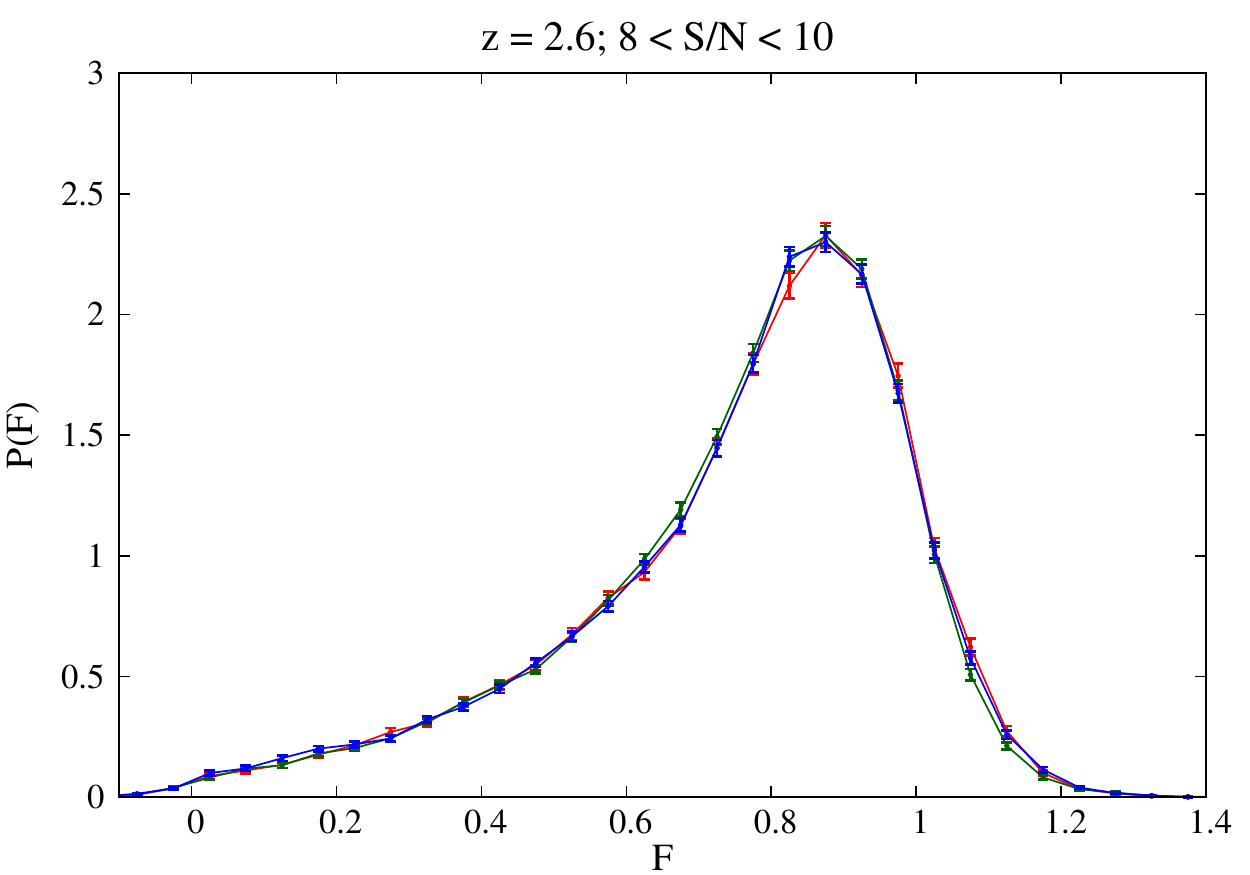}} 
\resizebox{140pt}{110pt}{\includegraphics{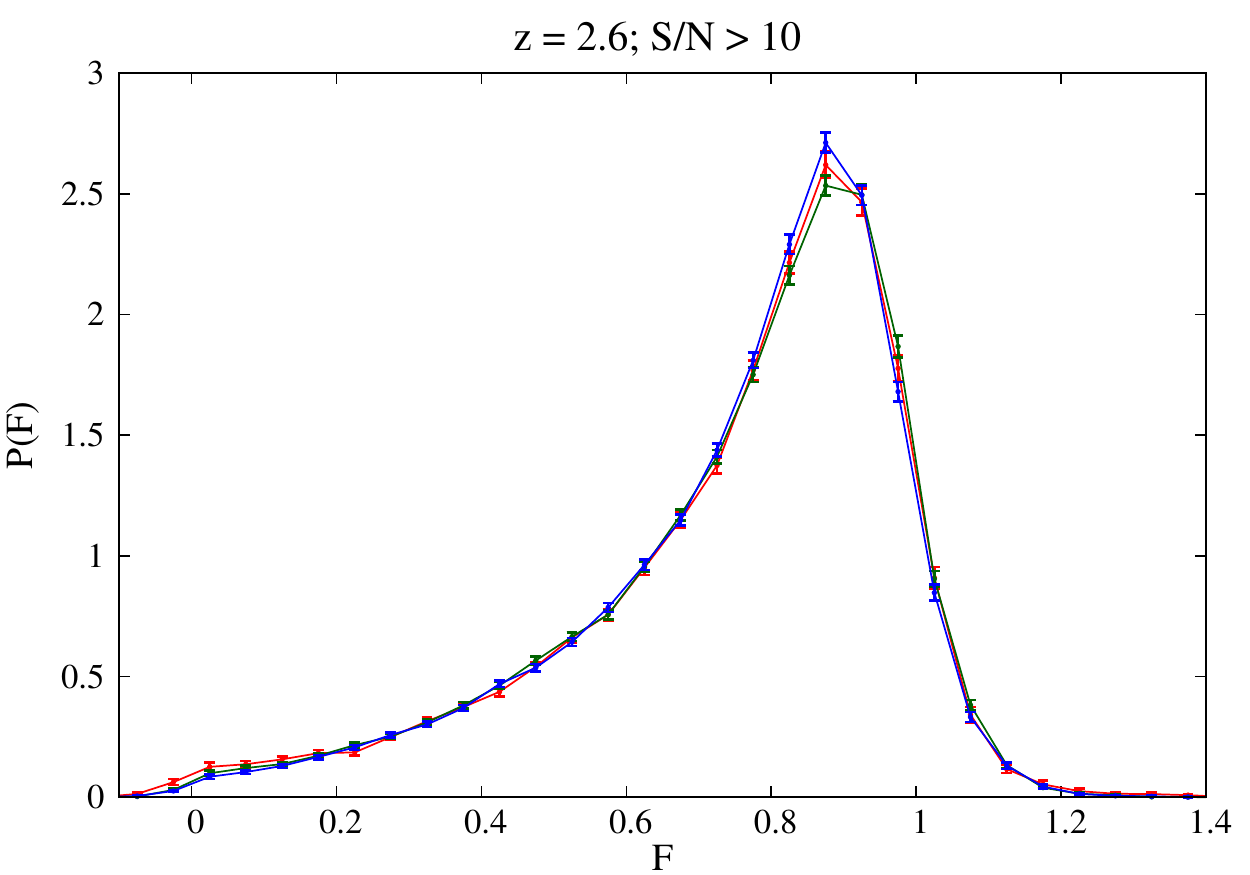}} 

\resizebox{140pt}{110pt}{\includegraphics{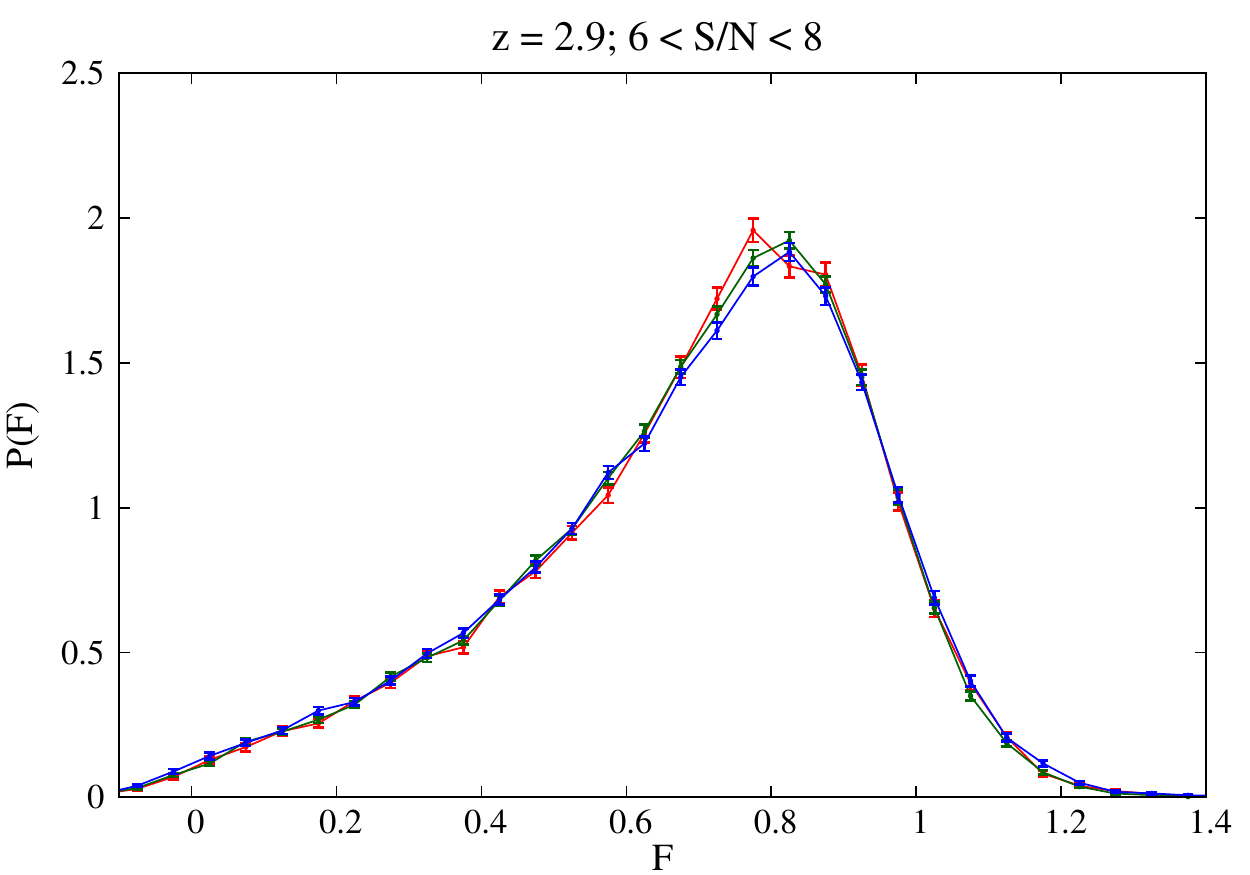}} 
\resizebox{140pt}{110pt}{\includegraphics{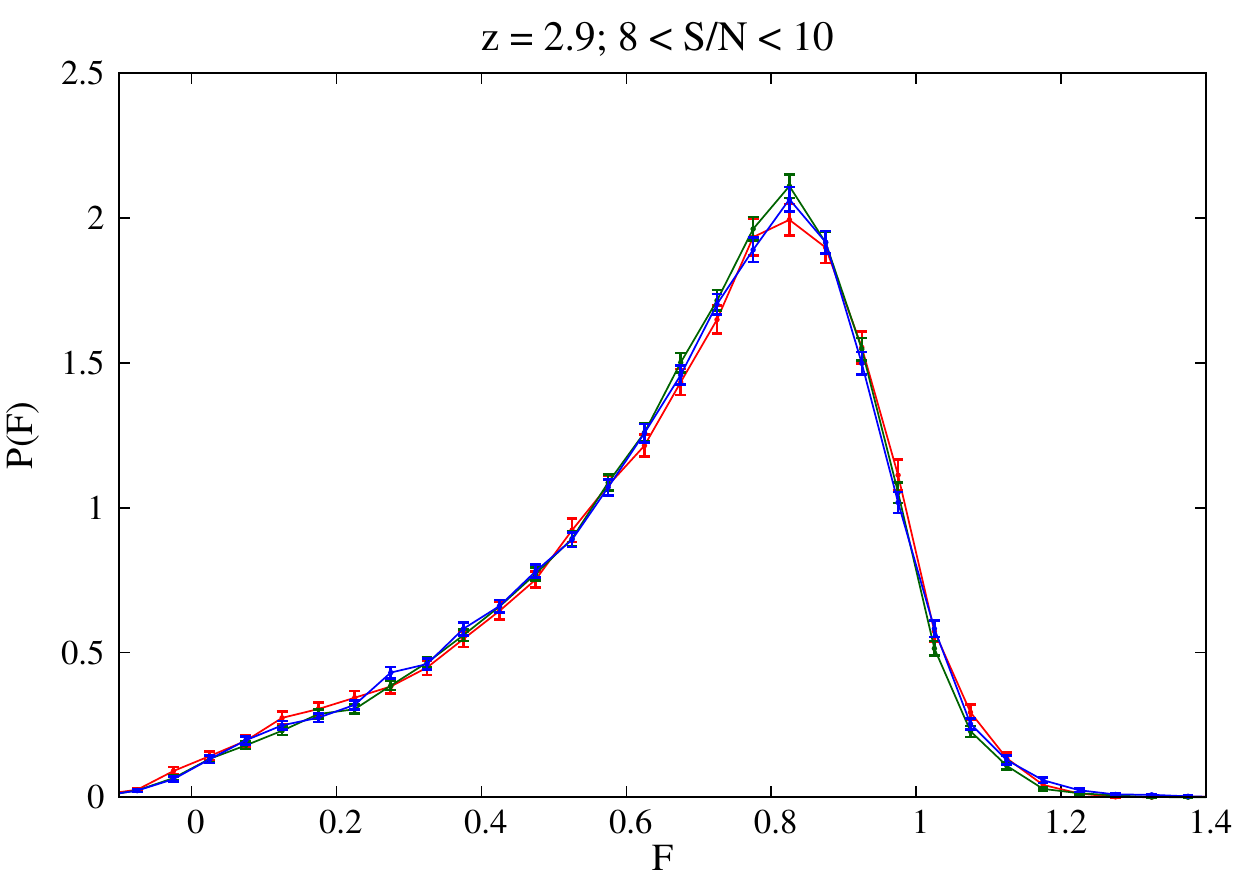}} 
\resizebox{140pt}{110pt}{\includegraphics{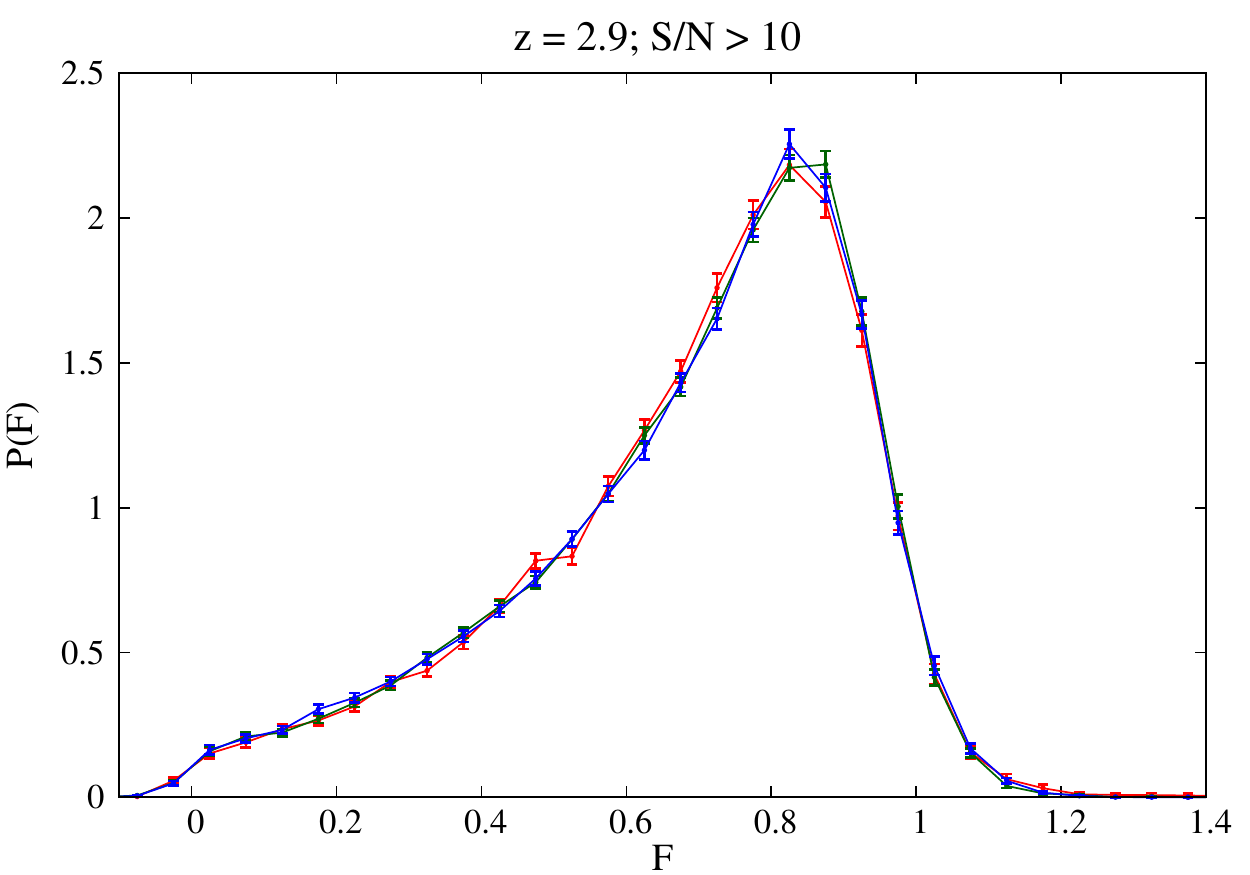}} 
\end{center}
\caption{\footnotesize\label{fig:comparison-PDF1}Comparison of PDF of the Lyman-$\alpha$ transmitted flux in different
patches for patch selection 1 (left of Fig.~\ref{fig:skycut}). The color codes represent the PDF's from the corresponding
patches. Along the rows we plot PDF's for different redshifts and along the columns we plot for different SNR. The error 
on the PDF is estimated through bootstrap resampling over $100\mathring{A}$ data chunks.} 
\end{figure*}

\begin{figure*}[!htb]
\begin{center} 
\resizebox{140pt}{110pt}{\includegraphics{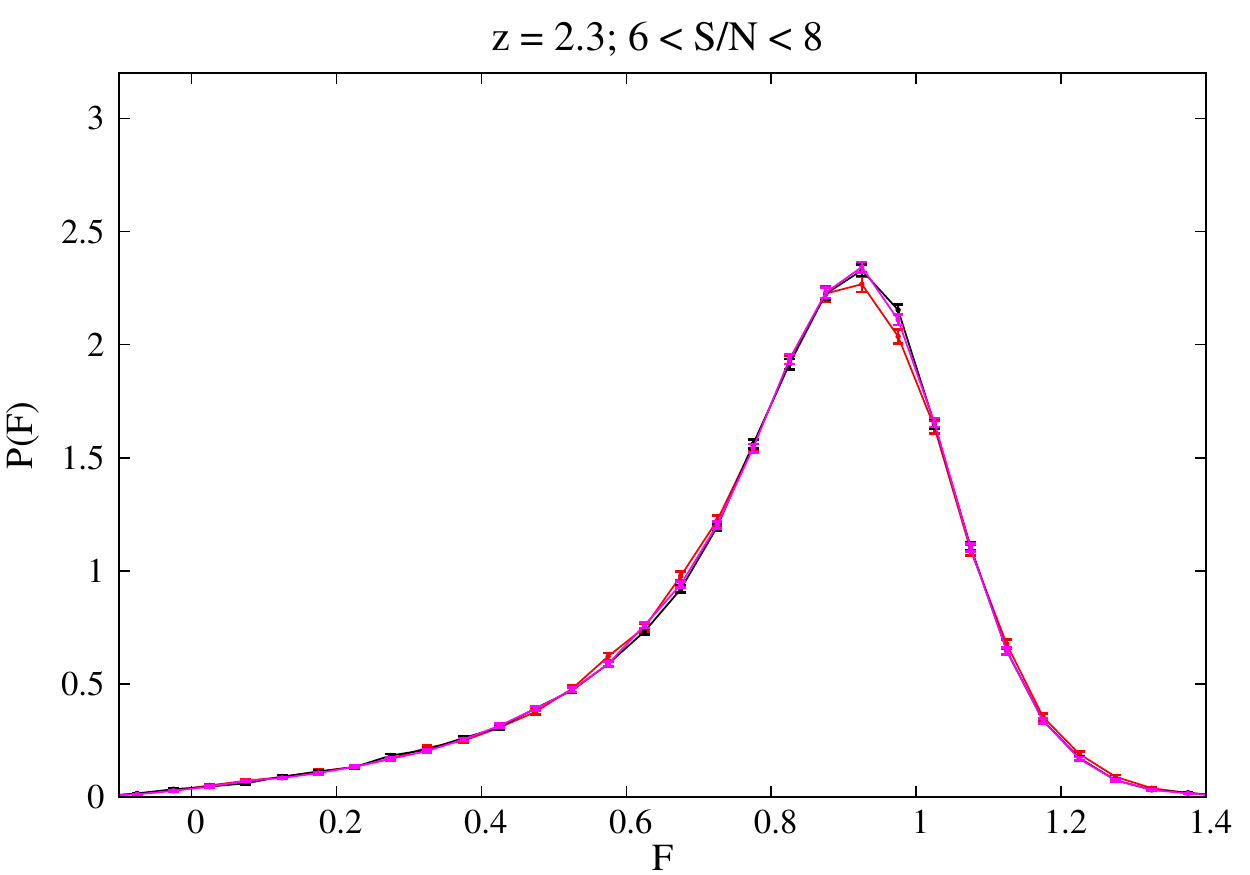}} 
\resizebox{140pt}{110pt}{\includegraphics{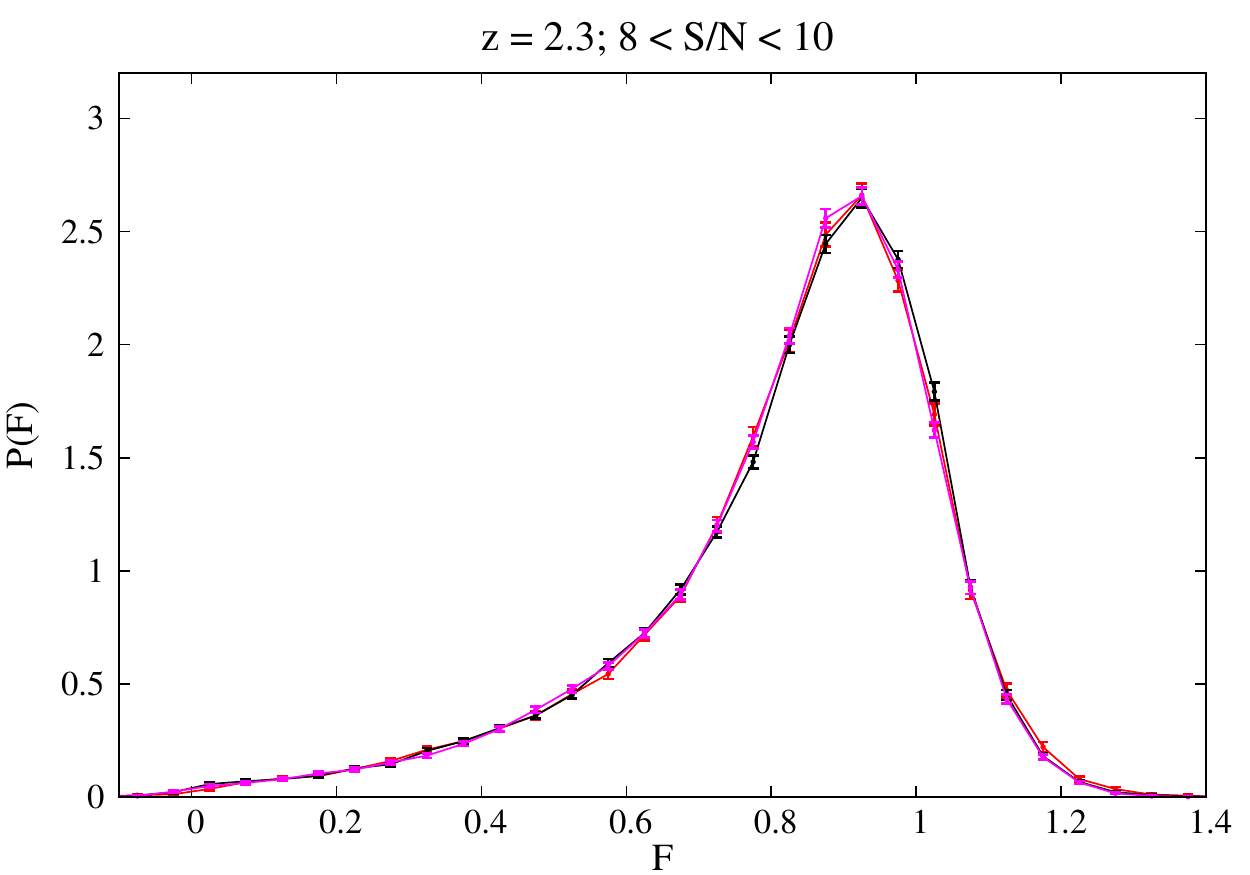}} 
\resizebox{140pt}{110pt}{\includegraphics{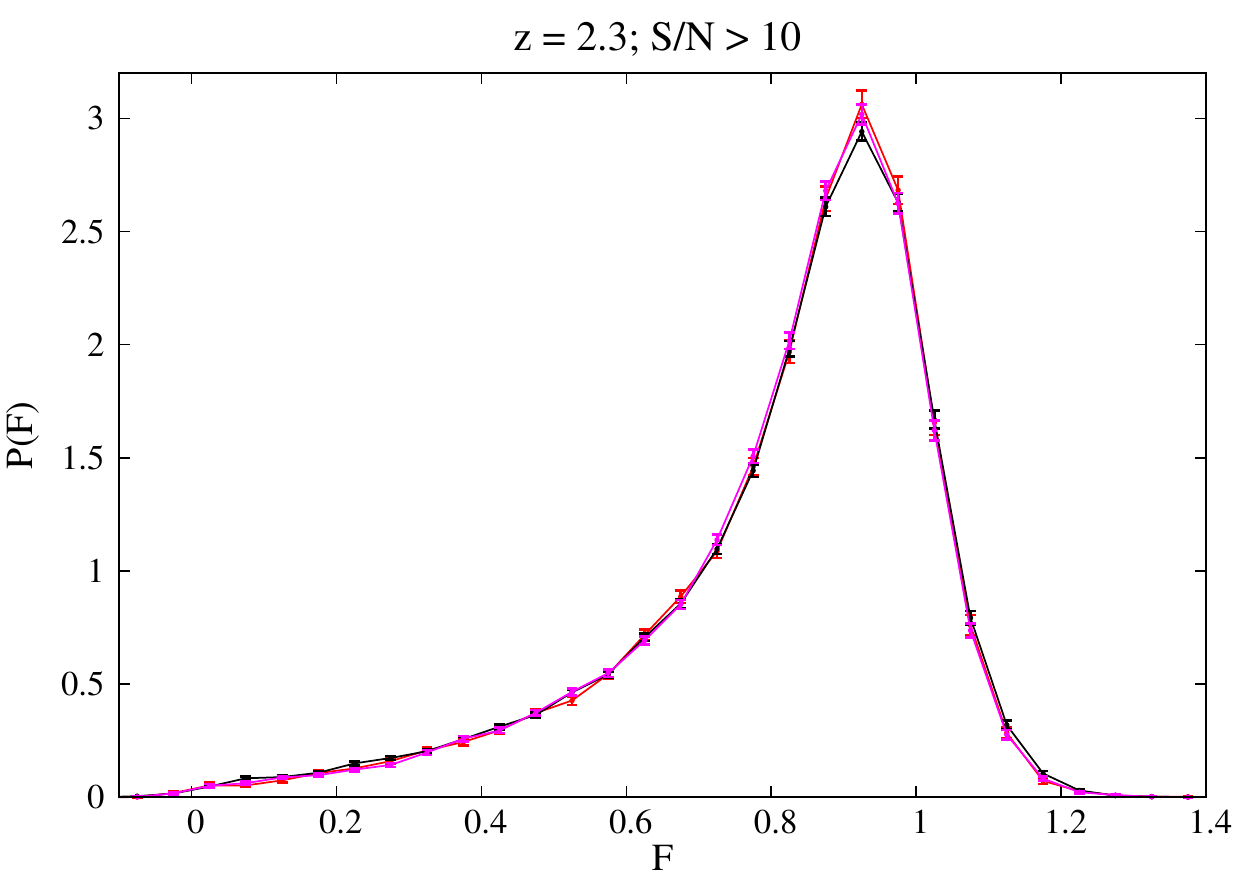}} 

\resizebox{140pt}{110pt}{\includegraphics{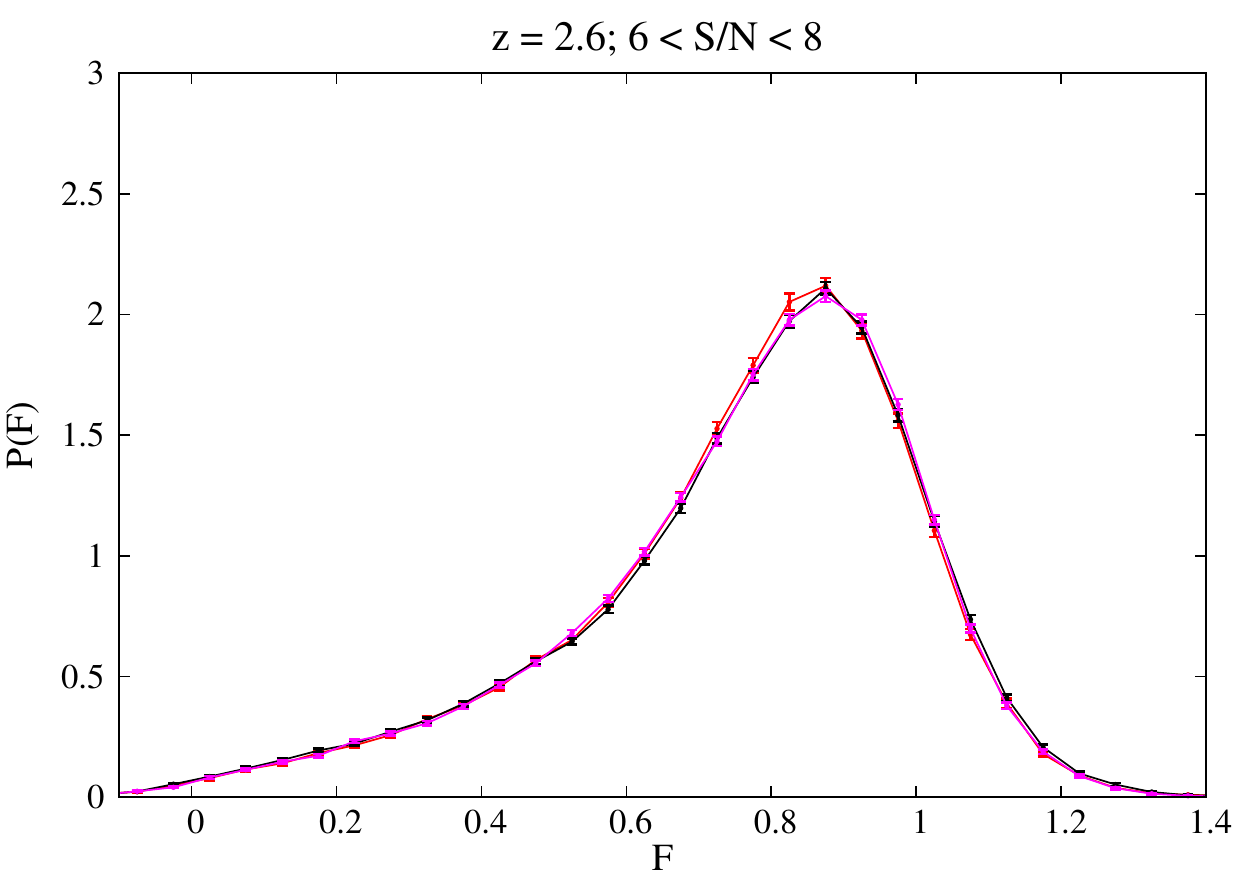}} 
\resizebox{140pt}{110pt}{\includegraphics{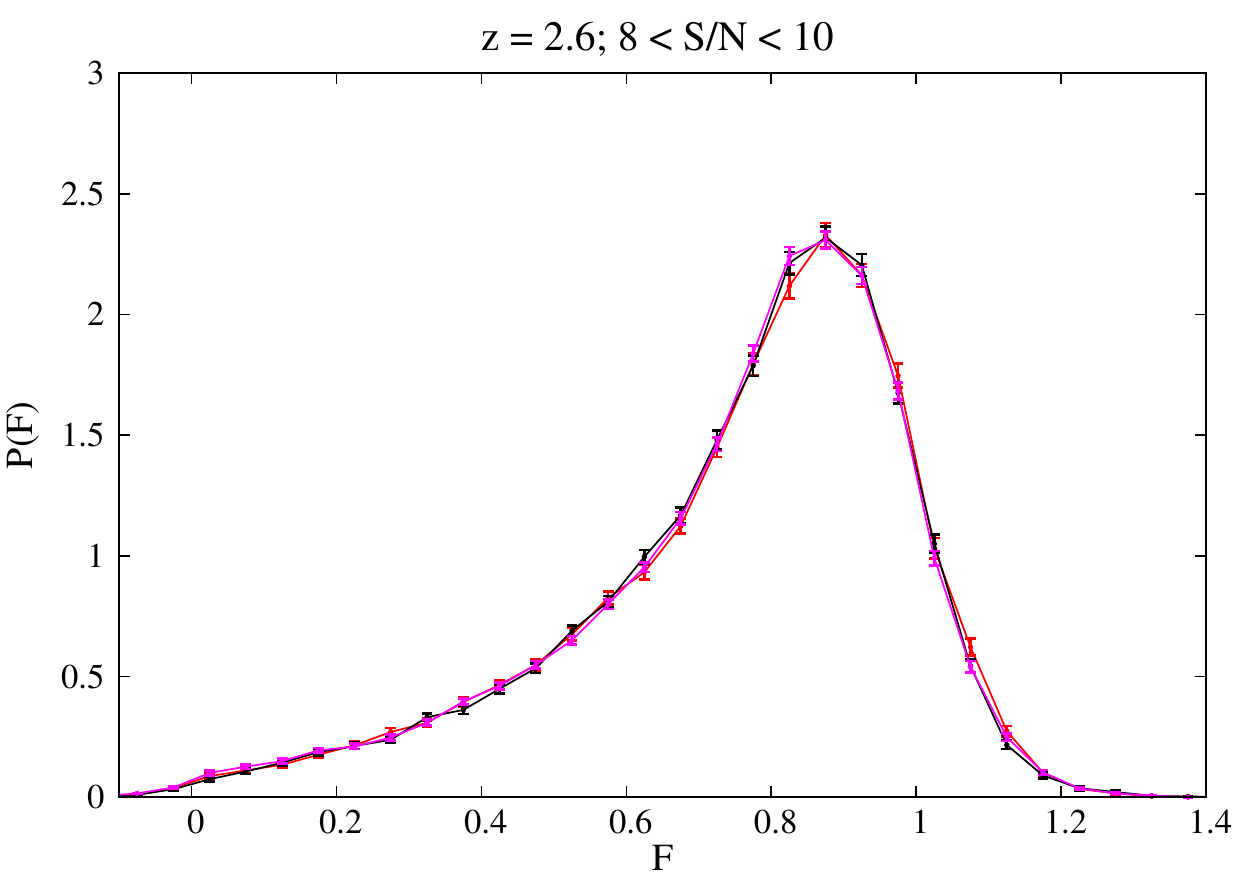}} 
\resizebox{140pt}{110pt}{\includegraphics{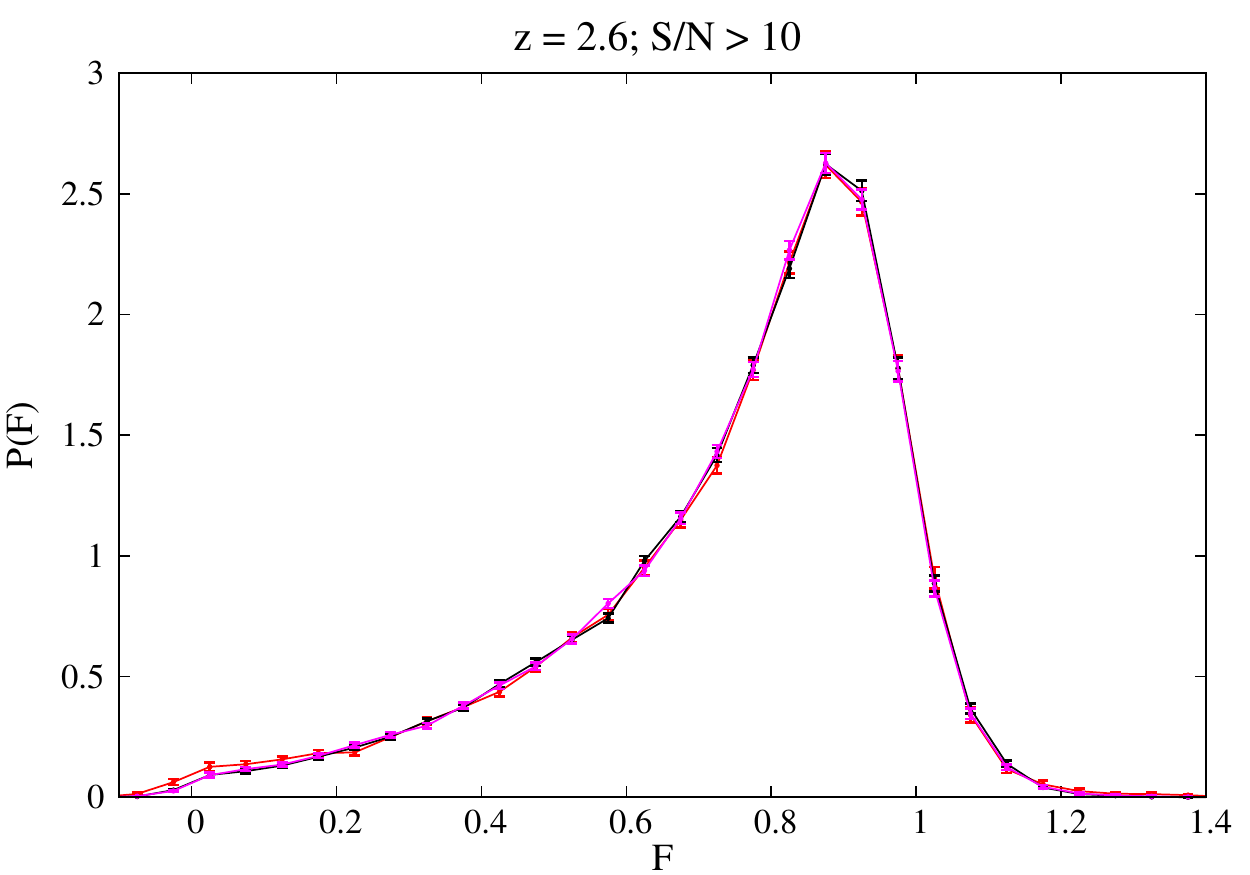}} 

\resizebox{140pt}{110pt}{\includegraphics{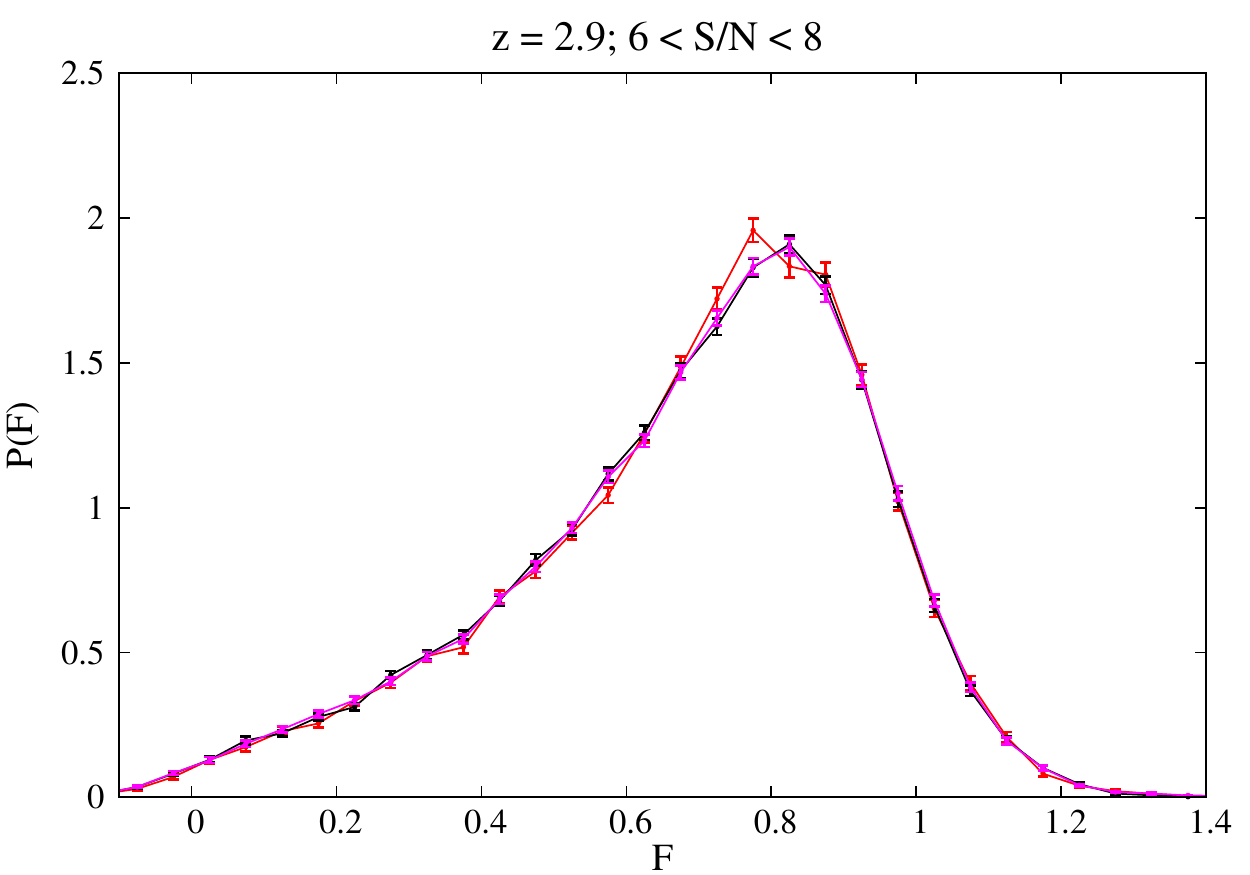}} 
\resizebox{140pt}{110pt}{\includegraphics{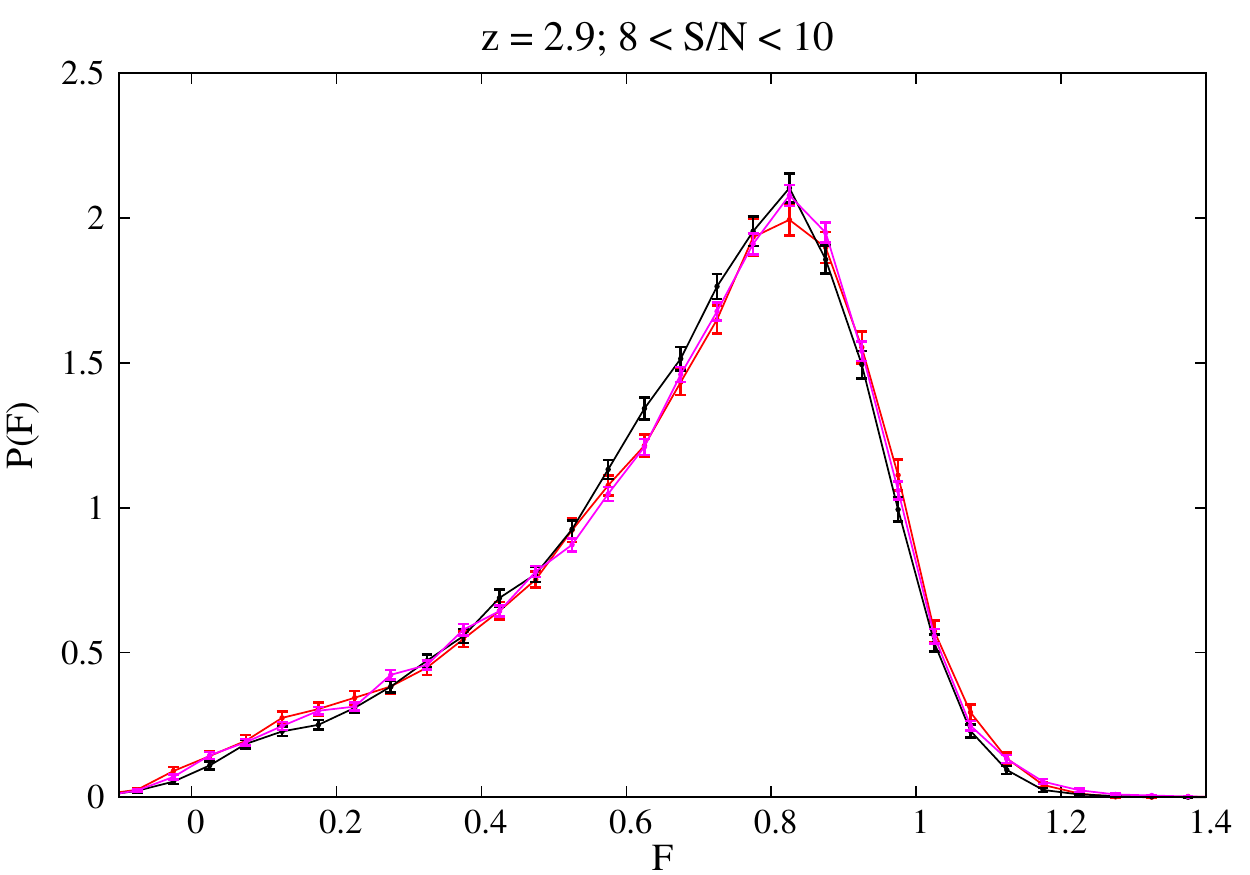}} 
\resizebox{140pt}{110pt}{\includegraphics{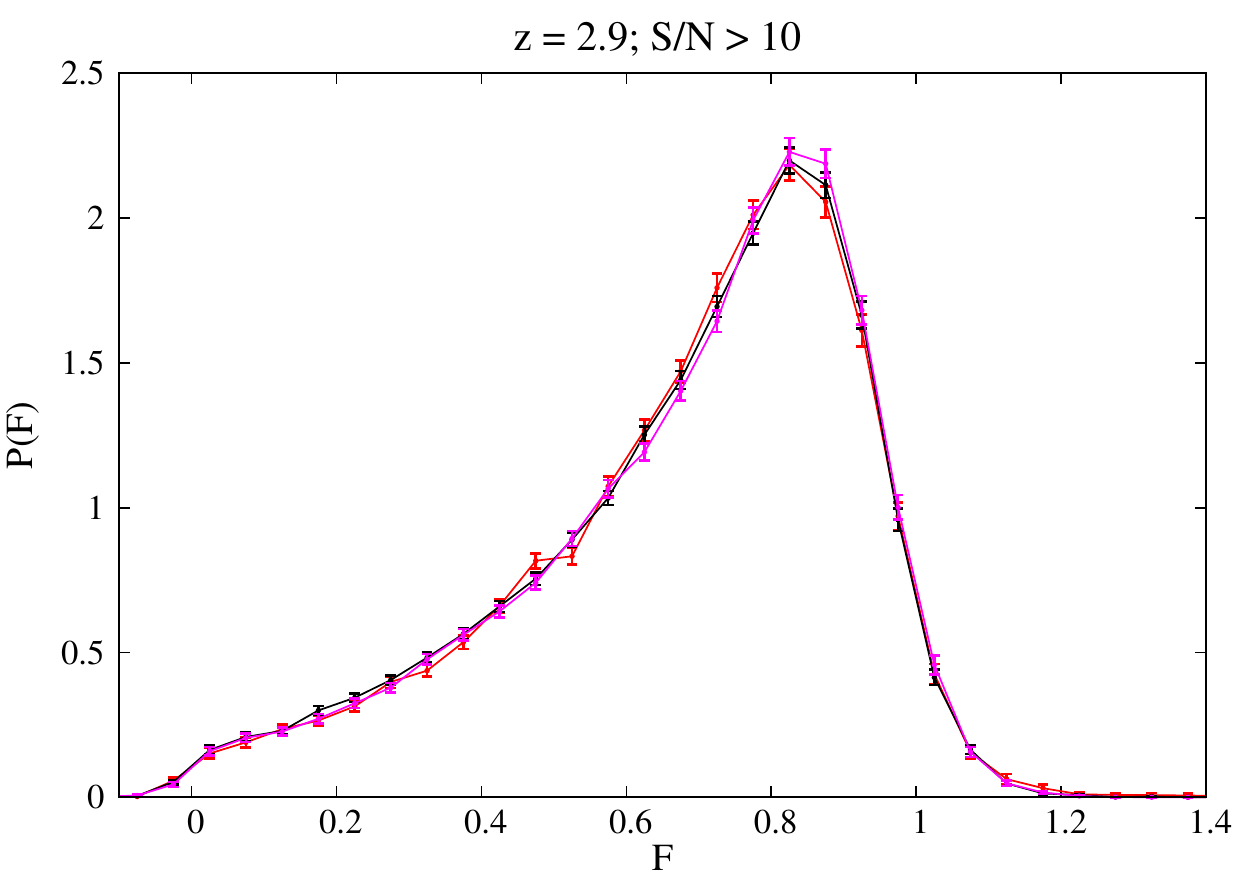}} 
\end{center}
\caption{\footnotesize\label{fig:comparison-PDF2} Comparison of PDF of the Lyman-$\alpha$ transmitted flux in different
patches for patch selection 2 (right of Fig.~\ref{fig:skycut}). The color codes represent the PDF's from the corresponding
patches. Along the rows we plot PDF's for different redshifts and along the columns we plot for different SNR. The error 
on the PDF is estimated through bootstrap resampling over $100\mathring{A}$ data chunks.} 
\end{figure*}

To start with, we provide the flux PDF with the errors in Fig.~\ref{fig:comparison-PDF1} for patch selection 1 (corresponding
to the selection in the left of Fig.~\ref{fig:skycut}) and in Fig.~\ref{fig:comparison-PDF2} for patch selection 2 (the right
selection in the Fig.~\ref{fig:skycut}). The colors of the PDF's correspond to different patches (as shown in the Fig.~\ref{fig:skycut}). 
The PDF's are calculated in the flux range [-0.2-1.5] in 34 bins. Theoretically 
the transmitted flux should be within 0 and 1 (for the complete and no absorption respectively) but due to the noise in 
observations and the bias in the continuum estimations, the transmitted flux might be obtained outside the theoretical 
boundary. The PDF's are normalized such that the total area under each PDF is 1. In both the figures, we provide the PDF for different redshifts and  
different SNR. As we mentioned in the previous section, due to less number of quasars in the higher 
SNR bins, the errors on the flux PDFs are comparable to that of the lower SNR bins. In the plots of the PDF, the 
difference between the PDF's in {\it good, better} and {\it best} SNR bins are evident. For the higher SNR bins 
the flux PDF are more sharp around the peak. It is interesting to note that at the same SNR and at the same redshift,
the PDF of transmitted flux from different patches are very similar. In some of the bins we find the flux from one patch is different 
from other ones. To quantify the difference, we next look for the different moments of the PDFs.

\begin{figure*}[!b]
\begin{center} 
\resizebox{140pt}{110pt}{\includegraphics{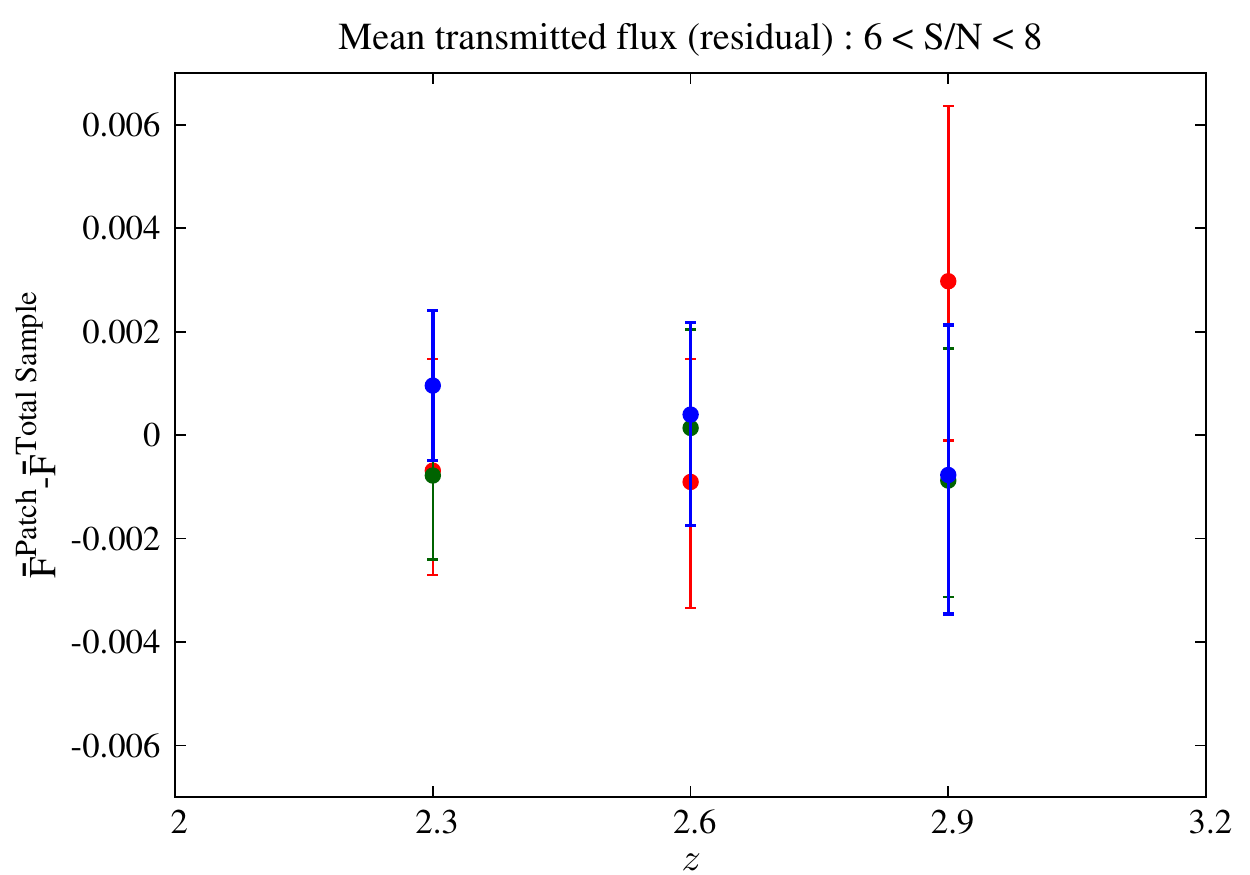}} 
\resizebox{140pt}{110pt}{\includegraphics{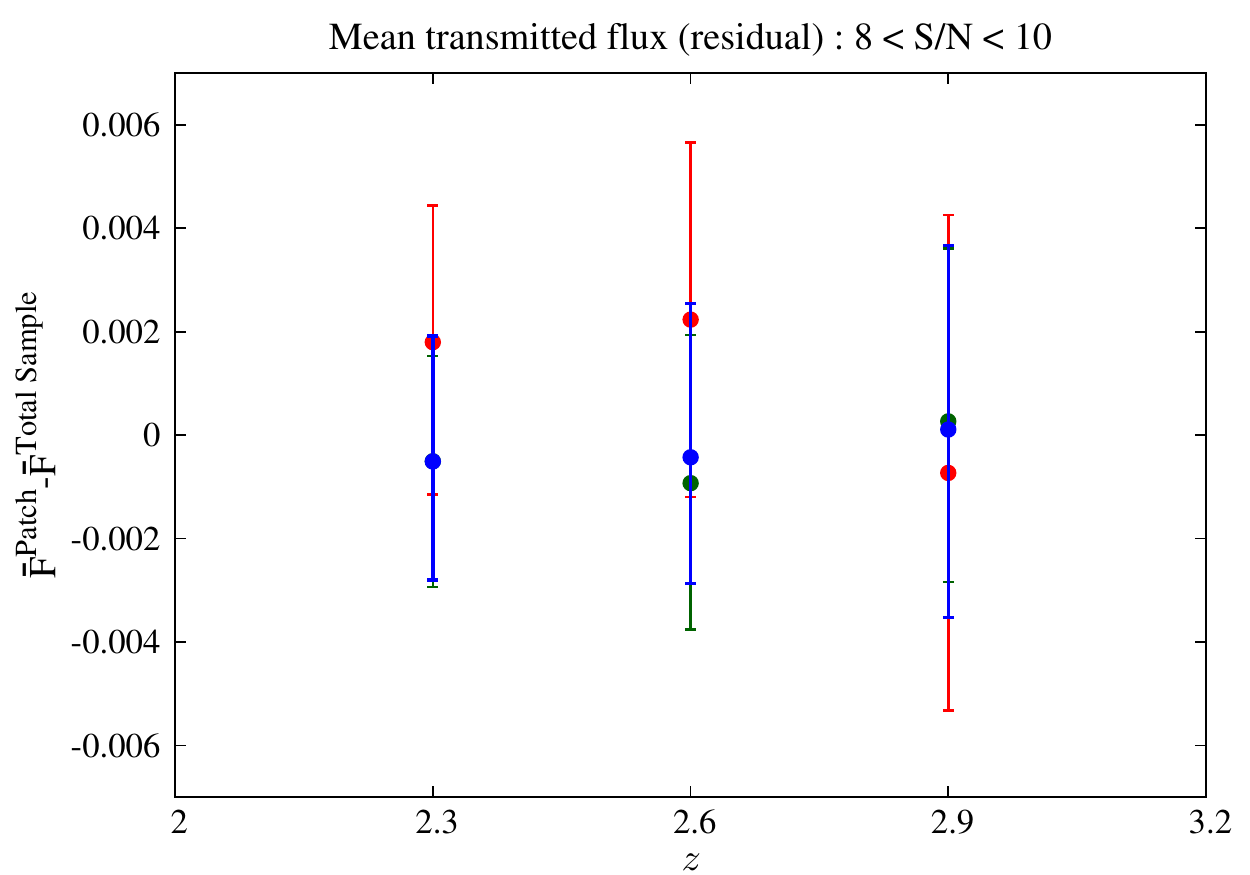}} 
\resizebox{140pt}{110pt}{\includegraphics{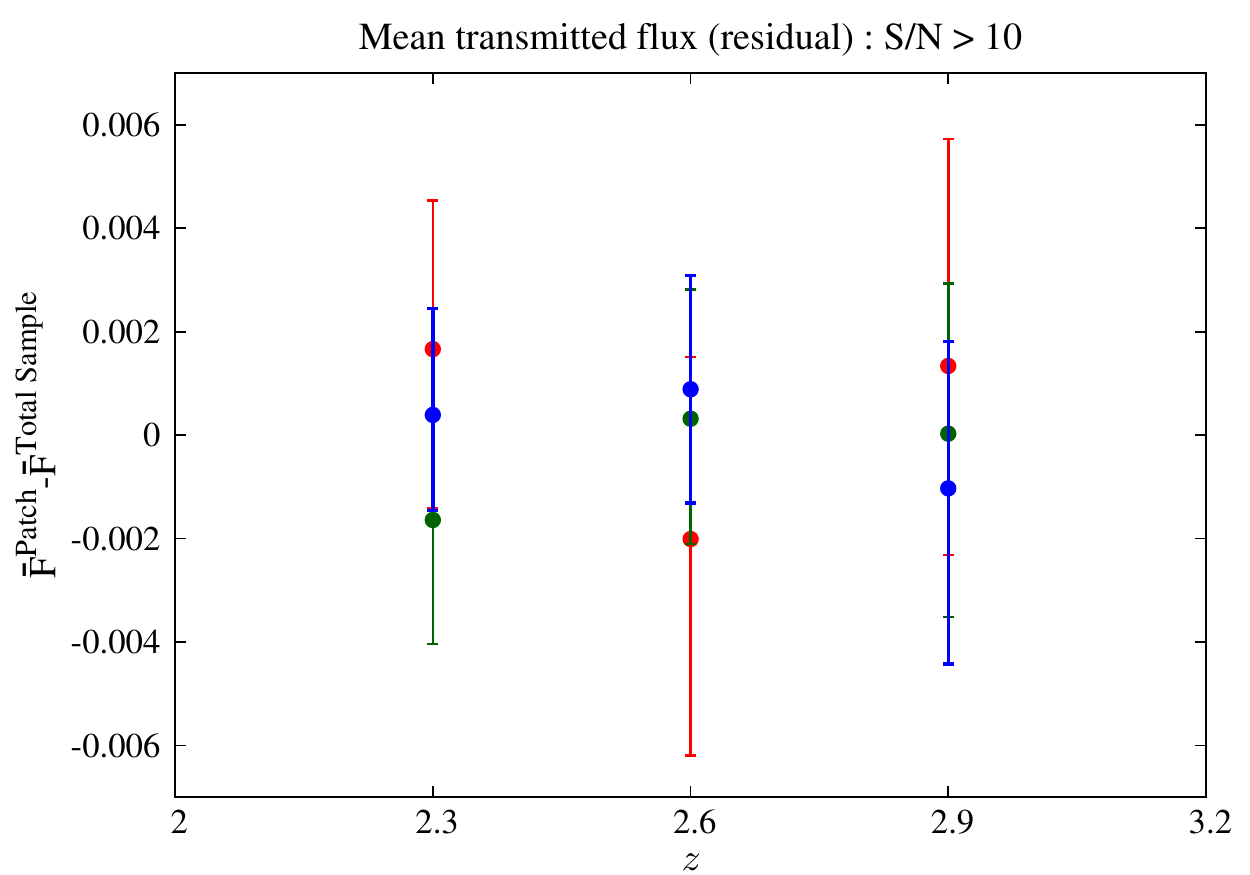}} 

\resizebox{140pt}{110pt}{\includegraphics{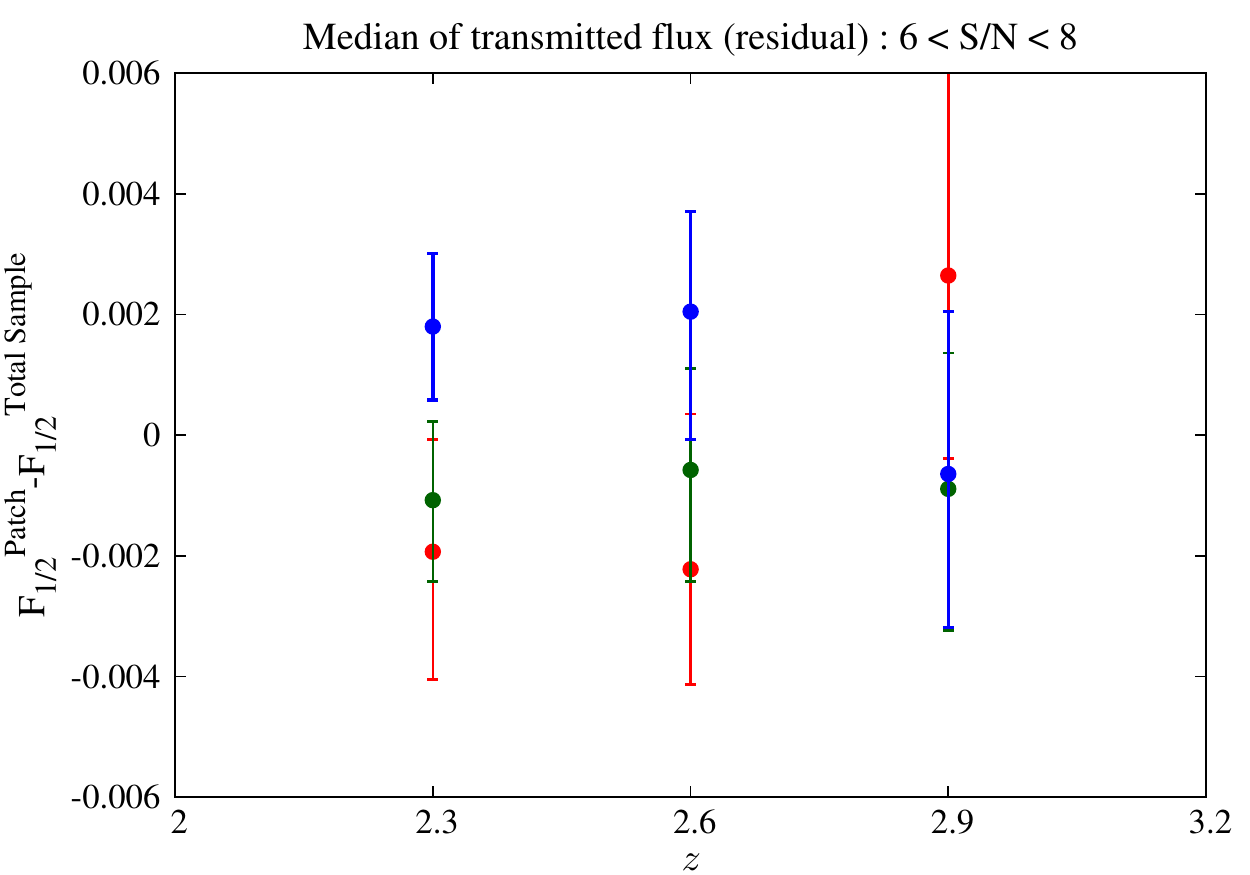}} 
\resizebox{140pt}{110pt}{\includegraphics{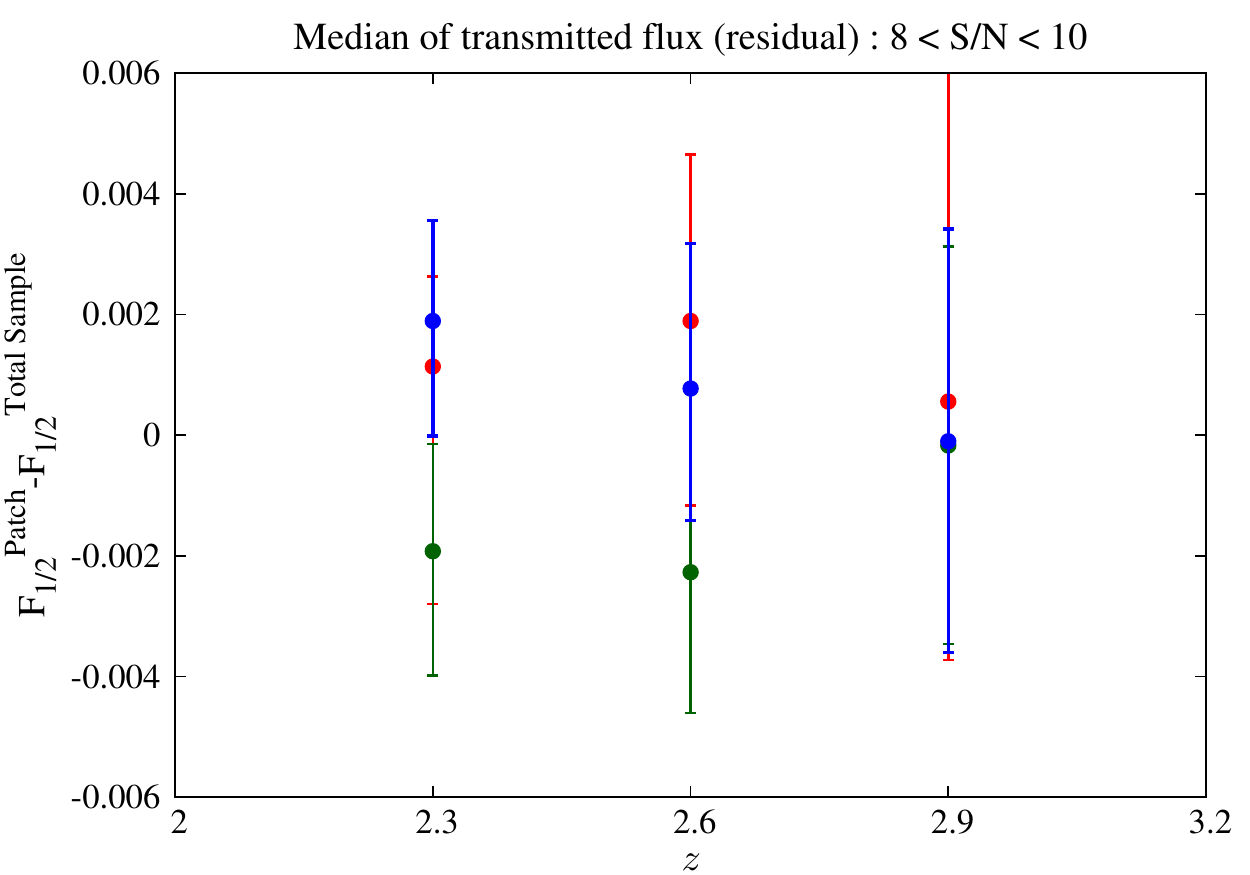}} 
\resizebox{140pt}{110pt}{\includegraphics{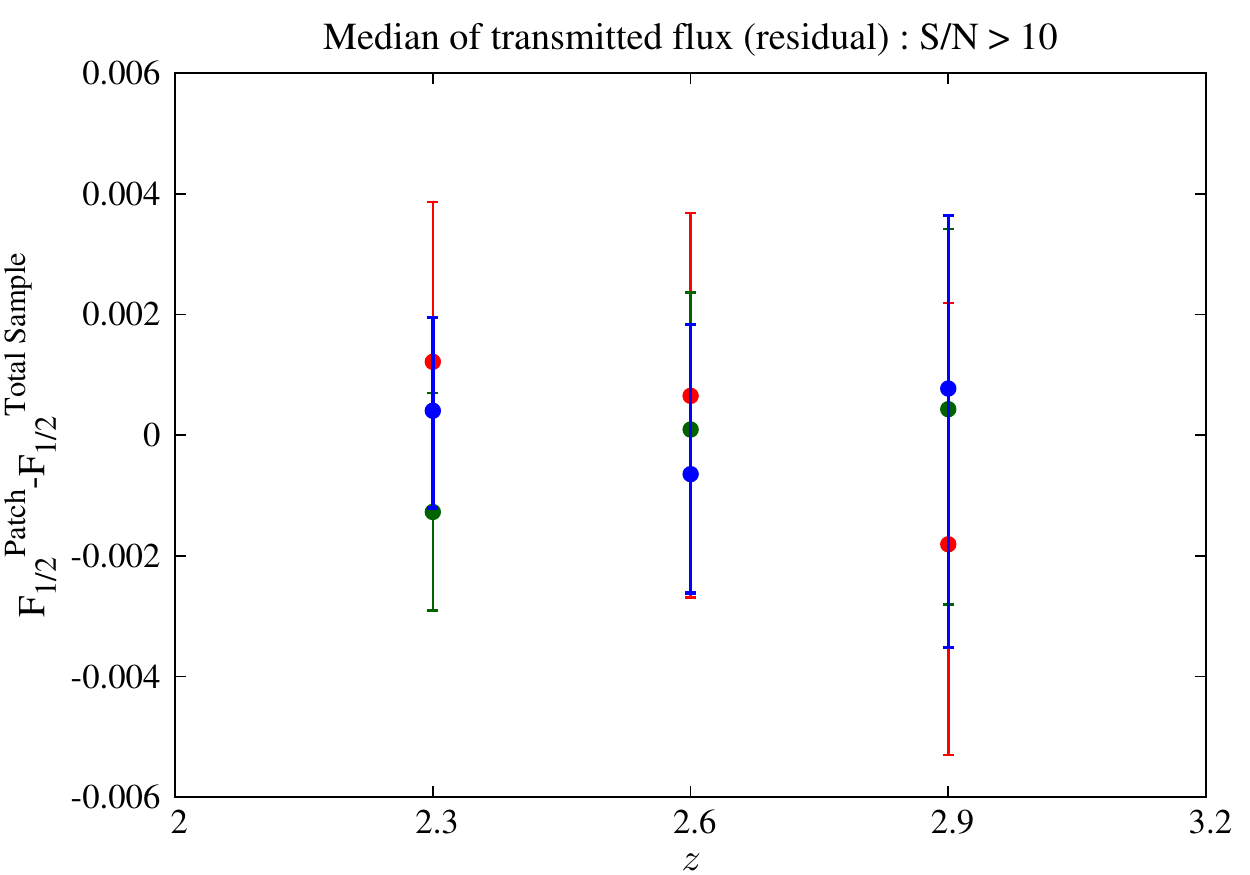}} 

\resizebox{140pt}{110pt}{\includegraphics{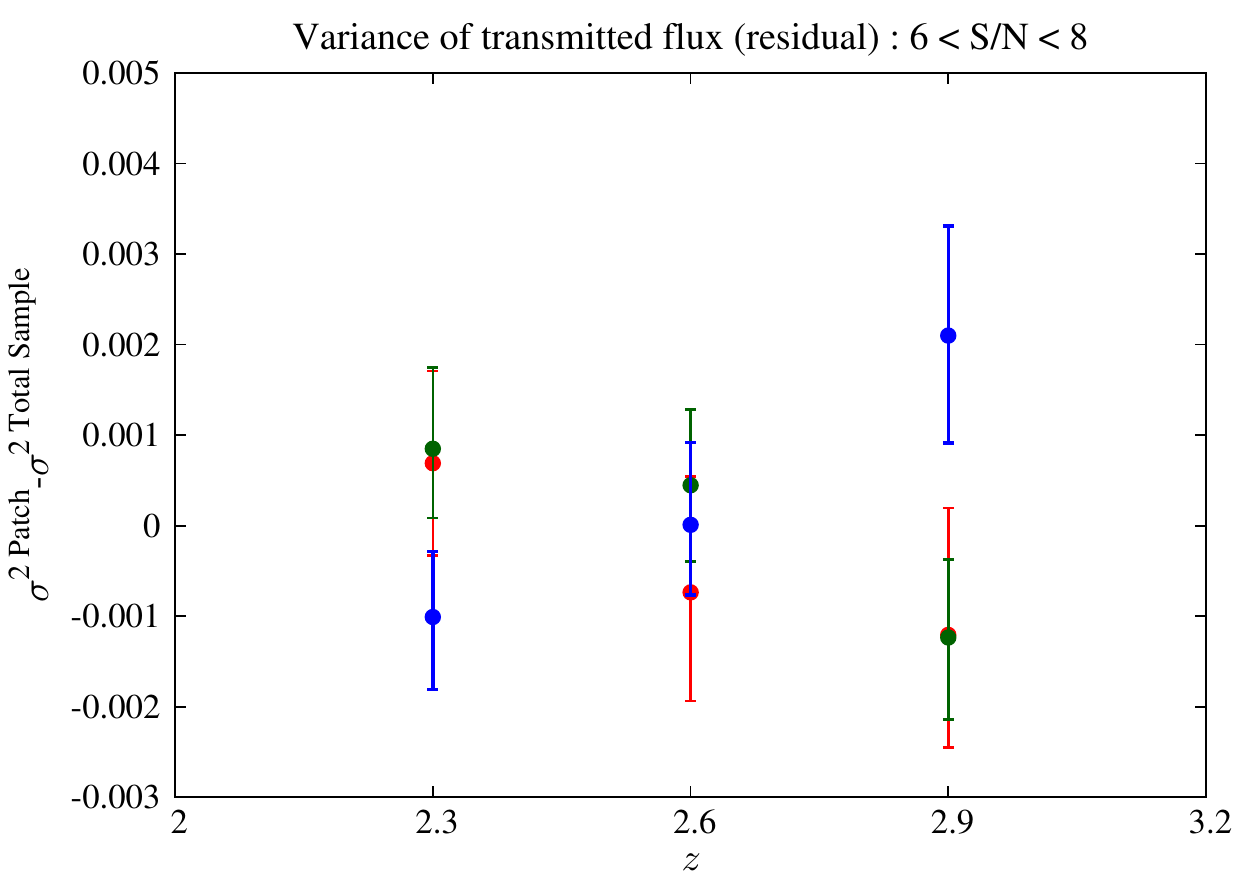}} 
\resizebox{140pt}{110pt}{\includegraphics{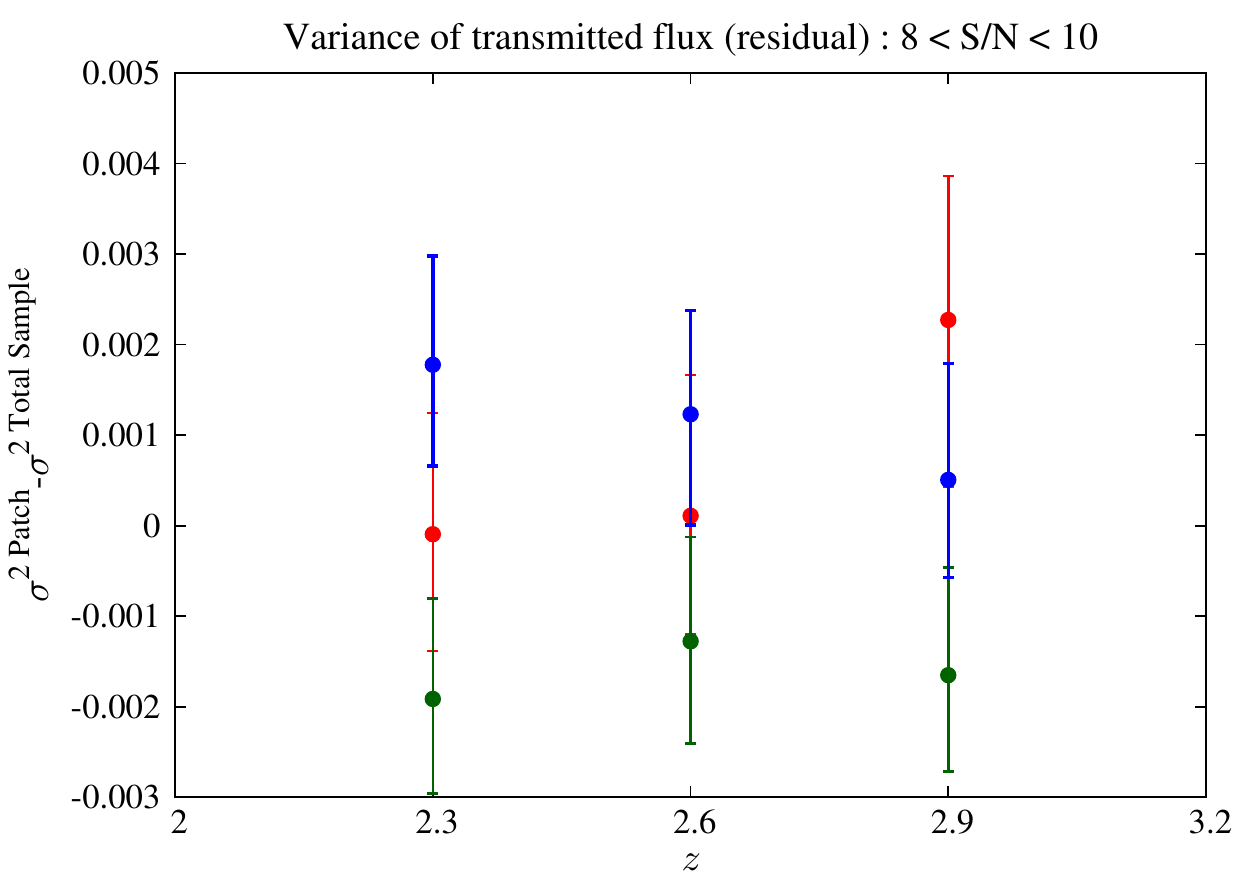}} 
\resizebox{140pt}{110pt}{\includegraphics{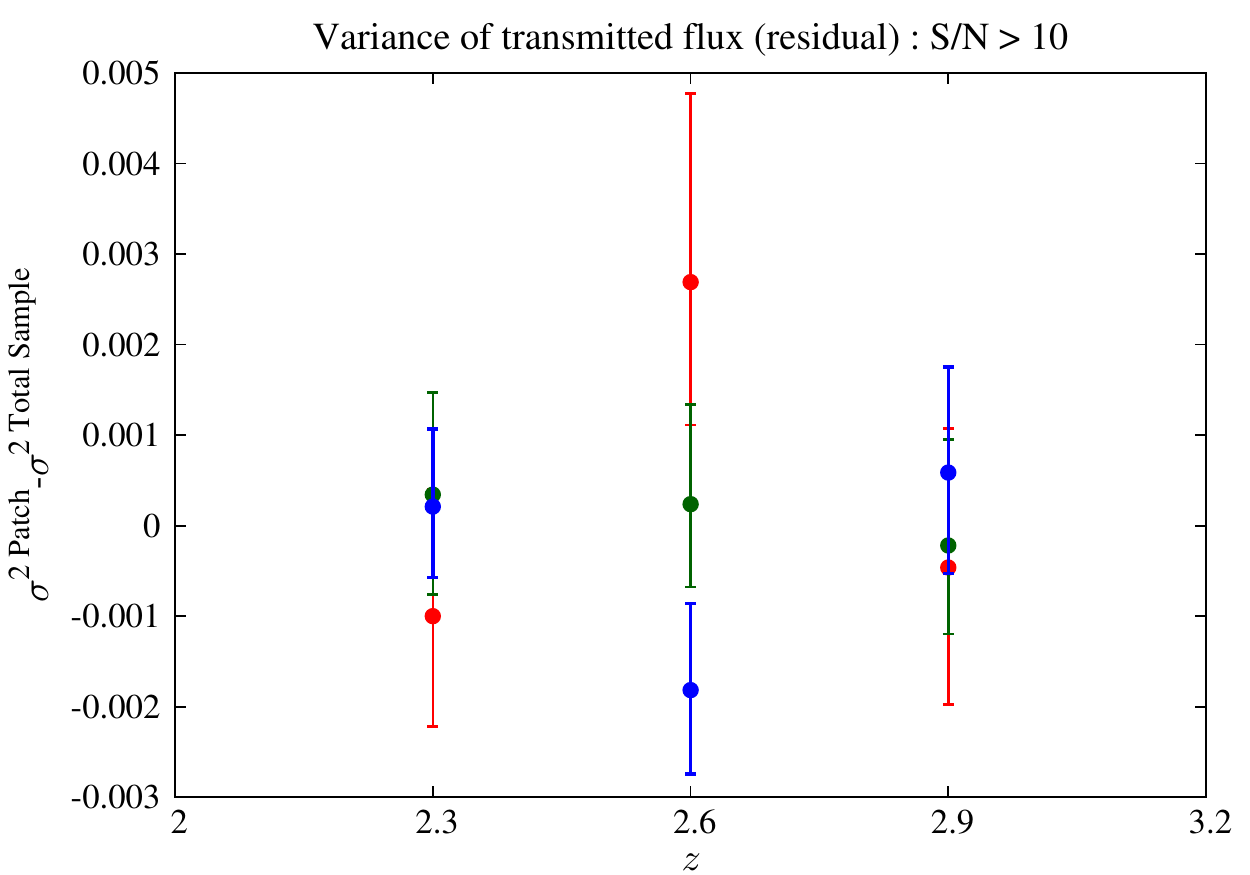}} 

\resizebox{140pt}{110pt}{\includegraphics{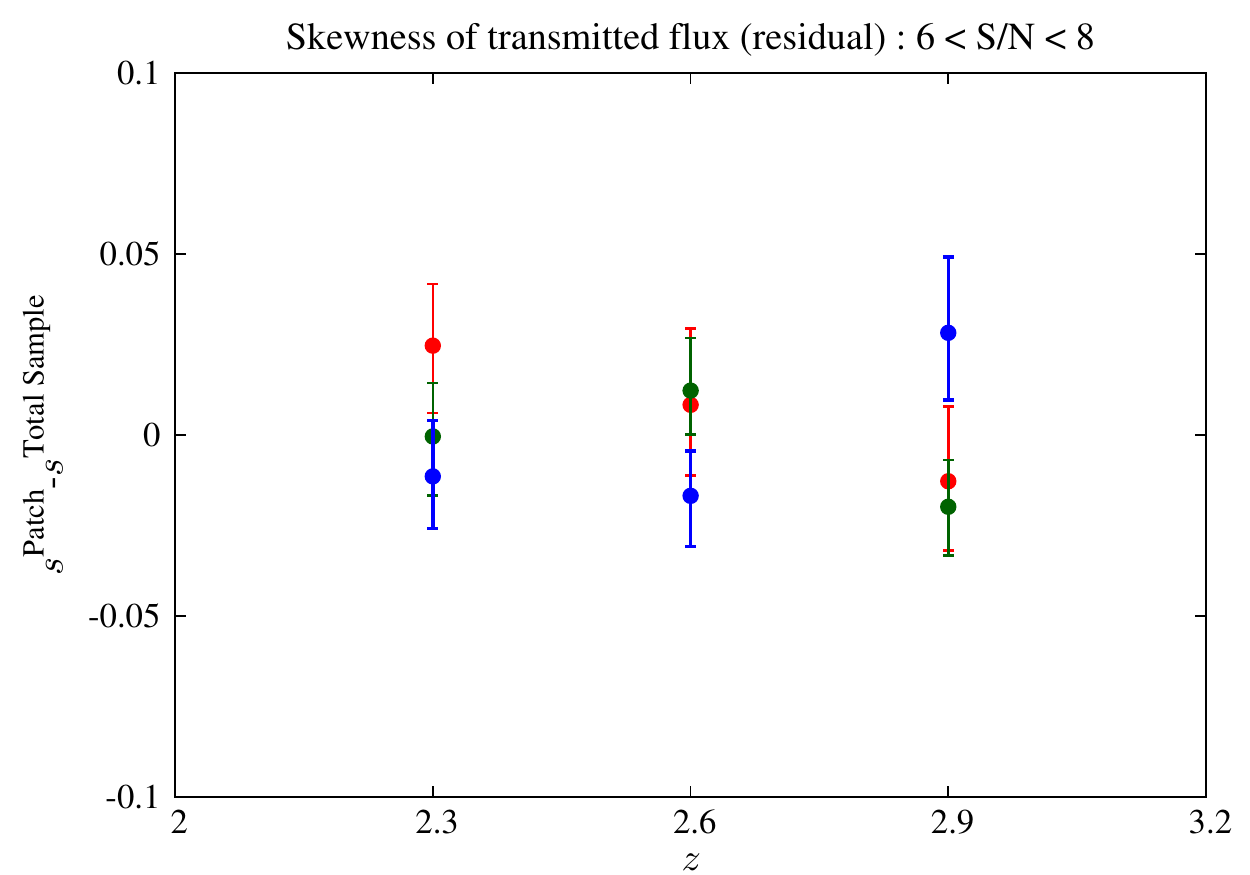}} 
\resizebox{140pt}{110pt}{\includegraphics{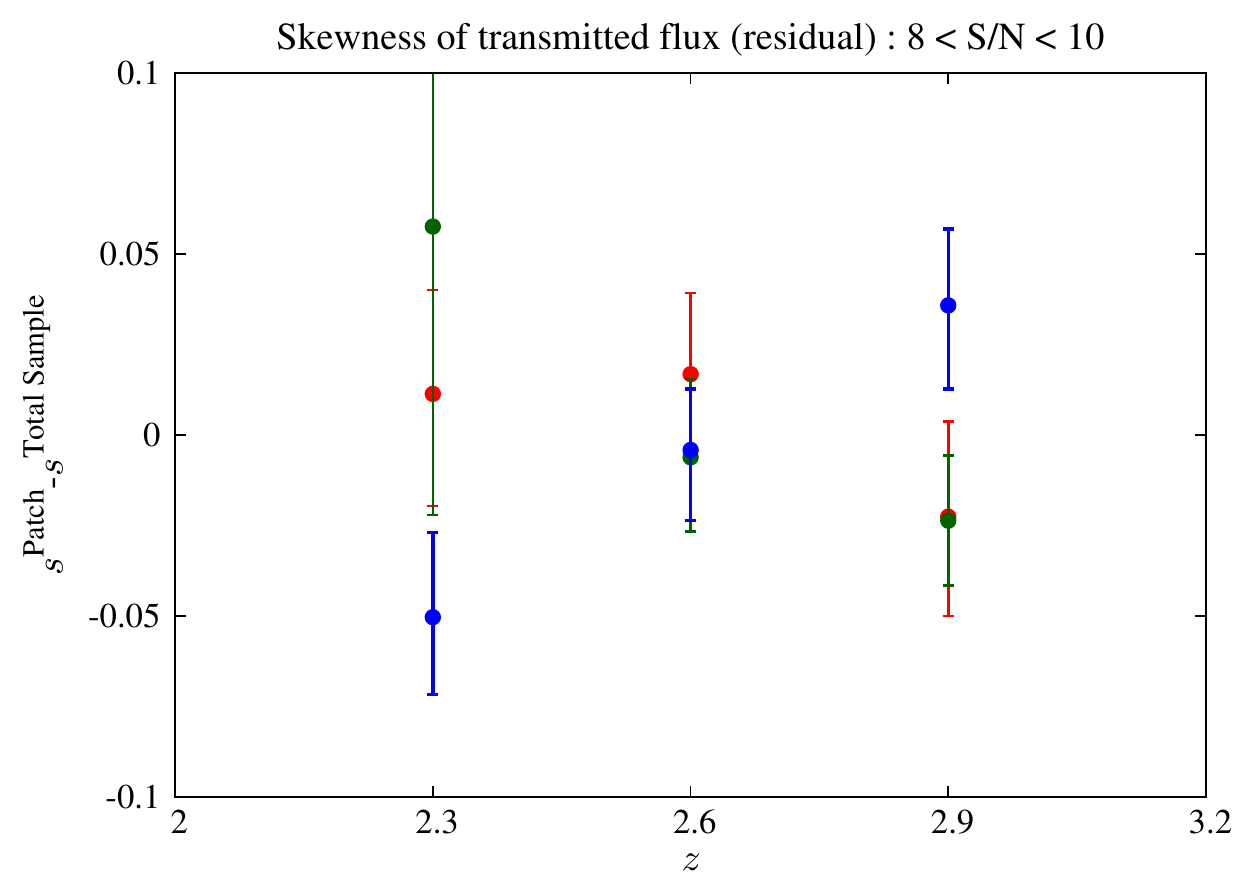}} 
\resizebox{140pt}{110pt}{\includegraphics{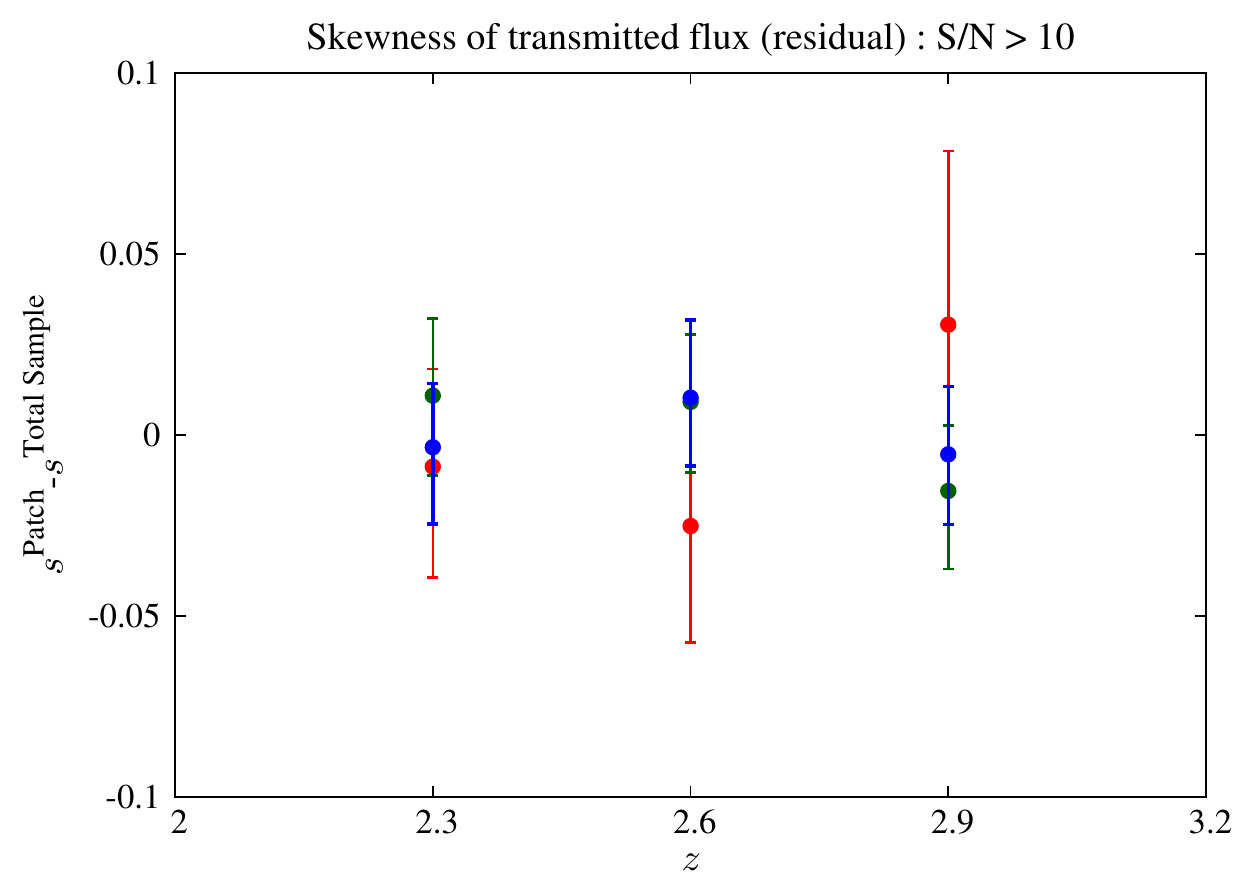}} 

\resizebox{140pt}{110pt}{\includegraphics{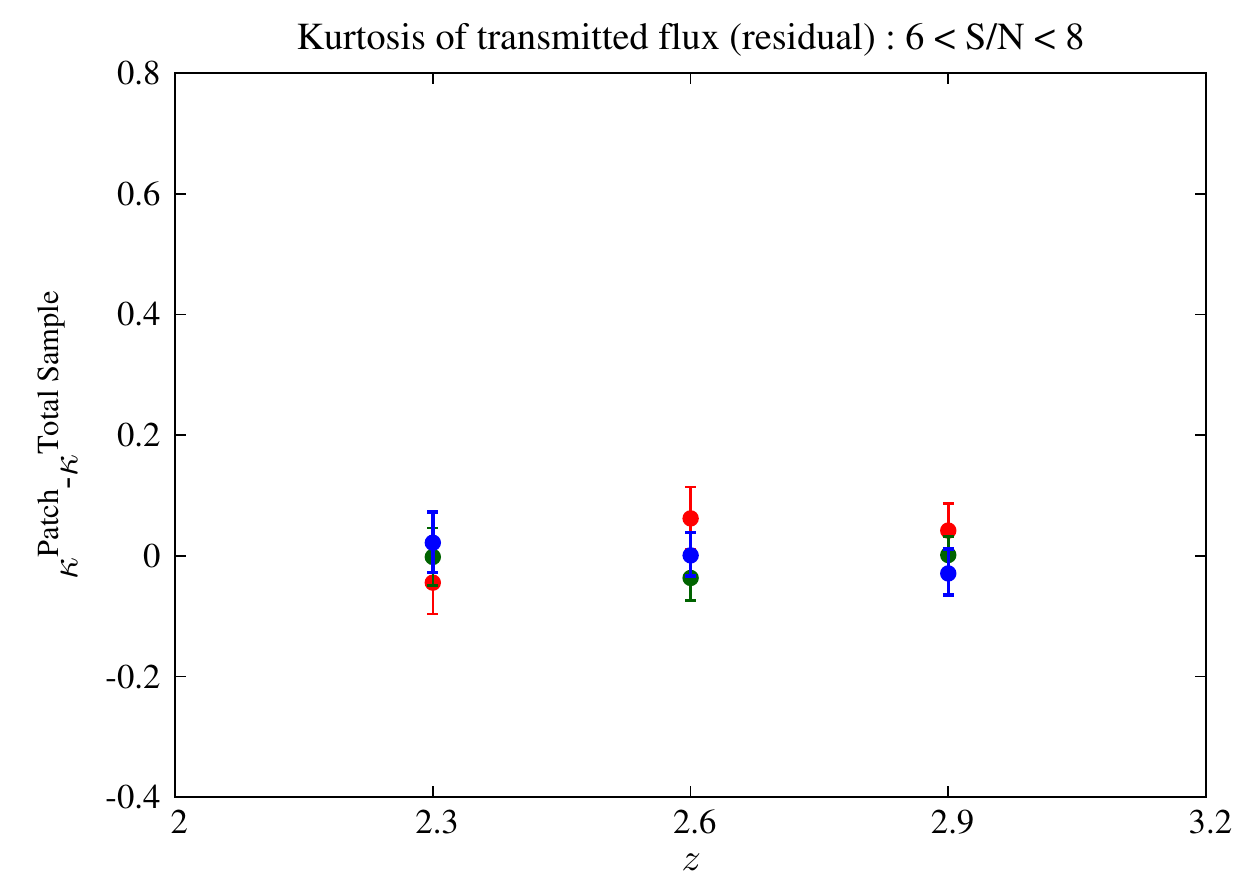}} 
\resizebox{140pt}{110pt}{\includegraphics{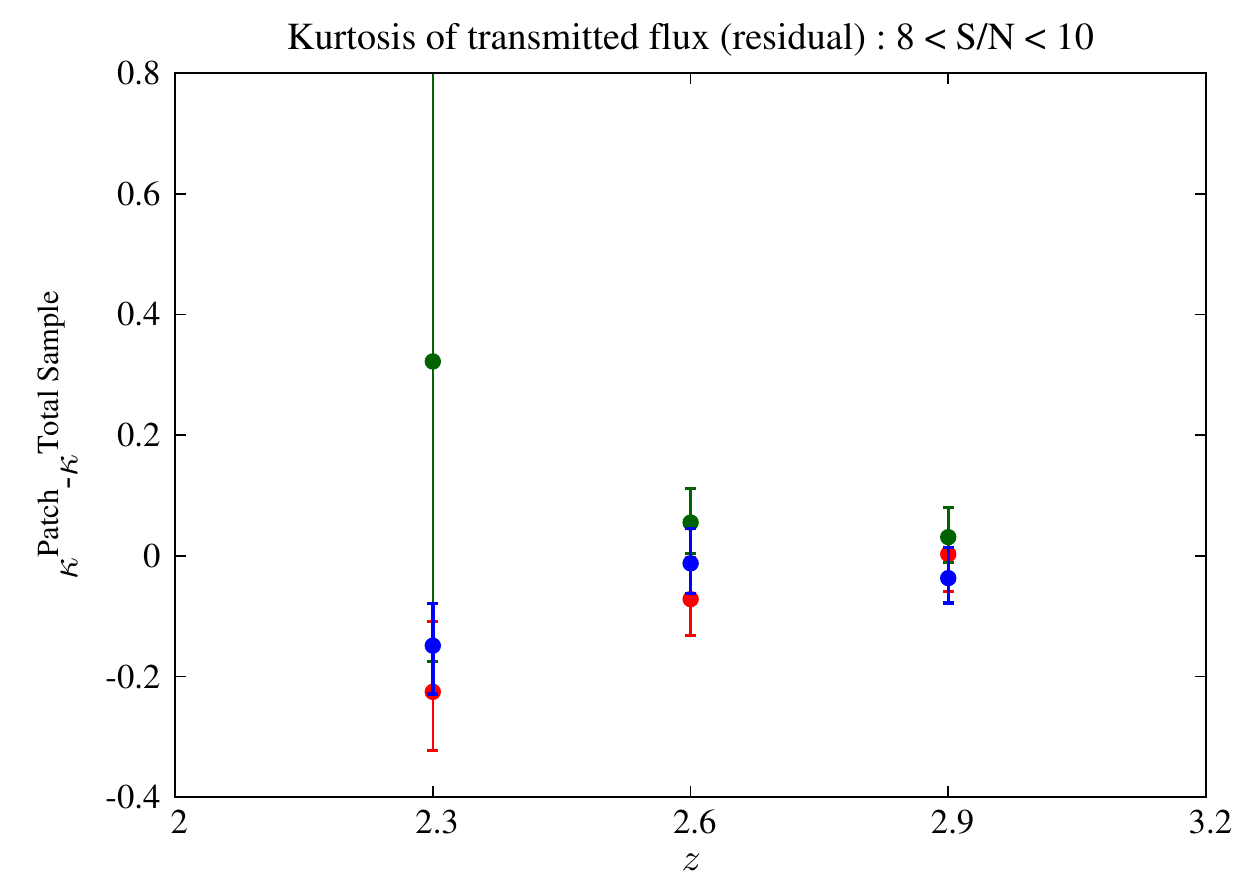}} 
\resizebox{140pt}{110pt}{\includegraphics{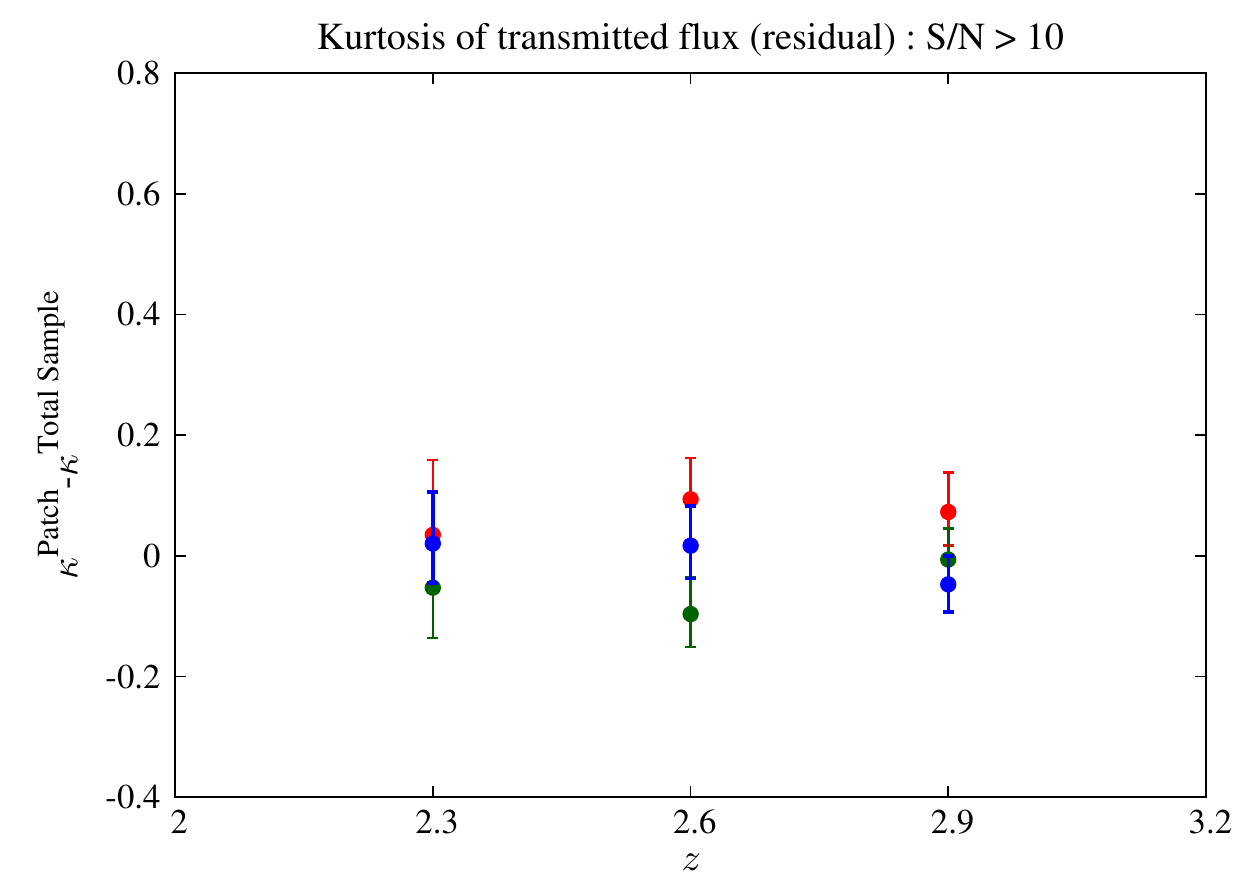}} 
\end{center}
\caption{\footnotesize\label{fig:comparison-1}First few statistical moments of the Lyman-$\alpha$ transmitted flux
PDF for different parts of the sky. For relative comparison we plot the residual moments from the total sample. The 
error bars correspond to the 1$\sigma$ bounds on that particular moment. This 
figure represents the comparison between the properties of the patches for patch selection 1 (left plot of Fig.~\ref{fig:skycut}).}
\end{figure*}

\begin{figure*}[!b]
\begin{center} 
\resizebox{140pt}{110pt}{\includegraphics{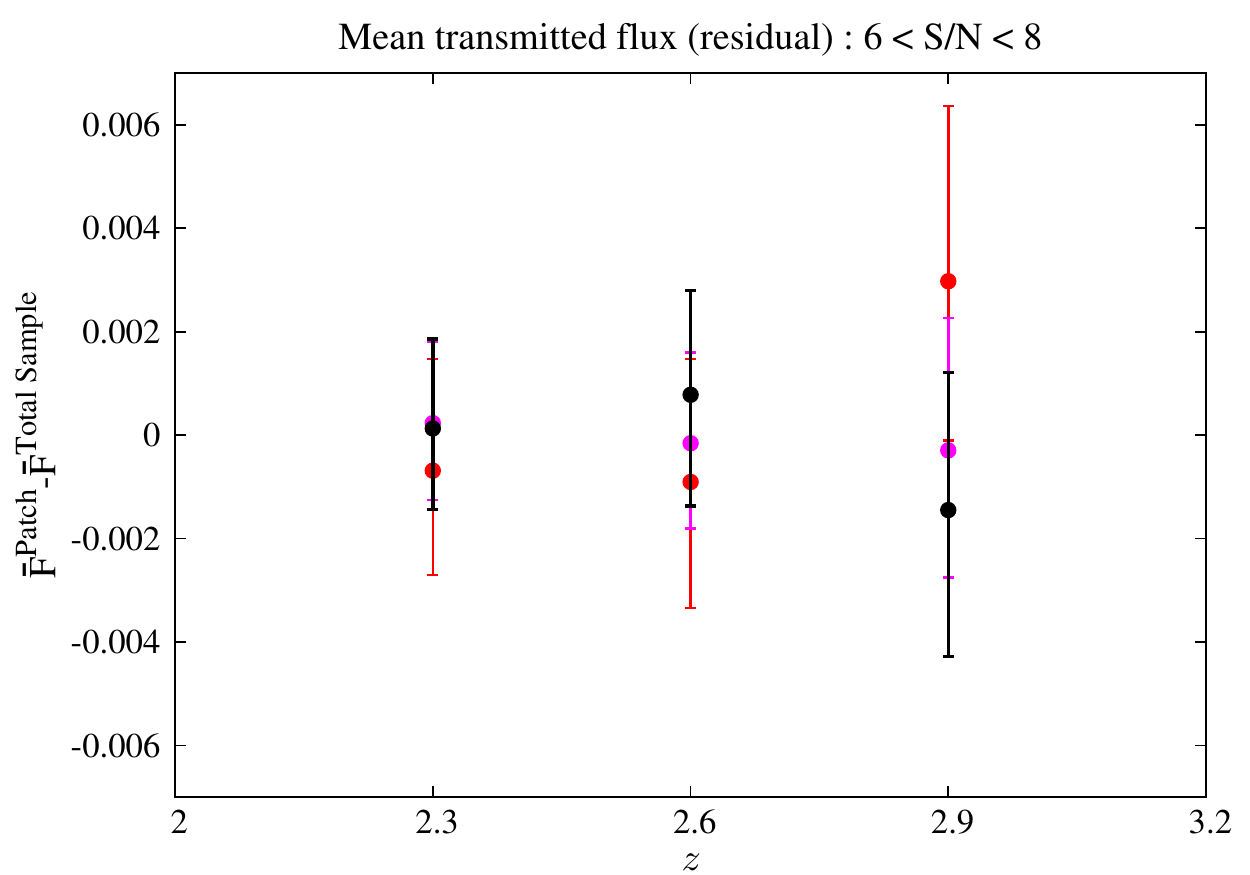}} 
\resizebox{140pt}{110pt}{\includegraphics{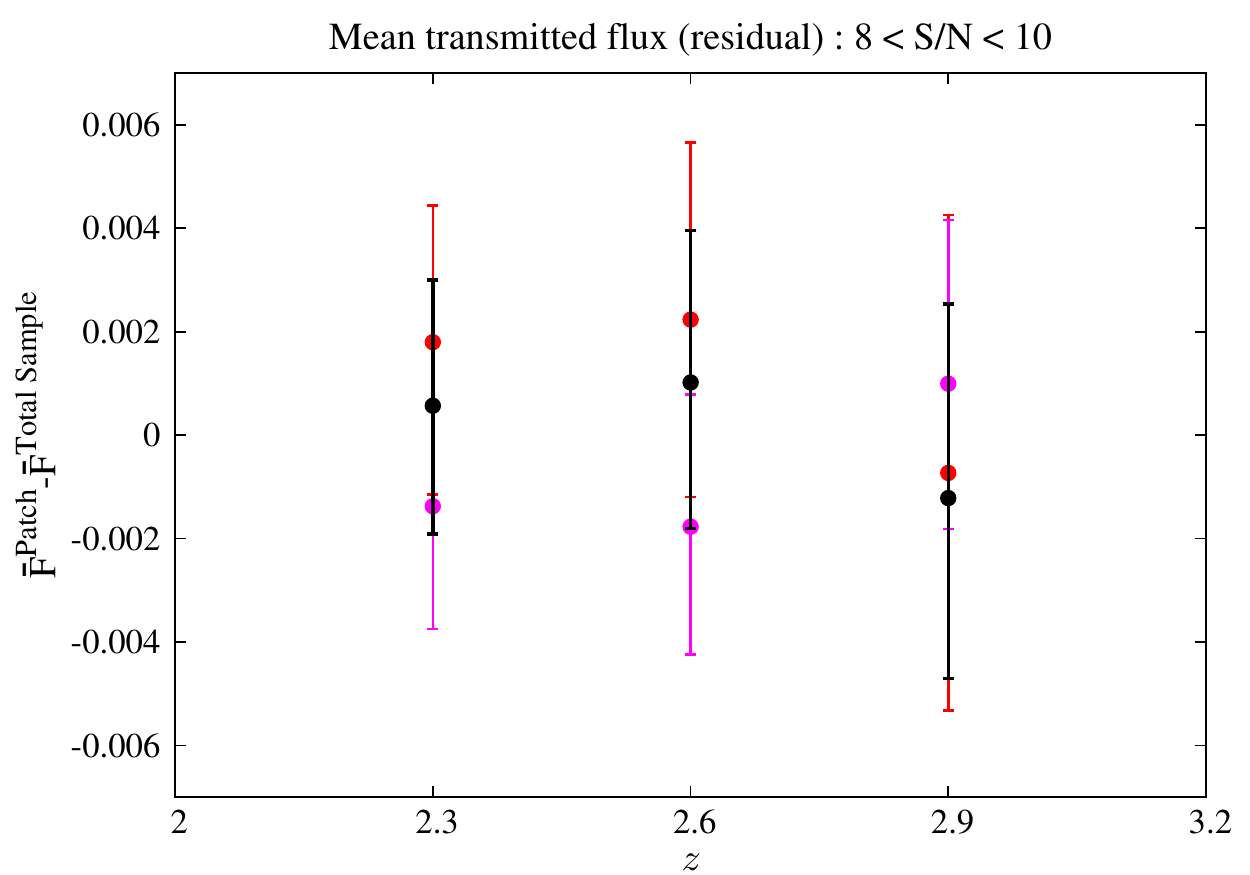}} 
\resizebox{140pt}{110pt}{\includegraphics{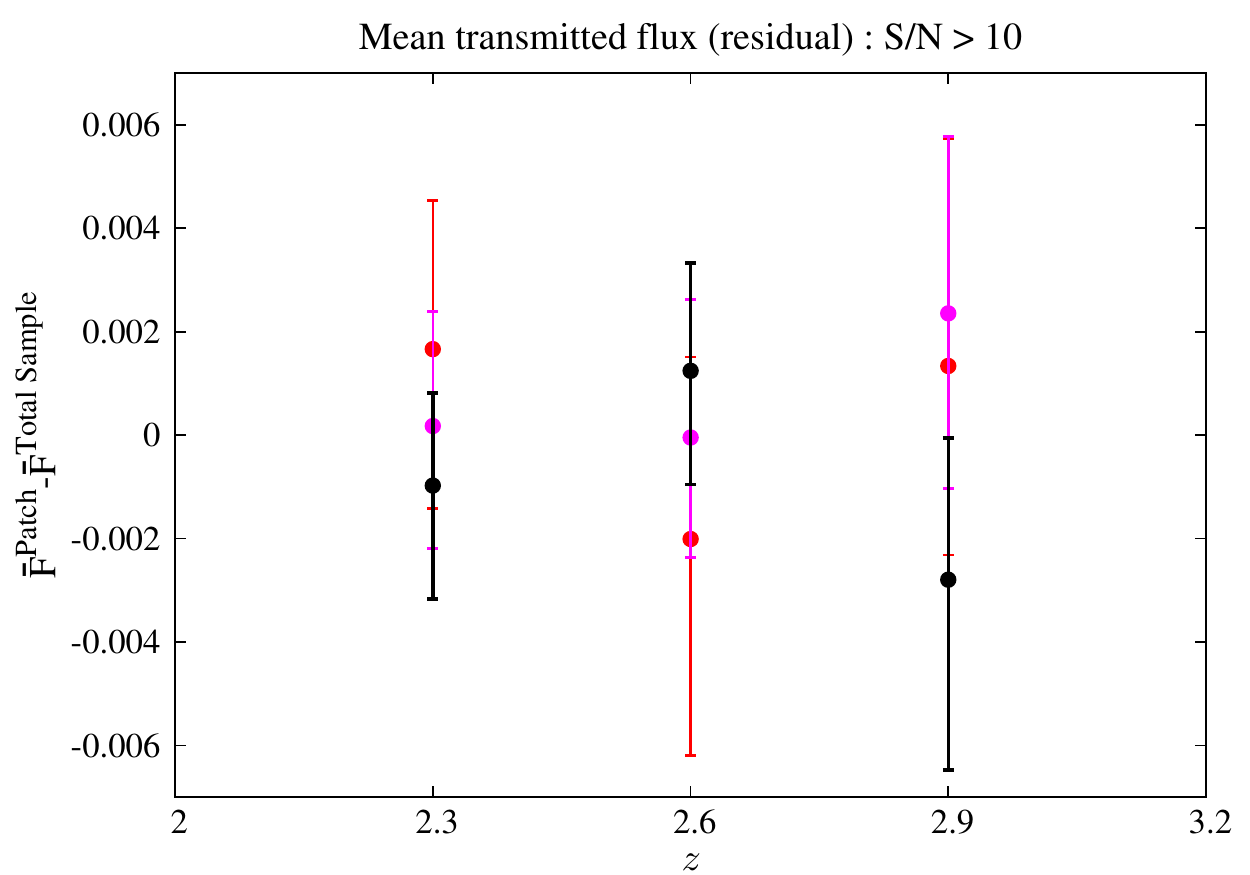}} 

\resizebox{140pt}{110pt}{\includegraphics{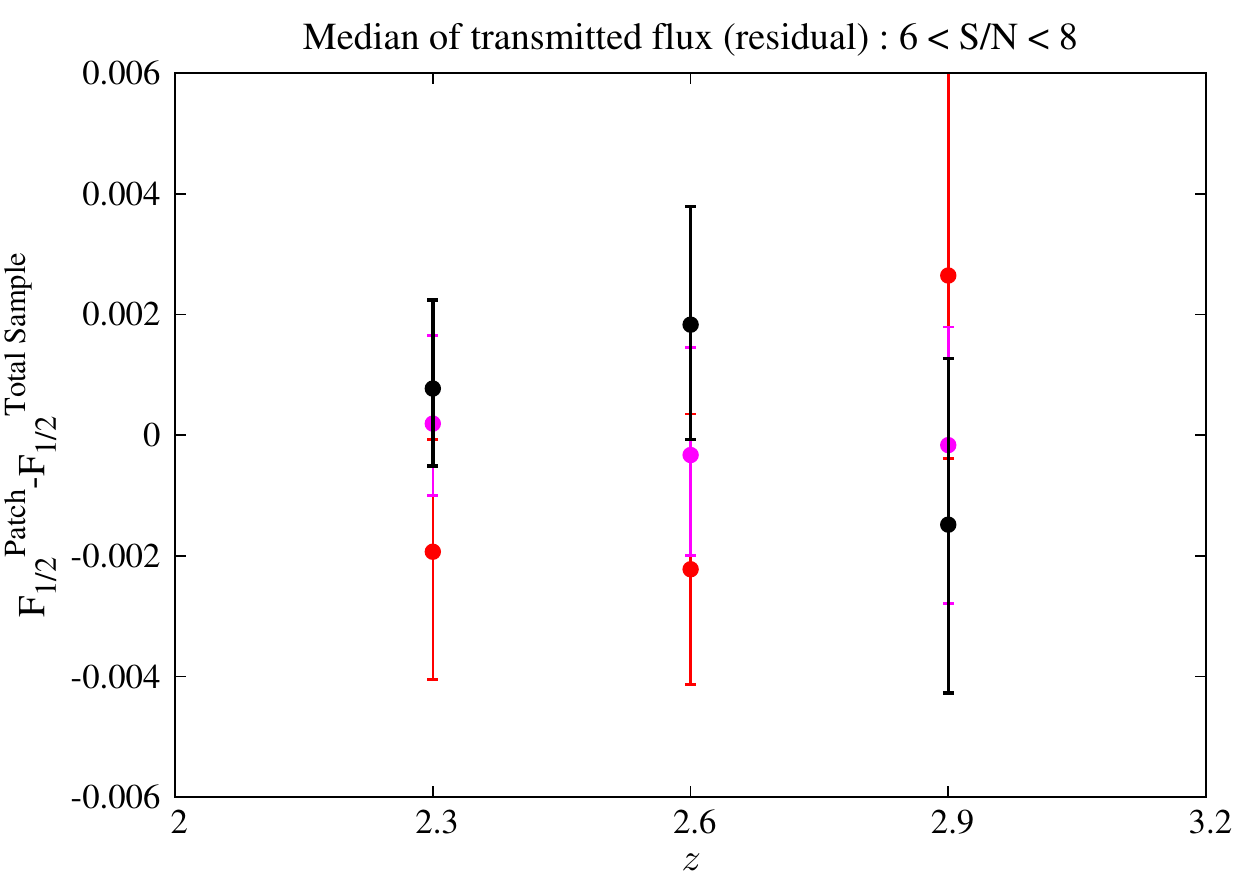}} 
\resizebox{140pt}{110pt}{\includegraphics{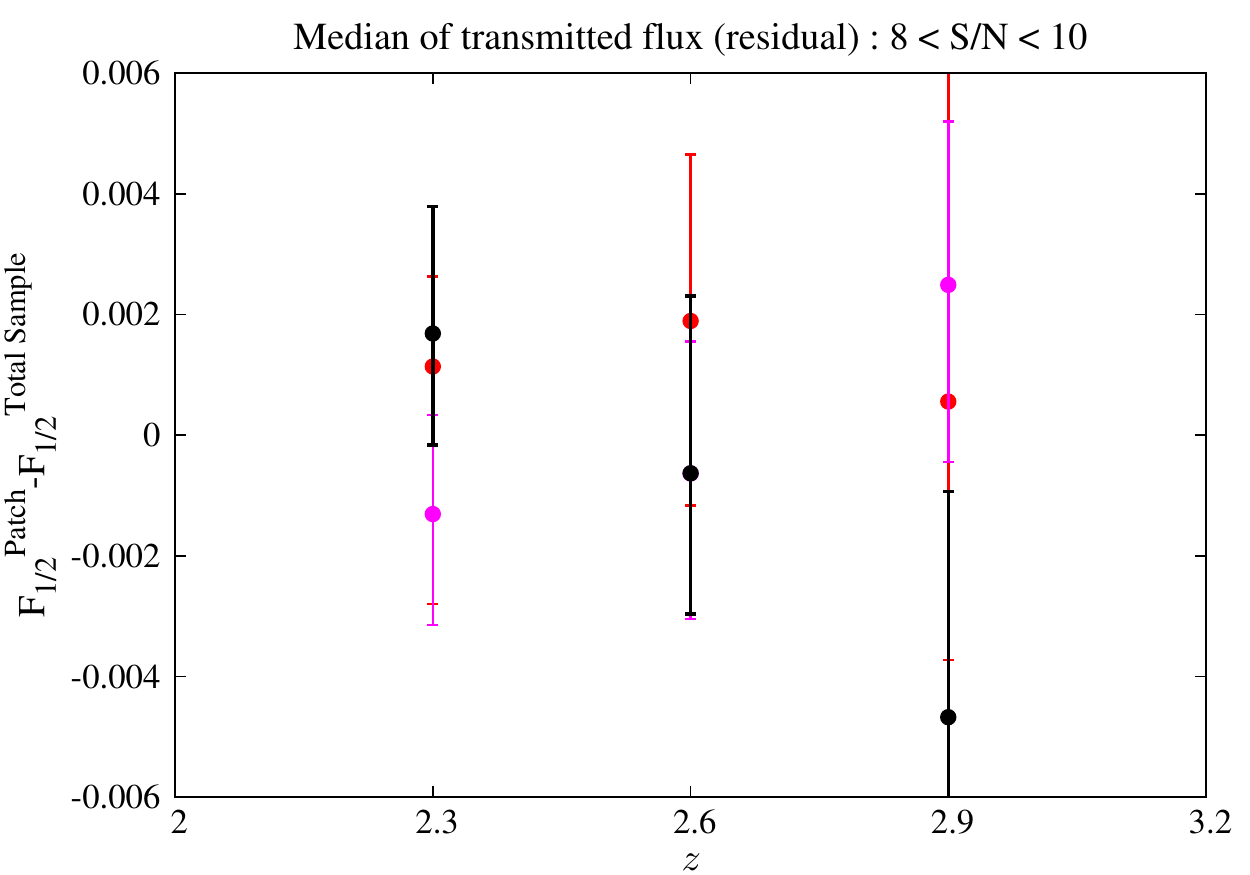}} 
\resizebox{140pt}{110pt}{\includegraphics{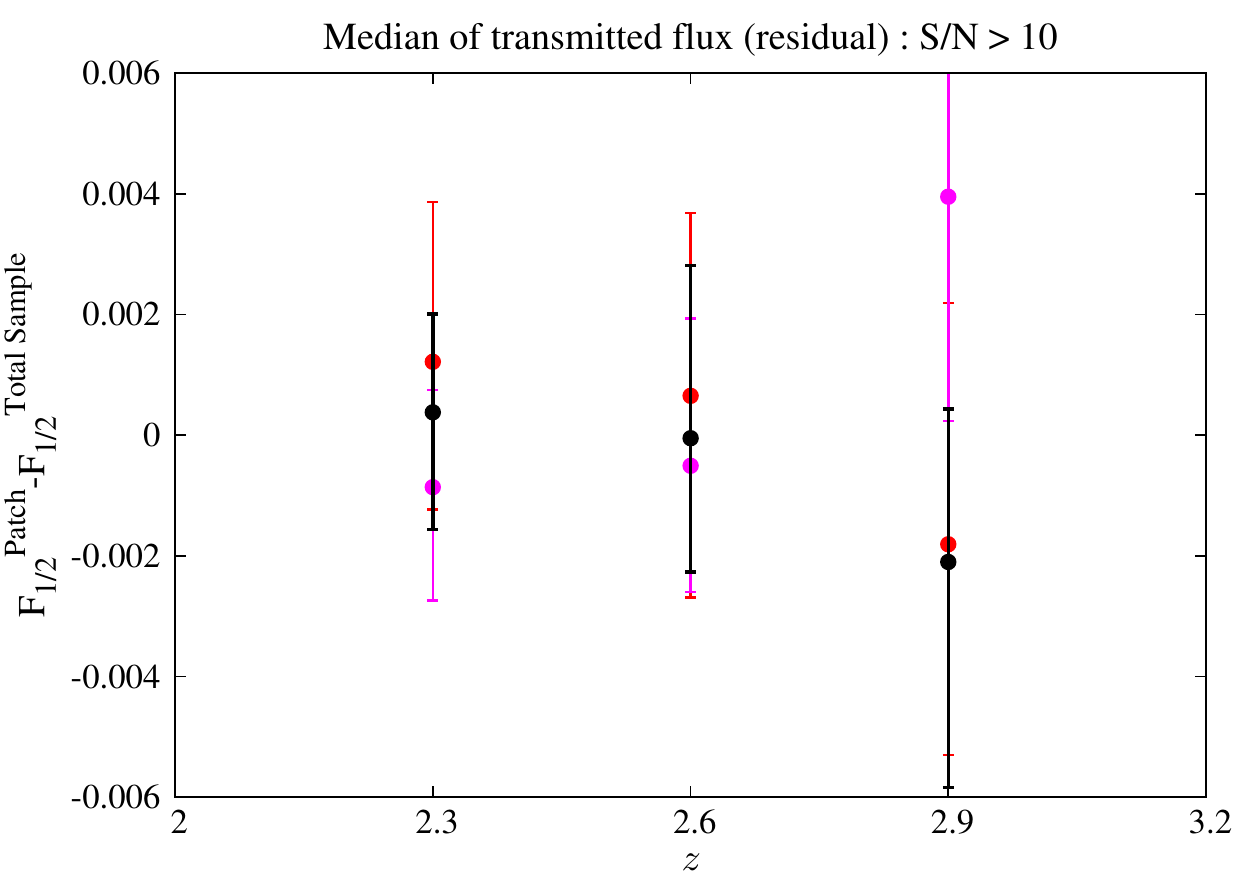}} 

\resizebox{140pt}{110pt}{\includegraphics{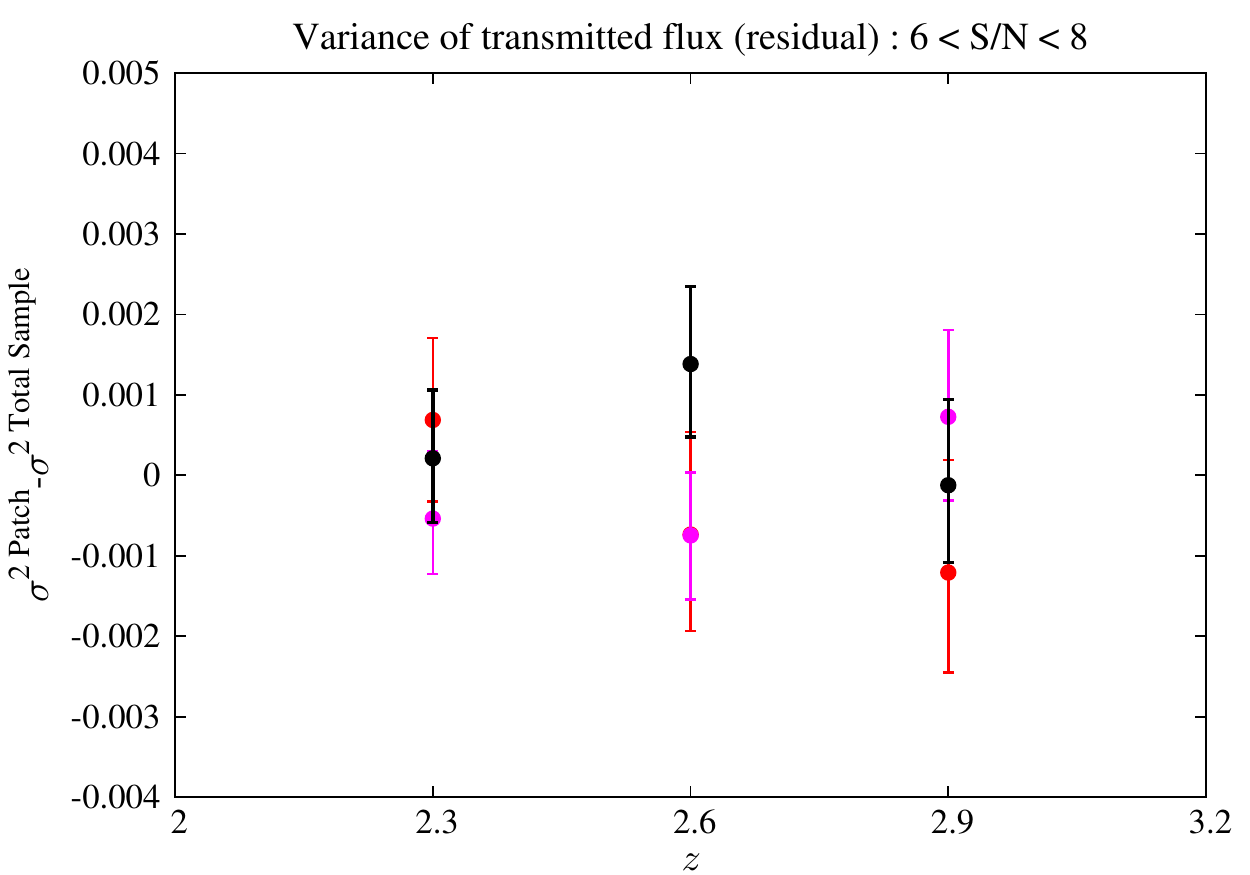}} 
\resizebox{140pt}{110pt}{\includegraphics{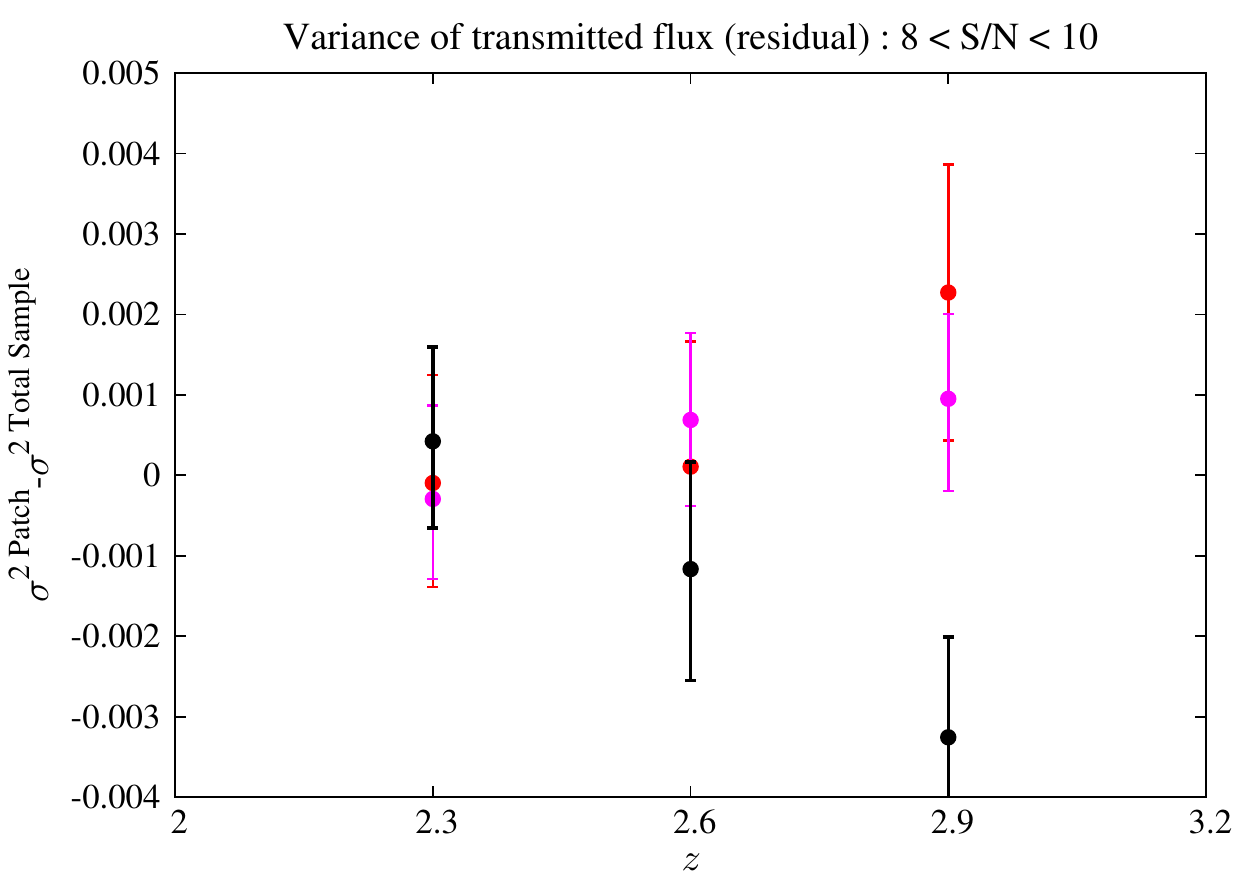}} 
\resizebox{140pt}{110pt}{\includegraphics{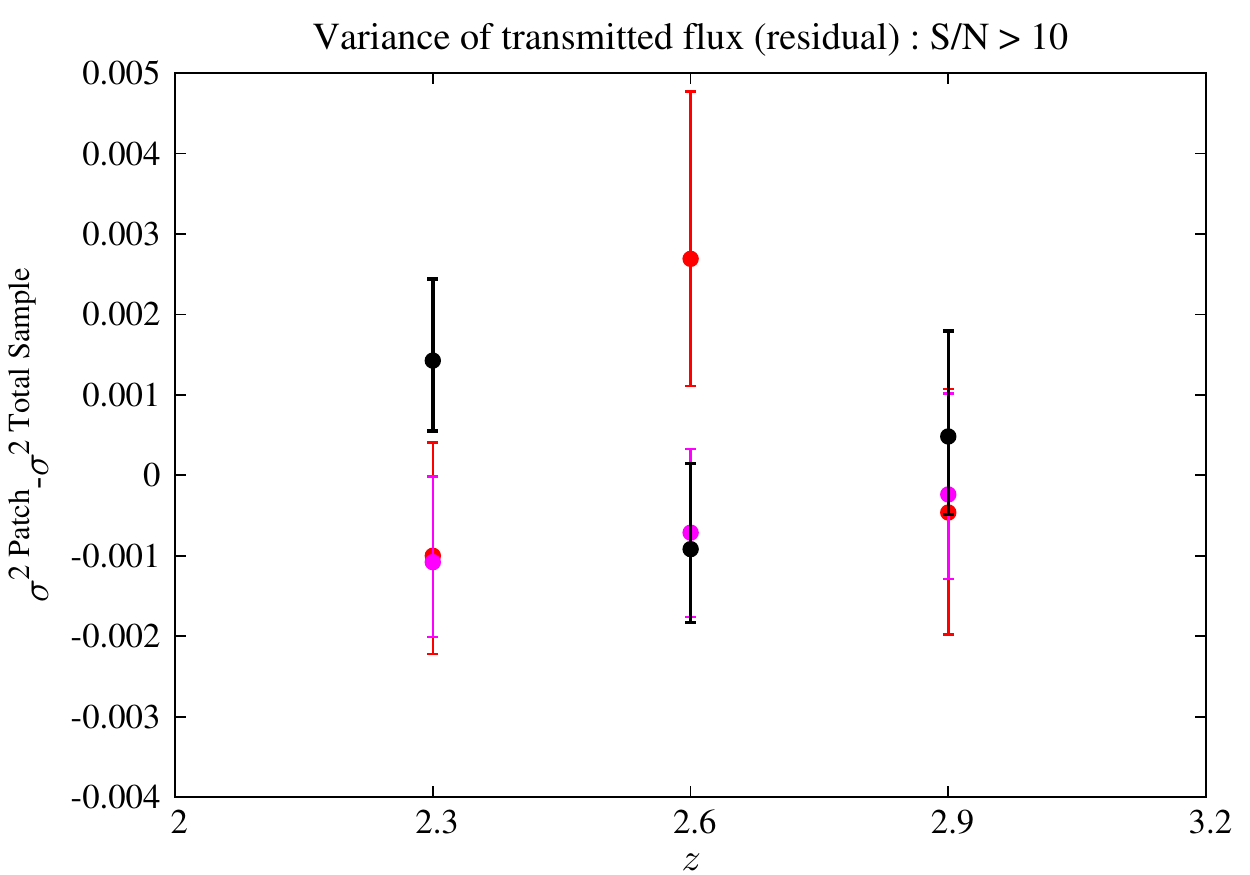}} 

\resizebox{140pt}{110pt}{\includegraphics{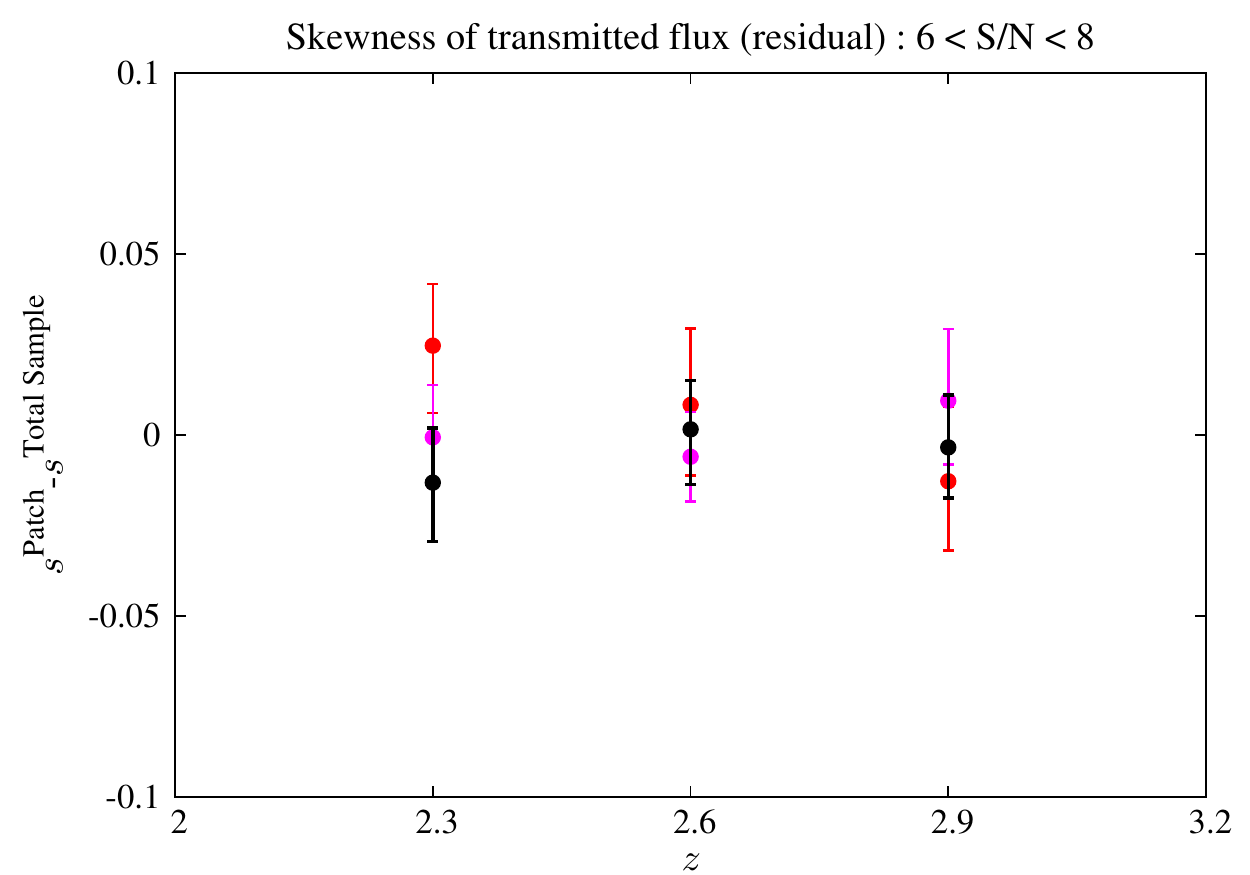}} 
\resizebox{140pt}{110pt}{\includegraphics{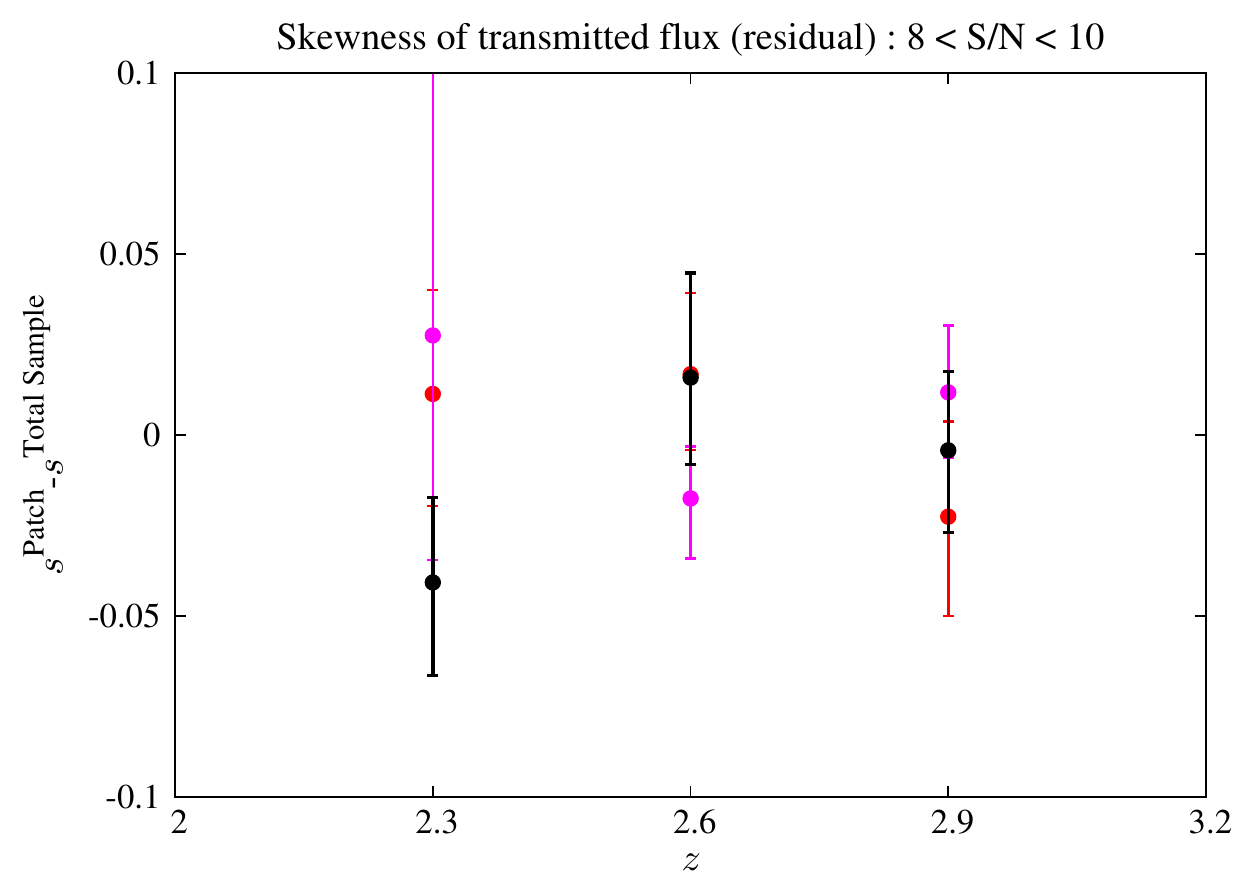}} 
\resizebox{140pt}{110pt}{\includegraphics{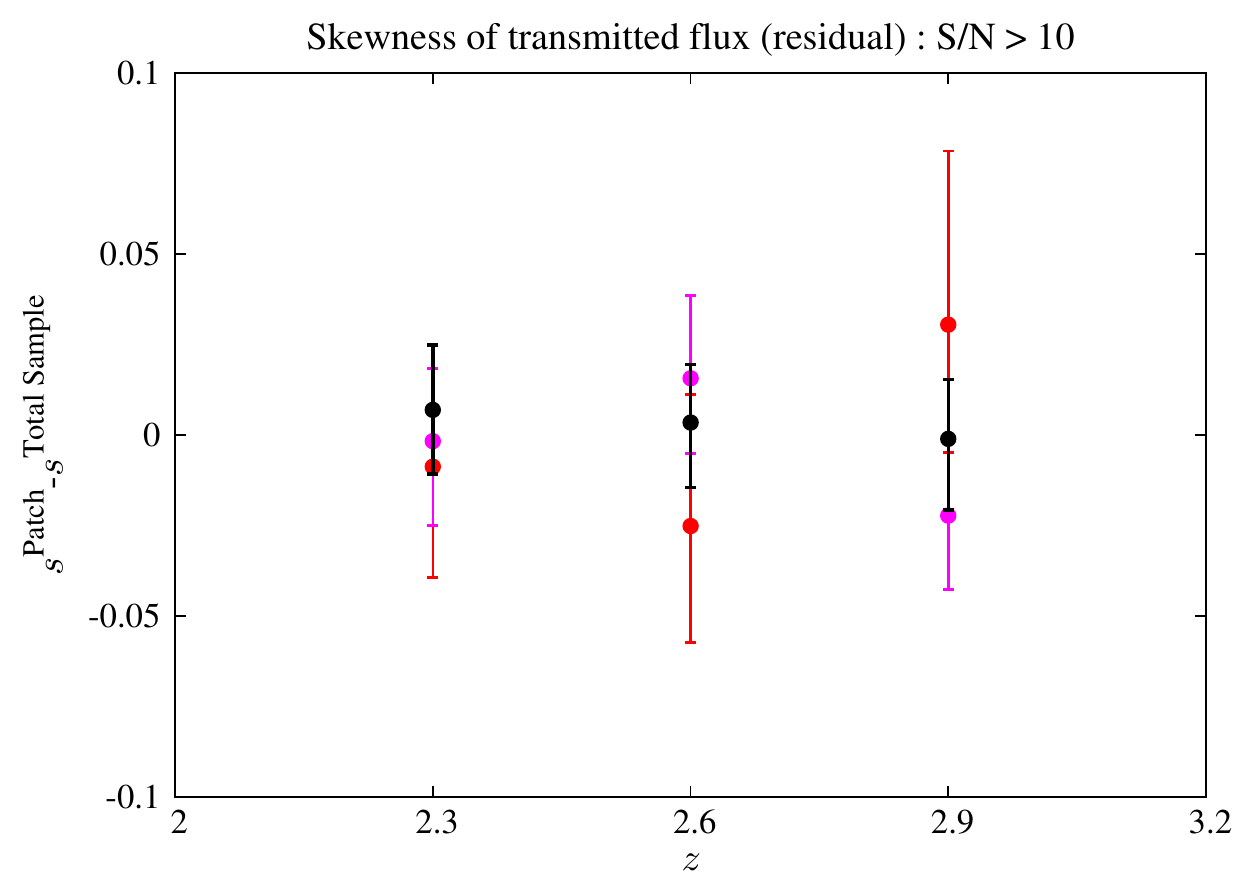}} 

\resizebox{140pt}{110pt}{\includegraphics{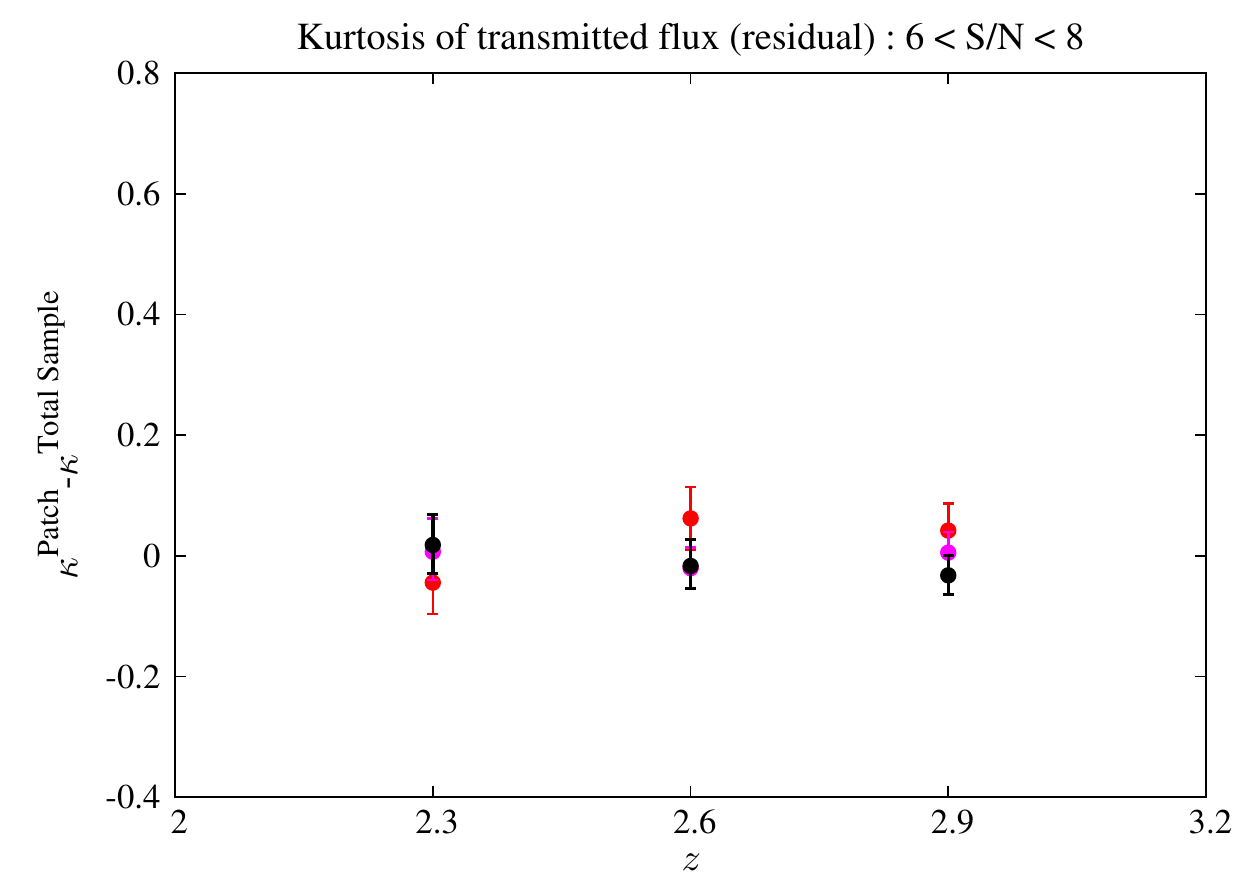}} 
\resizebox{140pt}{110pt}{\includegraphics{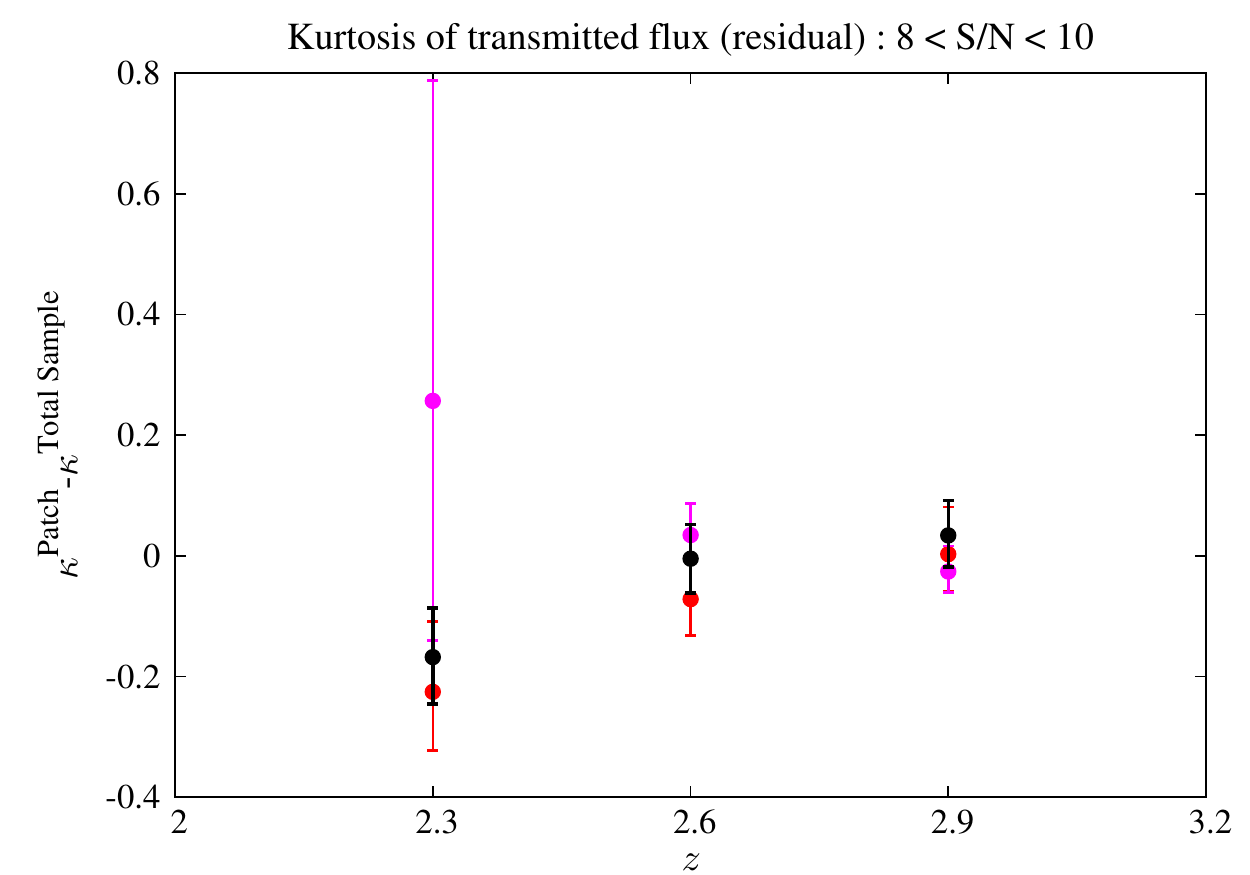}} 
\resizebox{140pt}{110pt}{\includegraphics{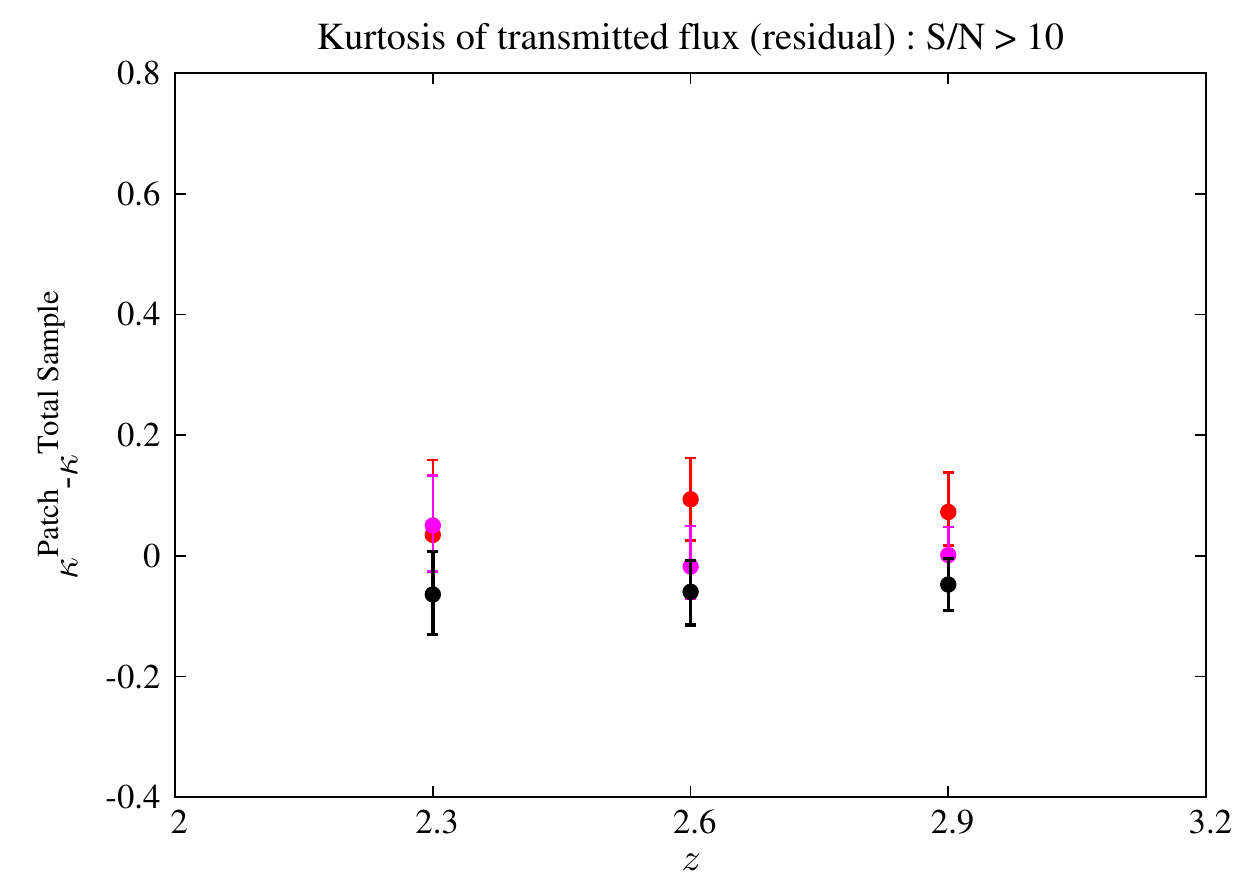}} 
\end{center}
\caption{\footnotesize\label{fig:comparison-2}First few statistical moments of the Lyman-$\alpha$ transmitted flux
PDF for different parts of the sky. For relative comparison we plot the residual moments from the total sample. The 
error bars correspond to the 1$\sigma$ bounds on that particular moment.This figure represents the comparison 
between the properties of the patches for patch selection 2 (right plot of Fig.~\ref{fig:skycut}).}
\end{figure*}

In this paper, we restrict ourselves to calculating up-to the fourth moment, {\it i.e.} till kurtosis of the PDF.
We calculate the mean ($\bar{F}$), the median ($F_{1/2}$), the variance ($\sigma^2$), the skewness ($s$) and 
the kurtosis ($\kappa$) of the flux PDFs in each bin. Since we have noticed in Fig.~\ref{fig:comparison-PDF1} and
~\ref{fig:comparison-PDF2} that the PDFs are very different in different redshifts, we report the residual of such 
quantities obtained in a patch {\it w.r.t.} the complete sample. In Fig.~\ref{fig:comparison-1} we plot the 
5 statistical quantities in residual space for the patch selection 1 and in Fig.~\ref{fig:comparison-2} we plot the 
same quantities for patch selection 2. As we are plotting in residual space, we find the moments are distributed about
the zero. We should mention that here the errors represent the uncertainties of the corresponding statistical 
moments obtained from the ECDFs of the moments. The plotted errors represent 1$\sigma$ uncertainties.

Note that almost all the statistical moments from different patches are comparable to each other within their 
associated uncertainties. We would like to stress that 
we find this consistency throughout redshift $2-3$ and at all SNRs. Hence our results are  {\it consistent with 
the isotropic distribution of neutral hydrogen} and also {\it consistent with isotropic absorption of photons by 
the hydrogen clouds in the IGM at different redshifts}. 

A closer examination shows that almost none of the patches contain any moment which is systematically 
higher/lower than the total sample during the time evolution. Hence the properties
in each patch is preserved during the matter dominated expansion of the Universe. 

We should mention that we have obtained some deviations from isotropy in our analysis. We find the residual moments 
for the blue patch and the red patch (for patch selection 1) do not agree at 2$\sigma$ level in cases. 
For example, we refer to the median mismatch (at $z=2.3$, $6\le{\rm SNR}<8$), the variance mismatches (at $z=2.9$, $6\le{\rm SNR}<8$ and 
at $z=2.6$, ${\rm SNR}\ge10$) and skewness mismatches (at $z=2.9$ at $8\le{\rm SNR}<10$). Similarly, for patch selection 2
we find such deviations at 2$\sigma$ level within the red and the black patches. Since, the number of such deviations are only a few 
and all the moments agree within 3$\sigma$, we do not report any statistically significant deviation. Moreover, 
we should note that due to less number of quasars, the red patch at the southern hemisphere contains the largest 
dispersion in the distribution of the statistical moments which in turn can lead to such fluctuations. Hence, in order to rule out or confirm 
any deviations between the blue and the red patches (or between black and red patches) we need more detection of 
Lyman-$\alpha$ forest in the southern galactic hemisphere. Future surveys will be able to provide 
significantly more quasar spectra with higher sky coverage, which are essential to extend this initiative beyond the 
flux PDF level. 

\clearpage

\section{Conclusions}~\label{sec:conclusions}
As a first approach to test the isotropy in the matter dominated epoch we have used the Lyman-$\alpha$ forest 
data in this paper. In order to remain independent from theoretical assumptions of the IGM, we have used only the 
observational data and compared the statistical properties of the PDF of the observed flux in different directions 
of the sky. Detection of large number high redshift quasar spectra by BOSS has enabled us to perform this test. Though we report   
the distribution of neutral hydrogen is consistent with isotropic Universe during $z\sim 2-3$, making any general claim about the isotropy 
of the Universe requires much more sky coverage.  
As we have mentioned before, our test of anisotropy is {\it partial} due to only 
3275 ${\rm deg}^2$ sky coverage. With a larger sky coverage we can address the precise direction and significance of 
the possible anisotropy, if present. With upcoming observations like e-BOSS~\cite{eboss} and DESI~\cite{desi} we expect to detect 
higher quality of Lyman-$\alpha$ forest data, significantly higher number of quasar spectra with larger sky coverage. Hence,
with the upcoming data the assumption of isotropy can be falsified with higher precision. Cross-correlating the  
data from different surveys will be also important to rule out any systematic effect. 

Any presence of anisotropy in the Lyman-$\alpha$ forest is interesting and it points towards the distribution of neutral
hydrogen, temperature-density relation and few other properties in the IGM. A straightforward extension of this topic 
would be to model the IGM using some semi-analytical modeling or simulations. With the modeling we can address if
the significance of such anisotropy being statistical or physical. Moreover, if any physical anisotropy is found, 
we need to examine the change in the properties of the IGM it refers to and search for the probable cause. 

We would like to conclude by mentioning that the major finding of our analysis is {\it consistency of the data with the isotropic Universe in the 
final stage of matter dominated epoch ($z\sim2-3$), and unfortunately we found no preferred direction in the Lyman-$\alpha$ forest to guide the travelers}. Though 
our result is a partial test, if similar tests on larger survey area 
also confirms the isotropy, that might hint towards the possibility that any late time anisotropy is probably caused by bulk flow and not an 
intrinsic anisotropy in the Universe.


\section{Acknowledgments}
We would like to thank Tapomoy Guha Sarkar for important discussions, suggestions and comments on the manuscripts. We would also 
like to thank Amir Aghamousa, Stephen Appleby, Eric Linder, Pat McDonald, Graziano Rossi and Tirthankar Roy Choudhury for their comments and suggestions. 
We thank Khee-Gan Lee for various clarifications regarding the BOSS analysis of the Lyman-$\alpha$ forest data. 
D.K.H. wish to acknowledge support from the Korea Ministry of Education, Science and Technology, Gyeongsangbuk-Do and Pohang 
City for Independent Junior Research Groups at the Asia Pacific Center for Theoretical Physics. A.S. would like to acknowledge
the support of the National Research Foundation of Korea (NRF-2013R1A1A2013795). We acknowledge
the use of data from SDSS III. Funding for SDSS-III has been provided by the Alfred P. Sloan Foundation, the Participating Institutions, 
the National Science Foundation, and the U.S. Department of Energy Office of Science. The SDSS-III web site is http://www.sdss3.org/.

SDSS-III is managed by the Astrophysical Research Consortium for the Participating Institutions of the SDSS-III Collaboration including the University of 
Arizona, the Brazilian Participation Group, Brookhaven National Laboratory, Carnegie Mellon University, University of Florida, the French Participation Group, 
the German Participation Group, Harvard University, the Instituto de Astrofisica de Canarias, the Michigan State/Notre Dame/JINA Participation Group, 
Johns Hopkins University, Lawrence Berkeley National Laboratory, Max Planck Institute for Astrophysics, Max Planck Institute for Extraterrestrial Physics, 
New Mexico State University, New York University, Ohio State University, Pennsylvania State University, University of Portsmouth, Princeton University, 
the Spanish Participation Group, University of Tokyo, University of Utah, Vanderbilt University, University of Virginia, University of Washington, and 
Yale University. 



\begin{thebibliography}{99}

\bibitem{Hinshaw:1996ut}
  G.~Hinshaw, A.~J.~Banday, C.~L.~Bennett, K.~M.~Gorski, A.~Kogut, C.~H.~Lineweaver, G.~F.~Smoot and E.~L.~Wright,
  Astrophys.\ J.\  {\bf 464} (1996) L25
  [astro-ph/9601061].

\bibitem{Spergel:2003cb}
  D.~N.~Spergel {\it et al.}  [WMAP Collaboration],
  Astrophys.\ J.\ Suppl.\  {\bf 148} (2003) 175
  [astro-ph/0302209].

\bibitem{Copi:2010na}
  C.~J.~Copi, D.~Huterer, D.~J.~Schwarz and G.~D.~Starkman,
  Adv.\ Astron.\  {\bf 2010} (2010) 847541
  [arXiv:1004.5602 [astro-ph.CO]].






\bibitem{de OliveiraCosta:2003pu}
  A.~de Oliveira-Costa, M.~Tegmark, M.~Zaldarriaga and A.~Hamilton,
  Phys.\ Rev.\ D {\bf 69} (2004) 063516
  [astro-ph/0307282].

\bibitem{Abramo:2006gw}
  L.~R.~Abramo, A.~Bernui, I.~S.~Ferreira, T.~Villela and C.~A.~Wuensche,
  Phys.\ Rev.\ D {\bf 74} (2006) 063506
  [astro-ph/0604346].

\bibitem{Land:2005ad}
  K.~Land and J.~Magueijo,
  Phys.\ Rev.\ Lett.\  {\bf 95} (2005) 071301
  [astro-ph/0502237].

\bibitem{Land:2006bn}
  K.~Land and J.~Magueijo,
  Mon.\ Not.\ Roy.\ Astron.\ Soc.\  {\bf 378} (2007) 153
  [astro-ph/0611518].

\bibitem{Rakic:2007ve}
  A.~Rakic and D.~J.~Schwarz,
  Phys.\ Rev.\ D {\bf 75} (2007) 103002
  [astro-ph/0703266].

\bibitem{Samal:2007nw}
  P.~K.~Samal, R.~Saha, P.~Jain and J.~P.~Ralston,
  Mon.\ Not.\ Roy.\ Astron.\ Soc.\  {\bf 385} (2008) 1718
  [arXiv:0708.2816 [astro-ph]].

\bibitem{Samal:2008nv}
  P.~K.~Samal, R.~Saha, P.~Jain and J.~P.~Ralston,
  Mon.\ Not.\ Roy.\ Astron.\ Soc.\  {\bf 396} (2009) 511
  [arXiv:0811.1639 [astro-ph]].


\bibitem{Eriksen:2007pc}
  H.~K.~Eriksen, A.~J.~Banday, K.~M.~Gorski, F.~K.~Hansen and P.~B.~Lilje,
  Astrophys.\ J.\  {\bf 660} (2007) L81
  [astro-ph/0701089].

\bibitem{Hoftuft:2009rq}
  J.~Hoftuft, H.~K.~Eriksen, A.~J.~Banday, K.~M.~Gorski, F.~K.~Hansen and P.~B.~Lilje,
  Astrophys.\ J.\  {\bf 699} (2009) 985
  [arXiv:0903.1229 [astro-ph.CO]].

\bibitem{Copi:2006tu}
  C.~Copi, D.~Huterer, D.~Schwarz and G.~Starkman,
  Phys.\ Rev.\ D {\bf 75} (2007) 023507
  [astro-ph/0605135].

\bibitem{Copi:2005ff}
  C.~J.~Copi, D.~Huterer, D.~J.~Schwarz and G.~D.~Starkman,
  Mon.\ Not.\ Roy.\ Astron.\ Soc.\  {\bf 367} (2006) 79
  [astro-ph/0508047].

\bibitem{Schwarz:2004gk}
  D.~J.~Schwarz, G.~D.~Starkman, D.~Huterer and C.~J.~Copi,
  Phys.\ Rev.\ Lett.\  {\bf 93} (2004) 221301
  [astro-ph/0403353].

\bibitem{Souradeep:2006dz}
  T.~Souradeep, A.~Hajian and S.~Basak,
  New Astron.\ Rev.\  {\bf 50} (2006) 889
  [astro-ph/0607577].

\bibitem{Ade:2013nlj}
  P.~A.~R.~Ade {\it et al.}  [Planck Collaboration],
  arXiv:1303.5083 [astro-ph.CO].

\bibitem{Akrami:2014eta}
  Y.~Akrami, Y.~Fantaye, A.~Shafieloo, H.~K.~Eriksen, F.~K.~Hansen, A.~J.~Banday and K.~M.~Górski,
  Astrophys.\ J.\  {\bf 784} (2014) L42
  [arXiv:1402.0870 [astro-ph.CO]].



\bibitem{Fernandez-Cobos:2013fda}
  R.~Fernández-Cobos, P.~Vielva, D.~Pietrobon, A.~Balbi, E.~Martínez-González and R.~B.~Barreiro,
  Mon.\ Not.\ Roy.\ Astron.\ Soc.\  {\bf 441} (2014) 2392
  [arXiv:1312.0275 [astro-ph.CO]].

\bibitem{Cai:2013lja}
  R.~G.~Cai, Y.~Z.~Ma, B.~Tang and Z.~L.~Tuo,
  Phys.\ Rev.\ D {\bf 87} (2013) 12,  123522
  [arXiv:1303.0961 [astro-ph.CO]].



\bibitem{Keenan:2009jh} 
  R.~C.~Keenan, L.~Trouille, A.~J.~Barger, L.~L.~Cowie and W.~-H.~Wang,
  Astrophys.\ J.\ Suppl.\  {\bf 186}, 94 (2010)
  [arXiv:0912.3090 [astro-ph.CO]].


\bibitem{Keenan:2012gr}
  R.~C.~Keenan, A.~J.~Barger, L.~L.~Cowie, W.~-H.~Wang, I.~Wold and L.~Trouille,
  Astrophys.\ J.\  {\bf 754} (2012) 131
  [arXiv:1207.1588 [astro-ph.CO]].



\bibitem{Keenan:2013mfa}
  R.~C.~Keenan, A.~J.~Barger and L.~L.~Cowie,
  Astrophys.\ J.\  {\bf 775} (2013) 62
  [arXiv:1304.2884 [astro-ph.CO]].

\bibitem{Whitbourn:2013mwa}
  J.~R.~Whitbourn and T.~Shanks,
  arXiv:1307.4405 [astro-ph.CO].



\bibitem{Frith:2003tb}
  W.~J.~Frith, G.~S.~Busswell, R.~Fong, N.~Metcalfe and T.~Shanks,
  Mon.\ Not.\ Roy.\ Astron.\ Soc.\  {\bf 345} (2003) 1049
  [astro-ph/0302331].
\bibitem{Busswell:2003ta}
  G.~S.~Busswell, T.~Shanks, P.~J.~Outram, W.~J.~Frith, N.~Metcalfe and R.~Fong,
  [astro-ph/0302330].

\bibitem{Frith:2005az}
  W.~J.~Frith, P.~J.~Outram and T.~Shanks,
  Mon.\ Not.\ Roy.\ Astron.\ Soc.\  {\bf 364} (2005) 593
  [astro-ph/0507215].

\bibitem{Frith:2004wd}
  W.~J.~Frith, T.~Shanks and P.~J.~Outram,
  Mon.\ Not.\ Roy.\ Astron.\ Soc.\  {\bf 361} (2005) 701
  [astro-ph/0411204].

\bibitem{Frith:2004tw}
  W.~J.~Frith, P.~J.~Outram and T.~Shanks,
  [astro-ph/0408011].
\bibitem{Appleby:2014lra} 
  S.~Appleby and A.~Shafieloo,
  JCAP {\bf 1410}, no. 10, 070 (2014)
  [arXiv:1405.4595 [astro-ph.CO]].

\bibitem{fdelta} 
  L.~Hui and N.~Y.~Gnedin,
  Mon.\ Not.\ Roy.\ Astron.\ Soc.\  {\bf 292}, 27 (1997)
  [astro-ph/9612232];
  N.~Y.~Gnedin and L.~Hui,
  Mon.\ Not.\ Roy.\ Astron.\ Soc.\  {\bf 296}, 44 (1998)
  [astro-ph/9706219, astro-ph/9706219];
  P.~McDonald, J.~Miralda-Escude, M.~Rauch, W.~L.~W.~Sargent, T.~A.~Barlow and R.~Cen,
  Astrophys.\ J.\  {\bf 562}, 52 (2001)
  [Astrophys.\ J.\  {\bf 598}, 712 (2003)]
  [astro-ph/0005553].
\bibitem{sdss}
See, {\tt http://www.sdss3.org/}
\bibitem{boss}
See, {\tt https://www.sdss3.org/surveys/boss.php}
  
\bibitem{Lyman-sample} 
  K.~G.~Lee, S.~Bailey, L.~E.~Bartsch, W.~Carithers, K.~S.~Dawson, D.~Kirkby, B.~Lundgren and D.~Margala {\it et al.},
  arXiv:1211.5146 [astro-ph.CO].

  \bibitem{Leeetal} 
  K.~G.~Lee, N.~Suzuki and D.~N.~Spergel,
  Astron.\ J.\  {\bf 143}, 51 (2012)
  [arXiv:1108.6080 [astro-ph.CO]].
  
\bibitem{bossigm} 
  K.~G.~Lee, J.~P.~Hennawi, D.~N.~Spergel, D.~H.~Weinberg, D.~W.~Hogg, M.~Viel, J.~S.~Bolton and S.~Bailey {\it et al.},
  Astrophys.\ J.\  {\bf 799}, no. 2, 196 (2015)
  [arXiv:1405.1072 [astro-ph.CO]].
\bibitem{Busca:2012bu} 
  N.~G.~Busca, T.~Delubac, J.~Rich, S.~Bailey, A.~Font-Ribera, D.~Kirkby, J.~M.~Le Goff and M.~M.~Pieri {\it et al.},
  Astron.\ Astrophys.\  {\bf 552}, A96 (2013)
  [arXiv:1211.2616 [astro-ph.CO]].

\bibitem{eboss}
See, {\tt https://www.sdss3.org/future/eboss.php}
\bibitem{desi}
See, {\tt http://desi.lbl.gov}
  
\end{thebibliography}
\end{document}